\newcommand\nodata{{...} }

\newcommand\etal{et~al.}

\newcommand\mdot{M$_{\odot}$~}
\newcommand\kms{\ifmmode {\rm\,km\,s^{-1}}\else${\rm\,km\,s^{-1}}$\fi}

\def\spose#1{\hbox to 0pt{#1\hss}}
\newcommand\simlt{\mathrel{\spose{\lower 3pt\hbox{$\mathchar"218$}}
     \raise 2.0pt\hbox{$\mathchar"13C$}}}
\newcommand\simgt{\mathrel{\spose{\lower 3pt\hbox{$\mathchar"218$}}
     \raise 2.0pt\hbox{$\mathchar"13E$}}}
\newcommand\aap{{\em A\&A}}

\newcommand\apjl{{\em ApJ}}
\newcommand\apjs{{\em ApJS}}

\documentclass[usenatbib,12pt]{mn2e}
\usepackage{rotating}
\usepackage{multicol}
\usepackage{rotating}
\usepackage{amssymb}
\usepackage{amsmath}
\usepackage{times}
\title[Distant radio galaxies from SUMSS and NVSS]{A search for distant radio galaxies from SUMSS and NVSS: III. radio spectral energy distributions  \& the $z-\alpha$ correlation}
\author[I.~J. Klamer \etal]{
\parbox[t]{\textwidth}{
Ilana J. Klamer$^{1,2}$, Ron D. Ekers$^{2}$, Julia J. Bryant$^1$, Richard W.\ Hunstead$^1$, Elaine M.\ Sadler$^1$, Carlos De Breuck$^3$ }
\vspace*{6pt} \\ 
$^1$ School of Physics, University of Sydney, NSW 2006, Australia.\\
$^2$ CSIRO Australia Telescope National Facility, P.O. Box 76, Epping, NSW 1710, Australia.\\
$^3$ European Southern Observatory, Karl Schwarzschild Stra\ss e 2, D-85748 Garching, Germany.\\
}
\pubyear{2005}
\begin{document}
\maketitle

\begin{abstract}

This is the third in a series of papers that present observations and results for a sample of 76 ultra-steep-spectrum radio sources designed to find galaxies at high redshift. Here we present multi-frequency radio observations, from the Australia Telescope Compact Array, for a subset of 37 galaxies from the sample. Matched resolution observations at 2.3, 4.8 and 6.2~GHz are presented for all galaxies, with the $z<2$ galaxies additionally observed at 8.6 and 18~GHz. New angular size constraints are reported for 19 sources based on high resolution 4.8 and 6.2~GHz observations. Functional forms for the rest-frame spectral energy distributions are derived: 89\% of the sample is well characterised by a single power law, whilst the remaining 11\% show some flattening toward higher frequencies: not one source shows any evidence for high frequency steepening. We discuss the implications of this result in light of the empirical correlation between redshift and spectral index seen in flux limited samples of radio galaxies. Finally, a new physical mechanism to explain the redshift -- spectral index correlation is posited: extremely steep spectrum radio galaxies in the local universe usually reside at the centres of rich galaxy clusters. We argue that if a higher fraction of radio galaxies, as a function of redshift, are located in environments with densities similar to nearby rich clusters, then this could be a natural interpretation for the correlation. We briefly outline our plans to pursue this line of investigation. 
\end{abstract}


\section{Introduction}\label{intro}

\indent Until the late 1970s the search for high redshift radio galaxies (HzRGs) was restricted to optical spectroscopy at the position of a radio source whose optical host galaxy is undetected on, or at the detection limit of, the Palomar Observatory Sky Survey (POSS) plates (e.g. \citealp{kri74,kri78,smi80,gun81,lai83,spin84,per84,spin85}). Indeed, the first galaxy discovered above redshift one was found in this way \citep{spin82}.\newline 

 The field was revolutionised when it was realised that the radio sources without optical counterparts on the POSS plates tended to have the steepest spectral indices $\alpha$ (where $S_{\nu}\propto \nu^{\alpha}$) measured between 178 and 1415~MHz \citep{tie79,blu79}. Since then most radio selected samples designed to find HzRGs have used some degree of spectral index selection. \newline

 Three explanations have been put forward in the literature to justify the trend for HzRGs to have steeper spectral indices; the so-called $z-\alpha$ correlation. The first is based on the observation that the spectral energy distributions (SEDs) of radio galaxies typically steepen toward higher frequencies. For a fixed set of observing frequencies, a radio galaxy at higher redshift will be sampled at higher rest-frame frequencies where, on average, it will exhibit progressively steeper spectral indices: a k-correction. It has also been suggested that the SEDs of HzRGs will be intrinsically steeper thanks to enhanced inverse Compton (IC) losses of the relativistic electron population against the cosmic microwave background (CMB) photons whose energy density increases as $(1+z)^4$ \citep{kro91}. These two explanations have been adopted in the literature over recent years as the main contributors to the correlation \citep{cdb00sample,cdb02wish,ped03,coh04,jar04}. The third explanation holds that the $z-\alpha$ correlation itself is indirect, reflecting an intrinsic correlation between radio luminosity and spectral index coupled to a Malmquist bias \citep{blu99a}. Regardless of the underlying reason for the correlation, the steep-spectrum technique has enjoyed tremendous success over the years, with at least 30 radio galaxies above $z=3$ found to date using some form of spectral index culling.\newline

In the first paper of this series (paper I; \citealt{cdb04}), we defined a sample of 76 high redshift radio galaxy candidates selected between 843 and 1400~MHz on the basis of their ultra steep radio spectral index (USS; $\alpha\leq-1.3$). In the second paper of the series (paper II; \citealt{cdb06}) we reported the results of optical spectroscopy for 52 sources in the sample which included 35 spectroscopic redshifts. The aim of this paper III is to better understand the radio spectral characteristics of our sample of USS radio galaxies, over a broad range of redshifts. To this end, we report multi-frequency radio observations from the Australia Telescope Compact Array (ATCA) and present rest-frame radio SEDs for 37 galaxies in the sample. \newline

This paper is organised as follows. Some relevant synchrotron physics is given in \S\ref{synchphysics}. The subset of SUMSS-NVSS USS radio galaxies we observed for this paper is defined in \S\ref{sourceselection}. Multi-frequency ATCA observations and data analysis follow in \S\ref{ATCAobservations}. The results of the observational program are presented in \S\ref{ATCAresults}, which includes rest-frame radio SEDs. The remainder of the paper is discussion based. \S\ref{discussion} and \S\ref{zedalphadiscussion} focus on the implications of our results for existing interpretations of the redshift -- spectral index correlation. Penultimately, in \S\ref{neighbours}, we speculate on an alternative physical mechanism for this correlation and concluding remarks are given in \S\ref{summary}. Throughout this paper, we assume a flat $\Lambda$-dominated cosmology with $H_0=71$~\kms Mpc$^{-1}$ and $\Omega_{\Lambda}=0.73$ \citep{spe03}. 

\section{Synchrotron Radiation}\label{synchphysics}
A relativistic electron spiralling around a magnetic field line emits synchrotron radiation expressed as:
\begin{equation}
\nu\,=\gamma^2\frac{eB}{2\pi m_o}
\label{1}
\end{equation}
Equation~\ref{1}, which relates the specific frequency $\nu$ of the synchrotron radiation for relativistic particles with rest mass $m_o$ in a magnetic field $B$ with a Lorentz factor $\gamma$, is the fundamental relationship for synchrotron radiation processes in astrophysics. The emission from an ensemble of electrons with energies between $E$ and $E + dE$ can be approximated by a delta function at a frequency given by Equation~\ref{1}. The synchrotron emissivity is:
\begin{equation}\begin{split}
\epsilon_\nu&= 4\pi j_\nu \\           
            &=\int \delta(x-E') P_\nu\, N(E)\, dE\, dx \\
	    &= P_\nu\, N(E')\, dE'
\end{split}
\label{2}
\end{equation} 
where $P_\nu$ is the power per unit frequency emitted by one electron from an ensemble of $N$ electrons per unit volume per unit solid angle with energies in the range from $E'$ to $E' + dE'$. $P_\nu$ can be determined from the relativistic generalization of the Larmor formula which describes the total power $P$ radiated by a point charge {\it{in vacuo}}:
\begin{equation}\begin{split}
P&=4\,\pi\,\int P_\nu\,d\nu\\
&=\frac{\mu_0 q^2 \gamma^6}{6 \pi c}(a^2 - |\frac{\vec{\pmb{v}} \times \vec{\pmb{a}}}{c}|^2)
\label{3}\end{split}\end{equation}
where $a$ is the acceleration and $\mu_0$ is the permeability of free space. For circular motion where $\alpha$ is the angle between the velocity and acceleration vectors:
\begin{equation}
P=\frac{c \sigma_T B^2\,\gamma^2}{\mu_0} \rm sin^2\,\alpha
\label{4}\end{equation}
where $\sigma_T$ is the Thomson scattering cross-section for electrons. The observed frequency spectrum of synchrotron emission is often characterized by a power law. This spectrum is interpreted as originating from a distribution of relativistic electrons with a power law energy distribution of the form: 
\begin{equation}
N(E) dE = N_0 E^{\Gamma} dE
\label{5}\end{equation}
where $\Gamma$ is a constant power law index and $N_0$ is the total electron number density. Substituting Equations~\ref{4} and \ref{5} into \ref{2}, and assuming an isotropic distribution of electrons so that ${\rm \langle \rm sin^2\alpha\rangle=\frac{2}{3}}$, the power per unit frequency per unit solid angle averaged over an emitting volume $V$ is given by:
\begin{equation}
\epsilon_\nu\;=\; \frac{m c^2}{6\pi}\; c \sigma_T \;\frac{B^2}{2 \mu_0} \; \frac{\gamma^3}{\nu} \; N_0 E^{\Gamma} \; V\, ,
\label{6}
\end{equation}
 with the relationship between $\gamma$, $B$ and $\nu$ being given by Equation~\ref{1}. Also, since $E=\gamma m_o c^2$, then from Equation~\ref{1} we can see that $E^{\Gamma} \propto \nu^{\frac{\Gamma}{2}}$. Therefore, the radiated power at a specified observing frequency obeys the proportionality:
\begin{equation}
P_\nu \propto \nu^{\alpha}, \; \; \alpha = \frac{\Gamma + 1}{2}
\label{7}\end{equation} 
where we define $\alpha$ to be the spectral index of the synchrotron radiation. 
\subsection{Electron Energy Losses}\label{synchtheor3}
We now consider energy losses to the initial electron energy distribution, which lead to a departure from Equation~\ref{5}. The general form of the energy loss $\varphi(E)$ can be represented by:
\begin{equation}
\varphi(E)\; = \; -\zeta -\eta E - \xi E^2
\label{8}\end{equation}
where $\zeta$ represents the electron energy losses which are essentially independent of $E$, $\eta$ represents the losses proportional to $E$ and $\xi$ the losses proportional to $E^2$. We consider synchrotron and inverse Compton cooling which contribute to $\xi$ and adiabatic losses which contribute to $\eta$. 
\subsubsection{Synchrotron Losses: the $\xi$ term}\label{synchlosses}
In a constant ambient magnetic field the radiating electrons will lose energy at a rate which is given by Equation~\ref{4}. The timescale for an electron with initial energy $E_2=\gamma_2 m_0\,c^2$ to lose energy to its final state $E_1=\gamma_1 m_0\,c^2$ is then
\begin{equation}\begin{split}
\tau&=\int_{E_1}^{E_2} \frac{1}{P}\; dE\\
    &=\frac{\mu_0}{c\;\sigma_T\;B^2}\; ( \frac{1}{\gamma_1}\; -\; \frac{1}{\gamma_2} )\:. \\
\label{9}\end{split}\end{equation}
In a magnetic field, this electron will radiate at continuously lower frequencies over this timescale (Equation~\ref{1}). For a continuous injection model \citep{kar62,car91}, the initial electron energy distribution will steepen from $-\Gamma$ to $-(\Gamma+1)$ above a critical, or `break' energy. This will result in the associated synchrotron SED remaining a power law but with a spectral steepening from $\alpha$ to $\alpha-\frac{1}{2}$ above a critical frequency which continuously decreases with increasing time. In the limiting case where there was a single injection of electrons, then the SED steepens exponentially when $t>\tau$. 
\subsubsection{Inverse Compton Scattering: the $\xi$ term}\label{IClosses}
If the energy density of the radiation field is high enough, the synchrotron emitting electrons will collide with the ambient photons, suffering energy losses as they impart their momentum to the radiation field. As for synchrotron losses, the rate of energy loss is again proportional to the square of the electron energy, meaning that the resultant synchrotron spectral index will again steepen from $\alpha$ to $\alpha-\frac{1}{2}$. The ambient photon field consists of the synchrotron photons themselves (in which case the process is called synchrotron self-Compton, SSC), the cosmic microwave background (CMB) photons, and the infra-red (IR) photons from the AGN and galaxy. Thus, considering inverse Compton cooling as well, Equation~\ref{9} becomes:
\begin{equation}\begin{split}
\tau&=\int_{E_1}^{E_2} \frac{1}{P}\; dE\\
    &=\frac{-\mu_0\; m_o\; c^2}{c\;\sigma_T\;(B^2+B_{\rm \textsc{ic}}(z)^2)}\int_{E_1}^{E_2} \frac{1}{E^2}\; dE\\
    &=\frac{\mu_0}{c\;\sigma_T\;(B^2+B_{\rm \textsc{ic}}(z)^2)}\; ( \frac{1}{\gamma_1}\; -\; \frac{1}{\gamma_2} ) \: ,\\
\label{10}\end{split}\end{equation}
\noindent where $B_{\rm \textsc{ic}}(z)$ is the equivalent inverse Compton magnetic field at redshift $z$. The ratio of the magnetic energy density 
\begin{equation}
U_B\;=\;\frac{B^2}{2\mu_0}
\label{11}\end{equation}
to the radiation energy density:
\begin{equation}\begin{split}
 U_{\rm rad} &= U_{\rm \textsc{cmb}}(z)\, +\, U_{\rm \textsc{ssc}}\, +\, U_{\rm \textsc{ir}}\\
             &= 4.1\; \times 10^{-13}(1 + z)^4 \, + \, U_{\rm \textsc{ssc}}\, + \, U_{\rm \textsc{ir}}, \\
\label{12}\end{split}
\end{equation}
in units of erg~cm$^{-3}$, will determine whether synchrotron losses or inverse Compton losses dominate the electron energy loss spectrum. 
\subsubsection{Adiabatic Expansion Losses: the $\eta$ term}\label{adiabaticlosses}

We follow closely the derivation of \citet{sch68}. An initial energy distribution of electrons described by $N(E_i)$ in a volume of radius $R_i$, undergoing adiabatic expansion to a volume of radius $R_f$ is governed by the invariance of $TV^{\Gamma-1}$, where $T$ is the temperature (energy), $V$ is the volume and $\Gamma=\frac{4}{3}$ is the usual adiabatic index for relativistic gas. Then,
\begin{equation}
N(E_f) = N(E_i)\;(\frac{R_i}{R_f})^{3(\Gamma-1)}, \\
\label{13}\end{equation}
where $N(E_f)$ is the energy distribution after expansion. Therefore, under adiabatic expansion, $N(E_f)$ will decrease relative to $N(E_i)$ by some linear expansion factor $l$. In the absence of tangled magnetic fields, conservation of magnetic flux (the third adiabatic invariant) gives,
\begin{equation}\begin{split}
B_i\; R_i^2 = B_f \; R_f^2.
\end{split}\label{14}\end{equation}
From Equations~\ref{1}, \ref{13} and \ref{14}, the synchrotron SED will be shifted to lower frequencies by a factor of $l^{4}$. In addition, the radiated power of an electron, as given by Equation~\ref{4}, is proportional to $\gamma^2\;B^2$. If the frequency of this radiation was fixed, then the amplitude of the radiated power would decrease by $l^{6}$, because $B^2$ decreases by $l^4$ and $\gamma^2$ decreases by $l^2$. However, since the frequency of the radiation decreases as $l^{4}$ (from Equation~\ref{1} where $B$ decreases by $l^2$ and $\gamma^2$ still decreases by $l^2$), then the actual radiated power at a fixed emission frequency decreases by $l^{2}$. This is a somewhat over-simplified model which we use to make the point that the flux density, at a given frequency, from the lobes of a radio galaxy will decrease over time at a rate which is governed by the pressure of the environment they are expanding into. Thus, a radio galaxy selected from a given flux-limited survey, whose lobes remain pressure-confined by a relatively high ambient gas density, may fall below the survey flux limit if it was, instead, subjected to larger adiabatic losses expanding into a more rarefied environment. For more comprehensive and realistic models of classical radio galaxies, we refer the reader to \citet{blu99a} and references therein.
\section{Target Selection}\label{sourceselection}
We selected 37 (of the 76) SUMSS-NVSS USS radio galaxies for follow-up multi-frequency radio observations. This subset consists of the 28 galaxies whose redshifts had been spectroscopically confirmed by 2004 September, and another nine galaxies without confirmed redshifts. Seven of the latter group have extremely faint host galaxies ($K\geq19.88$) and are likely to be at redshifts $z\geq4$ according to the $K-z$ relation for radio galaxies \citep{cdb02a,wil03a}. The final two sources without redshifts were included in the subset purely due to their favourable hour angle during the ATCA observations when extra observing time became available. The $K-z$ relation from \citealt{cdb02a} is used to estimate the redshifts of the nine galaxies without spectroscopically confirmed redshifts. 

\subsection{Revised 843\,MHz flux densities}\label{revised843}
Whilst determining the SEDs which are presented in \S\ref{SEDfits}, we noticed that almost all flux densities were well defined by a single power law except for those at 843~MHz from the SUMSS catalogue \citep{mau03} which were systematically larger than expected.  Re-measurement of these fluxes using the SUMSS images showed that the catalogued values were high by a few percent.\newline

The reason for this over-estimation of the flux density is now
understood. It {\it only} occurs
at the very northern edge of the SUMSS catalogue, which is precisely
where our USS sample has been selected, and arises from an incorrect
assumption inherent in the source fitting
algorithm (VSAD) which is optimised for a circular beam.   This error is
exacerbated for our sample because
the USS selection favours sources
with the largest flux over-estimate at 843MHz. On average, the
over-estimation is at the 4\% level and will be corrected in the next
release of the SUMSS catalogue. \newline


We have manually inspected the relevant SUMSS mosaics and revised the 843\,MHz flux densities for the 37 sources considered in this paper. The revised flux densities are listed in Column~5 of Table~2. A consequence of these adjustments is that several of the original USS sources now have spectral indices flatter than the $\alpha_{843}^{1400}=-1.3$ USS threshold. \newline

\section{Observations and data reduction}\label{ATCAobservations}

\begin{table*}
\caption{Journal of the multi-frequency radio observations obtained with the ATCA}
\label{ATCAjournal}
\begin{center}
\begin{tabular}{lllcccc}\hline
\multicolumn{1}{c}{Array} &\multicolumn{1}{c}{Epoch} & \multicolumn{1}{c}{Frequency 1}& \multicolumn{1}{c}{Frequency 2}& \multicolumn{1}{c}{Field of View$^{\dagger}$} & \multicolumn{1}{c}{Beam Size$^{\dagger}$}  \\
\multicolumn{1}{c}{}&\multicolumn{1}{c}{} & \multicolumn{1}{c}{GHz}& \multicolumn{1}{c}{GHz} & \multicolumn{1}{c}{arcmin} & \multicolumn{1}{c}{arcsec$\times$arcsec} \\
\hline
6A & 2004 December 10-12 & 4.800 & 6.208 & 9.9& $3\times6$\\
1.5A & 2005 April 16 & 2.368 &2.496 & 20.1 & $17\times32$\\
750A & 2005 April 29-30 & 4.800 & 6.208 & 9.9 & $17\times31$ \\
EW367 & 2005 June 12-14 & 8.568& 8.626& 5.5&$21\times39$\\ 
H168 & 2005 May 10-12 & 17.856 & 18.496 & 2.7 & $15\times15$ \\
\hline
\end{tabular}

$\dagger$Field of View (half power width of the primary beam) and Beam Size are quoted for Frequency 1.
\end{center}
\end{table*}
We used the ATCA to measure accurate radio positions and morphologies for the 16 highest redshift candidates in the sample: these included galaxies with (i) spectroscopic redshifts $z\geq2$, (ii) redshifts predicted from the $K-z$ relation. These sources are marked with an asterisk ($^*$) in Column~1 of Table~2. In order to measure total flux densities for all 37 sources in our sample, we further obtained matched angular resolution observations in the 2.4~GHz, 5~GHz, 8.6~GHz and 18~GHz bands; a journal of the radio observations is given in Table~\ref{ATCAjournal}. In all cases, we utilised the standard continuum correlator configuration which has dual frequency mode with two independent 128\,MHz bands (frequency 1 and 2 in Table~\ref{ATCAjournal}). For all frequencies, the flux density scale was determined by observations of the standard ATCA primary calibrator PKS~B1934--638. We observed PKS~B1921--293 to correct for the instrumental bandpass, and a suite\footnote{PKS~B1954--388, PKS~B2058--425, PKS~B2211--388, PKS~B2255--282, PKS~B0010--401, PKS~B0104--408, PKS~B0153--410, PKS~B0220--349} of compact bright active galactic nuclei (AGN) located in close proximity to the science targets to track variations in the atmospheric phase during the observations. 
\subsection{A note on observing strategy}\label{obserstrat}
We observed our sources using multiple short observations, making between four and eight snapshots per source spread over a range of hour angles. With such a large number of targets, telescope slewing time becomes appreciable so that some thought was given to the trade-off between optimising the uv-coverage and maximising the on-source time. The relatively large field of view at 2.3~GHz, and to a lesser extent at 4.8~GHz, results in many confusing sources within the primary beam. At these frequencies, the flux densities of the targets are still relatively bright and our experience has shown that it is optimal to maximise the uv-coverage to minimise effects of confusion. Therefore, for frequencies of 4.8~GHz and below, targets were observed for three minutes each and we observed a phase calibrator every 45 minutes. On the other hand, the number of confusing sources in the primary beam becomes negligible at frequencies above 6.2~GHz, but the target sources are now extremely faint and the atmospheric phases less stable. For these observations, we sacrificed uv-coverage for sensitivity by making fewer snapshots with longer integration times (10-15 minutes) and we observed a phase calibrator every 30 minutes. 
 
\subsection{Data reduction}
We employed the \textsc{miriad} radio interferometry data reduction package \citep{miriad} and used the same standard procedure to reduce all the observations. Initially, all data corrupted by radio frequency (RF) interference were removed; this was mostly a problem at 2.496~GHz, and to a lesser extent at 2.368~GHz. At 17.856~GHz and 18.496~GHz, we removed data when the rms path length through the atmosphere (measured from a two-element interferometer tracking a geo-stationary satellite) was sufficient to cause decorrelation during the individual scans (approximately 15\% of the time). The bandpass was determined along with the absolute flux density scale, and both solutions were applied to the secondary calibrators whose complex gains and phases were determined and in turn applied to the target sources. Continuum imaging was performed using multi-frequency synthesis \citep{sau94} prior to deconvolution. \newline 

As mentioned in \S\ref{obserstrat}, we were mostly limited by sidelobe confusion at frequencies below 6.2~GHz and limited by sensitivity above 6.2~GHz. We used images of the associated NVSS \citep{con98} field obtained from its online image server to constrain the deconvolution algorithm to the `real' sources in the field (as opposed to spurious sources due to phase errors or bright sidelobes), and accordingly restricted the deconvolution algorithm to tight regions surrounding each source. The sensitivity of the NVSS was sufficient to apply this constraint without biasing our results. This process is extremely time-consuming to do by-hand, and can also be difficult to reproduce systematically. As a result, an automated procedure has now been defined to select deconvolution (CLEAN) regions in ATCA images using the known positions of radio sources in the field. This is analogous to the \textsc{faces} task which was independently implemented into the \textsc{aips} data analysis package several years ago. \newline 

When the science target was unresolved, based on inspection of the
deconvolved images and the visibilities, and when there were no
confusing sources in the primary beam (usually at frequencies larger than
6.2 GHz), we determined the total flux density of the source from the peak of the `dirty' map, which is more accurate than the deconvolved value.
Otherwise, we measured total integrated flux densities from the deconvolved
images.

\section{Results}\label{ATCAresults}
Table~2 gives the results of these observations:\newline

\noindent {\bf Column~1:} NVSS source name. \newline
{\bf Column~2:} Redshift from paper~II.\newline
{\bf Column~3:} 1.4~GHz radio luminosity $L_{1.4}$, calculated using \newline
\begin{equation}
\frac{L_{1.4}}{\rm W~m^{-2}}=4\pi\,\frac{D_L^2(z)}{\rm cm^2}\,\frac{S_{1.4}}{\rm \mu Jy}\,10^{-36}(1+z)^{-(1+\alpha)},
\end{equation}
where $D_L(z)$ is the luminosity distance and $S_{1.4}$ is the flux density measured at an observing frequency of 1.4\,GHz.\newline

\noindent {\bf Column~4:} Spectral index between 843~MHz and 1.4~GHz.\newline
{\bf Column~5:} Total flux density measured at 843\,MHz.\newline
{\bf Column~6:} Total flux density at 1.4\,GHz from NVSS \citep{con98}.\newline
{\bf Column~7:} Total flux density measured at 2.4\,GHz.\newline
{\bf Column~8:} Total flux density measured at 4.8\,GHz.\newline
{\bf Column~9:} Total flux density measured at 6.2\,GHz.\newline
{\bf Column~10:}Total flux density measured at 8.6\,GHz.\newline
{\bf Column~11:}Total flux density measured at 18\,GHz.\newline

\begin{onecolumn}
\begin{sidewaystable}\small
\begin{minipage}{225mm}
{\bf Table 2.} List of 37 sources in this sample in order of increasing Right Ascension.\\
\label{radiojournal}
\begin{center}
\begin{tabular}{llcccccccccccccc}\hline
\multicolumn{1}{c}{(1)}&\multicolumn{1}{c}{(2)} & \multicolumn{1}{c}{(3)}&\multicolumn{1}{c}{(4)} & \multicolumn{1}{c}{(5)}&\multicolumn{1}{c}{(6)} & \multicolumn{1}{c}{(7)}&\multicolumn{1}{c}{(8)} & \multicolumn{1}{c}{(9)}&\multicolumn{1}{c}{(10)} & \multicolumn{1}{c}{(11)}\\
\multicolumn{1}{c}{Source}  & \multicolumn{1}{c}{$z$} &\multicolumn{1}{c}{$L_{1.4\textsc{gh}\rm z}$} &\multicolumn{1}{c}{$\alpha^{1400}_{843}$}& \multicolumn{1}{c}{S$_{843\textsc{mh}\rm z}$} & \multicolumn{1}{c}{S$_{1.4\textsc{gh}\rm z}$} &\multicolumn{1}{c}{S$_{2.4\textsc{gh}\rm z}$} &\multicolumn{1}{c}{S$_{4.8\textsc{gh}\rm z}$} &\multicolumn{1}{c}{S$_{6.2\textsc{gh}\rm z}$} &\multicolumn{1}{c}{S$_{8.6\textsc{gh}\rm z}$} &\multicolumn{1}{c}{S$_{18\textsc{gh}\rm z}$} \\
\multicolumn{1}{c}{}&\multicolumn{1}{c}{} &  \multicolumn{1}{c}{W~Hz$^{-1}$} &  &\multicolumn{1}{c}{mJy} &\multicolumn{1}{c}{mJy} &\multicolumn{1}{c}{mJy} &\multicolumn{1}{c}{mJy} &\multicolumn{1}{c}{mJy}& \multicolumn{1}{c}{mJy}& \multicolumn{1}{c}{mJy}\\
\hline
NVSS~J001339--322445$^{\ddagger}$& 0.2598$\pm$0.0003&   3.3$\times10^{25}$&  --0.93&  $248.4\pm12$&  $155.2\pm4.7$ & $82.5\pm4.1$&  $47.2\pm2.4$ & $37.4\pm1.9$& $26.1\pm1.3$&$10.7\pm2.1$\\                
NVSS~J002131--342225 & 0.249$\pm$0.001  &               3.8$\times10^{24}$&  --1.43&   $37.2\pm2$&   $18.0\pm1.0$ &  $7.6\pm1.2$&   $5.6\pm0.6$ &$4.8\pm0.5$&  $3.4\pm0.4$& \nodata     \\
NVSS~J002219--360728 & 0.364$\pm$0.001&                 7.8$\times10^{24}$&  --1.34&   $30.6\pm2$&   $15.5\pm0.7$ &  $8.5\pm0.4$&  $5.5\pm0.3$ & $4.2\pm0.3$ &  $3.0\pm0.3$&  $1.6\pm0.3$\\
NVSS~J002402--325253$^*$& 2.043$\pm$0.002  &            2.9$\times10^{27}$&  --1.34&  $81.7\pm4$& $41.3\pm1.3$ & $26.0\pm2.0$& $12.5\pm0.6$ &$10.5\pm0.6$&\nodata& \nodata              \\        
NVSS~J002627--323653$^{\ddagger}$ & 0.43$\pm$0.01 &                  3.8$\times10^{25}$&  --1.25&  $83.5\pm4$&   $44.4\pm1.7$ & $19.8\pm1.0$& $9.6\pm0.8$ &$7.8\pm0.5$ & $4.4\pm0.4$&  $2.0\pm0.4$  \\
NVSS~J011606--331241 & 0.352$\pm$0.001  &               1.1$\times10^{25}$&  --1.33&   $43.3\pm2$&   $22.1\pm0.8$ & $9.0\pm2.0$ & $3.6\pm0.3$ &$2.3\pm0.3$ &  $1.6\pm0.3$& $0.9\pm0.5$   \\
NVSS~J012904--324815& 0.1802$\pm$0.0003&   3.8$\times10^{24}$&  --1.39&   $82\pm4$&   $40.5\pm1.7$ & $16.0\pm4.0$ &  $8.5\pm2.0$&$4.4\pm1.0$&  $3.6\pm0.5$&   $1.3\pm0.3$ \\                
NVSS~J015232--333952$^{\ddagger}$ & 0.618$\pm$0.001&                 4.6$\times10^{26}$&  --1.05&  $420.5\pm21$&  $247.3\pm8.7$ &$157.0\pm8.0$ &  $95\pm5$&$77\pm4$ & $61\pm3$&  $23.4\pm4.7$  \\
NVSS~J015324--334117$^{\ddagger}$& 0.1525$\pm$0.0004&   1.3$\times10^{24}$&  --1.21&   $36.6\pm2$&   $19.8\pm0.8$ & $10.0\pm2.0$ & $8.4\pm1.0$&$7.0\pm1.0$&  $5.7\pm0.6$& $2.4\pm0.5$   \\
NVSS~J015544--330633$^{\ddagger}$& 1.048$\pm$0.002  &   2.6$\times10^{26}$&  --0.91&   $61.8\pm2$&   $39.0\pm1.6$ & \nodata &   $13.0\pm1.0$& $10.0\pm0.5$&  $7.7\pm0.3$& $3.2\pm0.5$ \\        
NVSS~J021308--322338& 3.976$\pm$0.001 &                 9.3$\times10^{27}$&  --1.48&  $63.7\pm3$ &   $30.0\pm1.0$ & $19.5\pm1.5$& $10.3\pm1.0$ &$8.5\pm0.6$&\nodata& \nodata            \\
NVSS~J030639--330432& 1.201$\pm$0.001  &                3.9$\times10^{26}$&  --1.69&   $63.7\pm3$&   $27.0\pm0.9$ & $11.5\pm1.0$ & $3.7\pm0.3$ & $2.4\pm0.3$ & $1.3\pm0.2$& $0.9\pm0.5$  \\
NVSS~J202026--372823$^{\ddagger}$& 1.431$\pm$0.001  &                6.4$\times10^{26}$&  --1.27&   $70.0\pm4$&   $36.8\pm1.2$ & \nodata &   $9.8\pm0.5$&$6.4\pm0.3$& $4.5\pm0.3$&  $1.7\pm0.6$   \\
NVSS~J202140--373942$^{\ddagger}$ & 0.185$\pm$0.001 &                2.0$\times10^{24}$&  --1.18&   $35.4\pm2$&   $19.5\pm1.1$ & $11.0\pm1.0$&   $6.0\pm1.0$& $5.6\pm0.8$& $4.6\pm0.8$&   $2.2\pm0.4$\\
NVSS~J202945--344812$^{\ddagger}$& 1.497$\pm$0.002 &    9.5$\times10^{26}$&  --1.22&   $97.5\pm5$&   $52.5\pm2.0$ & $32.0\pm1.6$&  $17.1\pm1.0$&$13.4\pm0.7$&  $11.2\pm0.6$&$3.5\pm0.7$  \\  
NVSS~J204420--334948$^*$&$>4^{\dagger}$ &             $>$7.7$\times10^{{27}{\dagger}}$&  --1.52&  $44.9\pm2$&  $20.8\pm0.8$ &  $10.0\pm0.5$& $3.3\pm0.3$&$2.0\pm0.3$& \nodata& \nodata \\ 
NVSS~J213510--333703$^*$& 2.518$\pm$0.001  &            2.1$\times10^{27}$&  --1.45&   $47\pm2$& $22.5\pm0.8$ & $10.7\pm0.5$& $5.6\pm0.3$ &$4.1\pm0.3$&  \nodata& \nodata             \\        
NVSS~J225719--343954 & 0.726$\pm$0.001 &                1.3$\times10^{26}$&  --1.62&   $85.1\pm4$&   $37.5\pm1.2$ & $15.3\pm1.0$ & $3.7\pm0.2$&$2.5\pm0.2$&  $1.6\pm0.2$ & $0.7\pm0.2$  \\
NVSS~J230035--363410$^*$&$4.0^{\dagger}$ &              7.0$\times10^{{27}{\dagger}}$&  --1.66&   $35.3\pm2$&  $15.2\pm0.7$ &  $9.1\pm0.5$ & $3.8\pm0.3$&$2.7\pm0.2$&\nodata & \nodata \\
NVSS~J230123--364656$^*$& 3.220$\pm$0.002  &            4.6$\times10^{27}$&  --1.57&   $44.3\pm2$&   $20.0\pm0.8$ &  $8.5\pm0.5$ &$4.1\pm0.5$&$2.7\pm0.5$  &\nodata& \nodata          \\        
NVSS~J230527--360534$^*$&$>4^{\dagger}$ &             $>$9.2$\times10^{{27}{\dagger}}$&  --1.38&  $63.1\pm2$& $31.3\pm1.0$ & $16.0\pm2.0$& $5.8\pm0.2$&$3.8\pm0.2$&\nodata& \nodata\\
NVSS~J230954--365653$^{\ddagger}$$^*$&$>4^{\dagger}$ &              3.6$\times10^{{27}{\dagger}}$&  --1.09&  $33.5\pm2$& $19.3\pm1.5$ & \nodata& $5.8\pm0.3$&$3.8\pm0.3$&\nodata& \nodata  \\
NVSS~J231144--362215$^{\ddagger}$$^*$& 2.531$\pm$0.002 &             1.3$\times10^{27}$&  --1.23&   $33.7\pm2$&  $18.1\pm0.7$ & $10.9\pm1.0$&  $5.5\pm0.8$&$3.5\pm0.3$& \nodata& \nodata  \\
NVSS~J231317--352133$^{\ddagger}$$^*$& $1.2^{\dagger}$  &            1.8$\times10^{{26}{\dagger}}$&  --1.23&   $30.8\pm1$&  $16.5\pm0.7$ &  $8.8\pm0.7$& $4.0\pm0.2$&$3.2\pm0.2$  &\nodata& \nodata \\        
NVSS~J231338--362708& 1.838 $\pm$0.002  &               6.7$\times10^{26}$&  --1.54&   $36.3\pm1$&   $16.6\pm0.7$ &  $7.6\pm0.3$&  $2.4\pm0.2$ & $2.0\pm0.3$ &  $1.0\pm0.1$ & \nodata\\
NVSS~J231402--372925$^{\ddagger}$$^*$& 3.450$\pm$0.005 &             2.5$\times10^{28}$&  --1.22&  $241.3\pm12$& $129.9\pm3.9$ & $71\pm4$& $38.0\pm2.0$ & $27.0\pm2.0$  &\nodata& \nodata    \\        
NVSS~J231727--352606$^{\ddagger}$$^*$& 3.874$\pm$0.002 &             1.5$\times10^{28}$&  --1.19&  $108.5\pm5$& $59.2\pm1.8$ & $31.4\pm2.0$ & $12.4\pm0.6$&$8.3\pm0.4$  &\nodata& \nodata     \\  
NVSS~J232058--365157$^{\ddagger}$& $1.3^{\dagger}$ &                 7.2$\times10^{{26}{\dagger}}$&  --1.28& $98.5\pm5$& $51.4\pm1.6$ & $34.5\pm2.0$ &$16.5\pm0.8$&$13.3\pm0.7$&$9.3\pm0.6$& $3.5\pm1.2$\\                
NVSS~J232100--360223$^*$& 3.320$\pm$0.005 &             4.8$\times10^{27}$&  --1.65&   $34.8\pm2$& $15.1\pm0.7$ &  $6.6\pm0.5$& $3.2\pm0.3$&$1.9\pm0.3$ &\nodata& \nodata                \\  
NVSS~J232219--355816$^*$&$>4^{\dagger}$ &             $>$1.4$\times10^{{28}{\dagger}}$&  --1.84&  $58.6\pm3$&  $23.1\pm0.8$ &  $14.0\pm1.0$& $5.8\pm0.3$&$4.4\pm0.3$ &\nodata& \nodata \\        
NVSS~J232408--353547$^{\ddagger}$ & 0.2011$\pm$0.0004&               2.0$\times10^{24}$&  --0.99&   $27\pm2$&   $16.3\pm1.2$ &  $8.5\pm1.5$&   $6.7\pm0.8$&$5.2\pm1.0$&  $5.1\pm0.6$&   $2.2\pm0.7$ \\
NVSS~J232602--350321$^{\ddagger}$& 0.293$\pm$0.001&     4.5$\times10^{24}$&  --1.07&   $28.9\pm2$&   $16.8\pm1.2$ & \nodata &   $7.3\pm0.5$&$6.6\pm0.7$&  $5.8\pm0.6$&  $2.0\pm0.4$      \\
NVSS~J232651--370909$^*$& 2.357$\pm$0.003  &            2.8$\times10^{27}$&  --1.35&  $64.3\pm3$ & $32.5\pm1.1$ & $15.6\pm 0.8$& $6.8\pm0.3$ &$4.2\pm0.3$&  \nodata& \nodata             \\        
NVSS~J234137--342230$^*$&$>4^{\dagger}$ &             $>$4.6$\times10^{{27}{\dagger}}$&  --1.31&  $33.9\pm1$&  $17.4\pm1.0$ & $6.8\pm1.0$&  $3.6\pm0.3$&$2.5\pm0.3$&\nodata& \nodata   \\
NVSS~J234145--350624$^{\ddagger}$ & 0.644$\pm$0.001 &                3.7$\times10^{27}$&  --1.10& $3190\pm160$& $1823.2\pm54.7$ & $1062\pm53$& $490\pm25$ & $353\pm18$& $242\pm12$& $82\pm16$     \\
NVSS~J234904--362451$^{\ddagger}$& 1.520$\pm$0.003  &                9.3$\times10^{26}$&  --1.23&   $86.6\pm3$&   $46.5\pm1.8$ & $28.2\pm1.4$&   $17.1\pm0.9$ & $13.7\pm0.7$& $9.4\pm1.0$& $4.3\pm0.9$\\
NVSS~J235137--362632$^{\ddagger}$$^*$&$>4^{\dagger}$ &             $>$8.1$\times10^{{27}{\dagger}}$&  --1.19&  $68.5\pm3$&   $37.4\pm1.5$ & $18.8\pm1.5$& $8.7\pm0.5$&$5.8\pm0.3$ &\nodata&\nodata  \\        

\hline
\end{tabular}
\begin{flushleft}
$\dagger$Redshifts are estimated using $K(8'') = 4.633\,\rm{log}_{10}(z)\,+\,17.066$ where $K(8'')$ is the K-band magnitude defined in an $8''$ aperture \citep{cdb02a}. Since the $K-z$ relation is not known to hold beyond $z\sim4$, the six sources where the relation predicts redshifts above this are listed as $z>4$. \newline
$\ddagger$These sources no longer meet the original USS sample threshold of $\alpha\leq-1.3$ (see \S\ref{revised843}). \newline
$*$Sources also observed in 6A array at 4.8~GHz and 6.2~GHz. \end{flushleft}
\end{center}
\end{minipage}
\end{sidewaystable}
\end{onecolumn}

\twocolumn
\subsection{Host galaxy identification and radio structure}
Freak flooding occurred at the ATCA during our high spatial resolution observations on 2004 December 10. As a result, the atmospheric water vapour content sky-rocketed and the phase stability at 4.8~GHz and 6.2~GHz became extremely poor. Consequently, our observations on December 11 and 12 have not been included in the analysis as we could not salvage the data. Unfortunately, the signal-to-noise from the December 10 images is not adequate to determine reliable polarisation properties or accurate flux densities. However, they are of sufficient quality to measure the angular size and further constrain the positions of the host galaxies which have so far eluded identification (paper~II). Table~\ref{6cmsizes} lists the angular and linear sizes (for sources with spectroscopic redshifts) derived from these observations, or a limit if the sources remained unresolved. An asterisk (*) in column 1 of Table~2 shows the subset of sources in Table~\ref{6cmsizes}. 

\setcounter{table}{2}
\begin{table}
\caption{\small Size constraints derived from the high resolution 4.8~GHz and 6.2~GHz observations of the 16 highest redshift candidates (see \S\ref{sourceselection}) on 2004 December 10. Projected linear sizes are given for sources with spectroscopic redshifts from paper~II. Positive position angles of the observed radio axis are measured East from North.}
\label{6cmsizes}
\begin{center}
\begin{tabular}{lllll}\hline
\multicolumn{1}{c}{Source} & \multicolumn{1}{c}{LAS}& \multicolumn{1}{c}{PA}& type &\multicolumn{1}{c}{Linear Size$^{\dagger}$} \\
                           & \multicolumn{1}{c}{\arcsec}&\multicolumn{1}{c}{$^{\circ}$}& & kpc\\
\hline
NVSS~J002402--325253 & $<1$& \nodata& unresolved& $<9$\\
NVSS~J204420--334948 & $<2$ &\nodata& unresolved& \nodata\\
NVSS~J213510--333703 & $<1.5$ &\nodata & unresolved& $<12$\\
NVSS~J230035--363410 & $<1.5$ & \nodata & unresolved & \nodata \\
NVSS~J230123--364656 & $<1$ & \nodata & unresolved & $<8$\\
NVSS~J230527--360534 & $<1$ & \nodata & unresolved & \nodata\\
NVSS~J230954--365653 & 55.2$^*$ & 25$^*$ & resolved  &\nodata\\
NVSS~J231144--362215 & 15.4 & 64 & resolved  & 125\\
NVSS~J231317--352133 & $<1.5$&\nodata & unresolved & \nodata\\ 
NVSS~J231402--372925 & 4 & 61 & resolved & 28\\
NVSS~J231727--352606 & 4 & --42 &resolved  & 26\\
NVSS~J232100--360223 & $<1$&\nodata & unresolved & $<8$\\
NVSS~J232219--355816 & $<1$ & \nodata & unresolved & \nodata\\
NVSS~J232651--370909 & 7 & 13 & extended & 58\\
NVSS~J234137--342230 & $<2$ & \nodata & unresolved & \nodata\\
NVSS~J235137--362632 & $6$ & 10 & extended & \nodata\\
\end{tabular}

\begin{flushleft}
$^*$From paper~1.\newline
$^{\dagger}$Linear sizes are given for those sources with spectroscopic redshifts.\newline
\end{flushleft}
\end{center}
\end{table}

\subsection{Spectral Energy Distributions}\label{SEDfits}
For each source in the sample, we derived a functional form for its rest-frame SED by using a $1/\sigma^2$ weighted least squares fit to log($S_\nu/(1+z)$) versus log($\nu(1+z)$). We began by fitting each of the data sets with first-order polynomials. The goodness-of-fit of each model was determined using the $\chi^2$ test. Models were rejected at the 2\% level. The SEDs of 33 out of 37 sources were well characterised by a single power law. First-order polynomial fits were rejected for four sources in the sample. For each of these, we fitted the data sets with second-order polynomials and repeated the statistical tests. The results of the fits are given in Table~\ref{sedfits} and the SEDs are shown in Figure~\ref{sedplots}.

\begin{figure*}\begin{center}
\includegraphics[width=8cm]{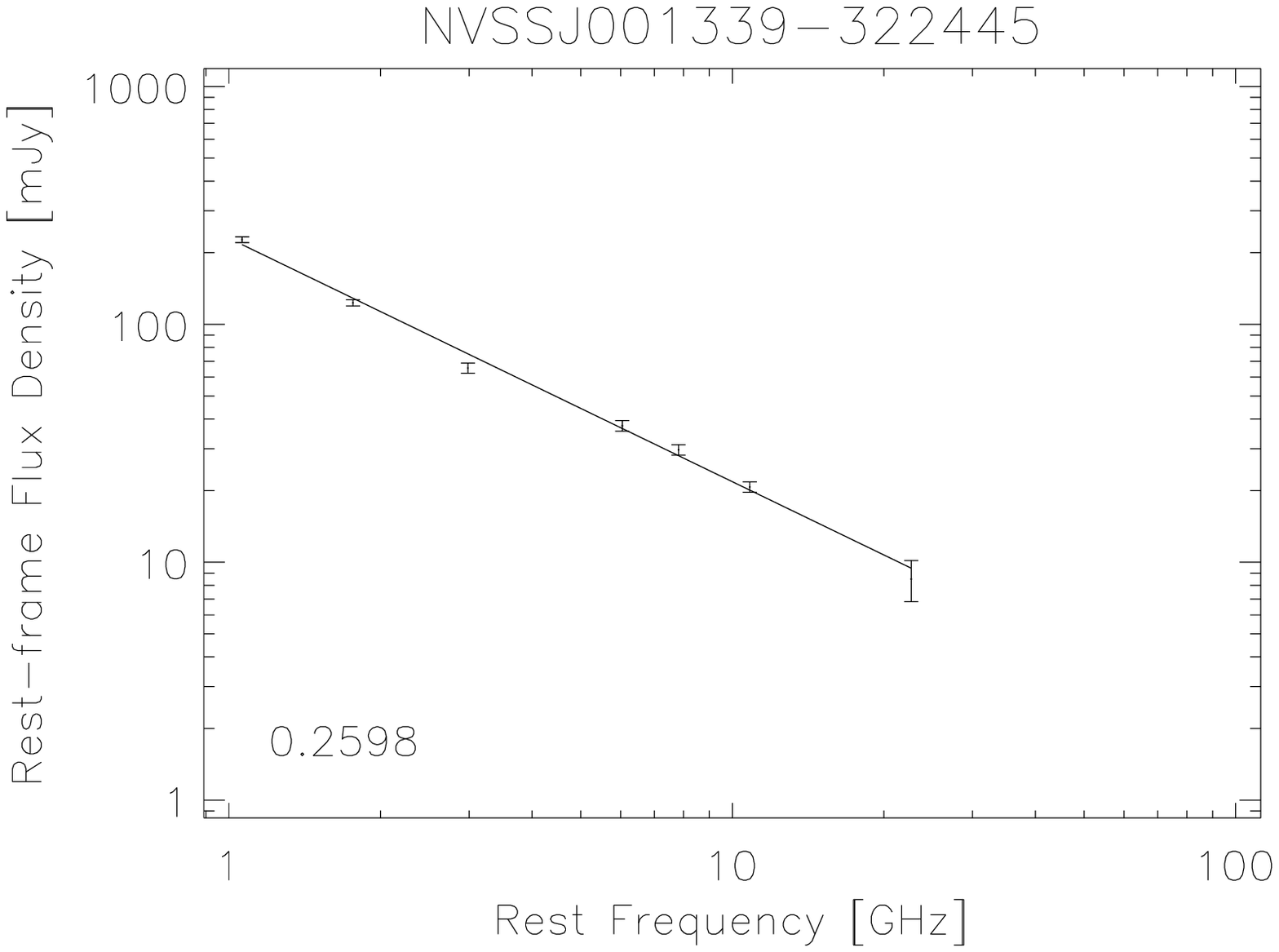} 
\includegraphics[width=8cm]{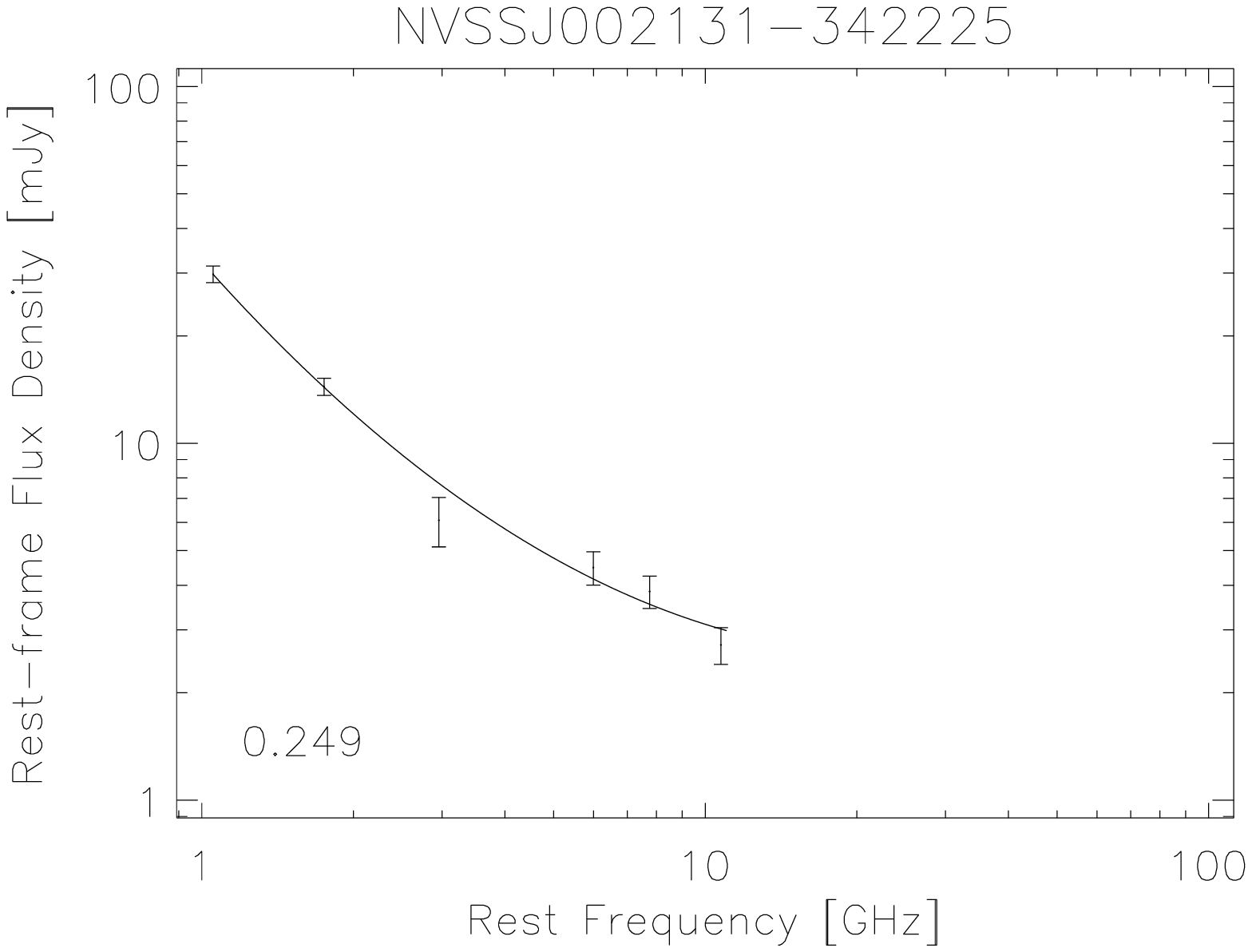} 

\includegraphics[width=8cm]{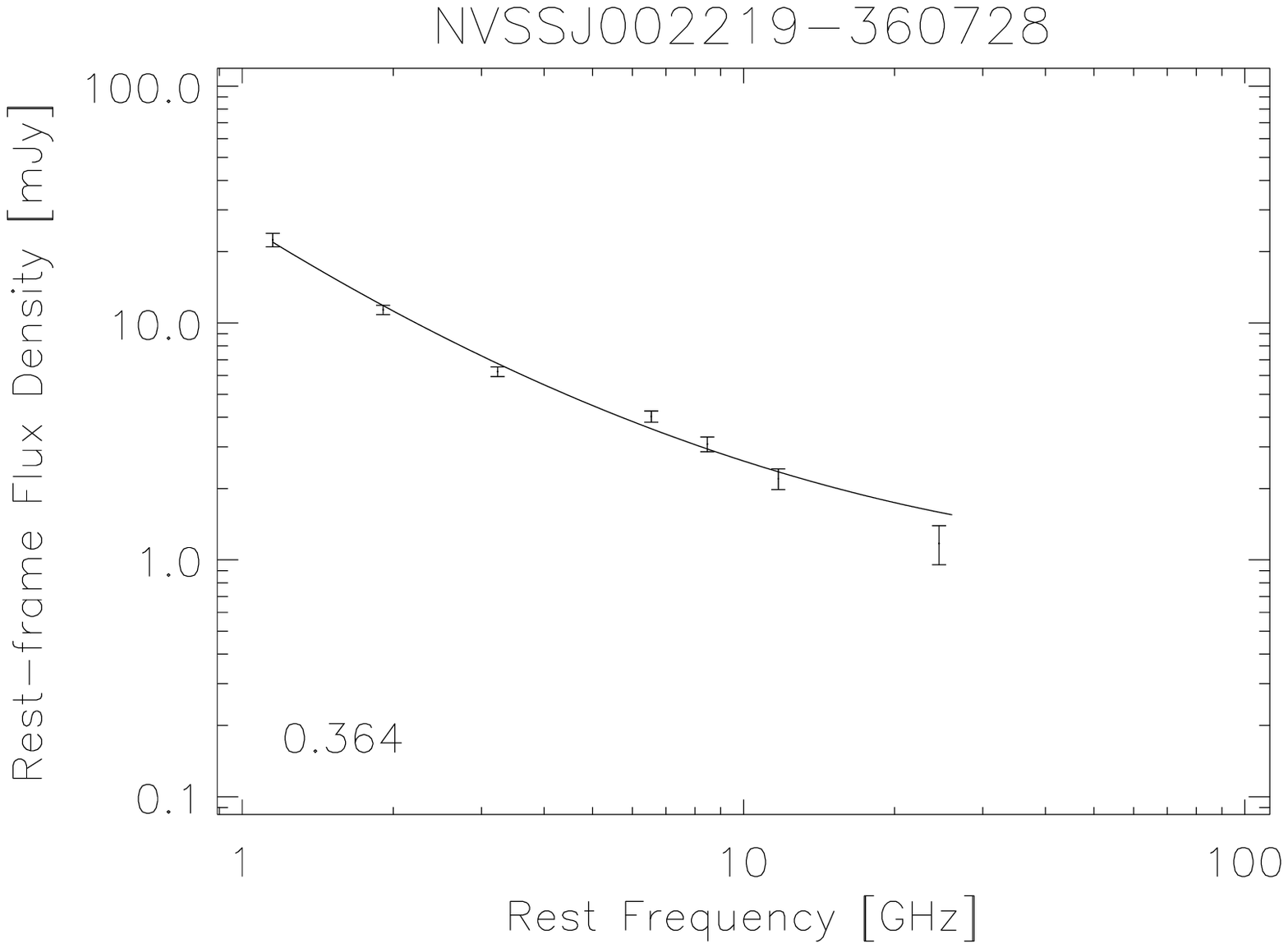} 
\includegraphics[width=8cm]{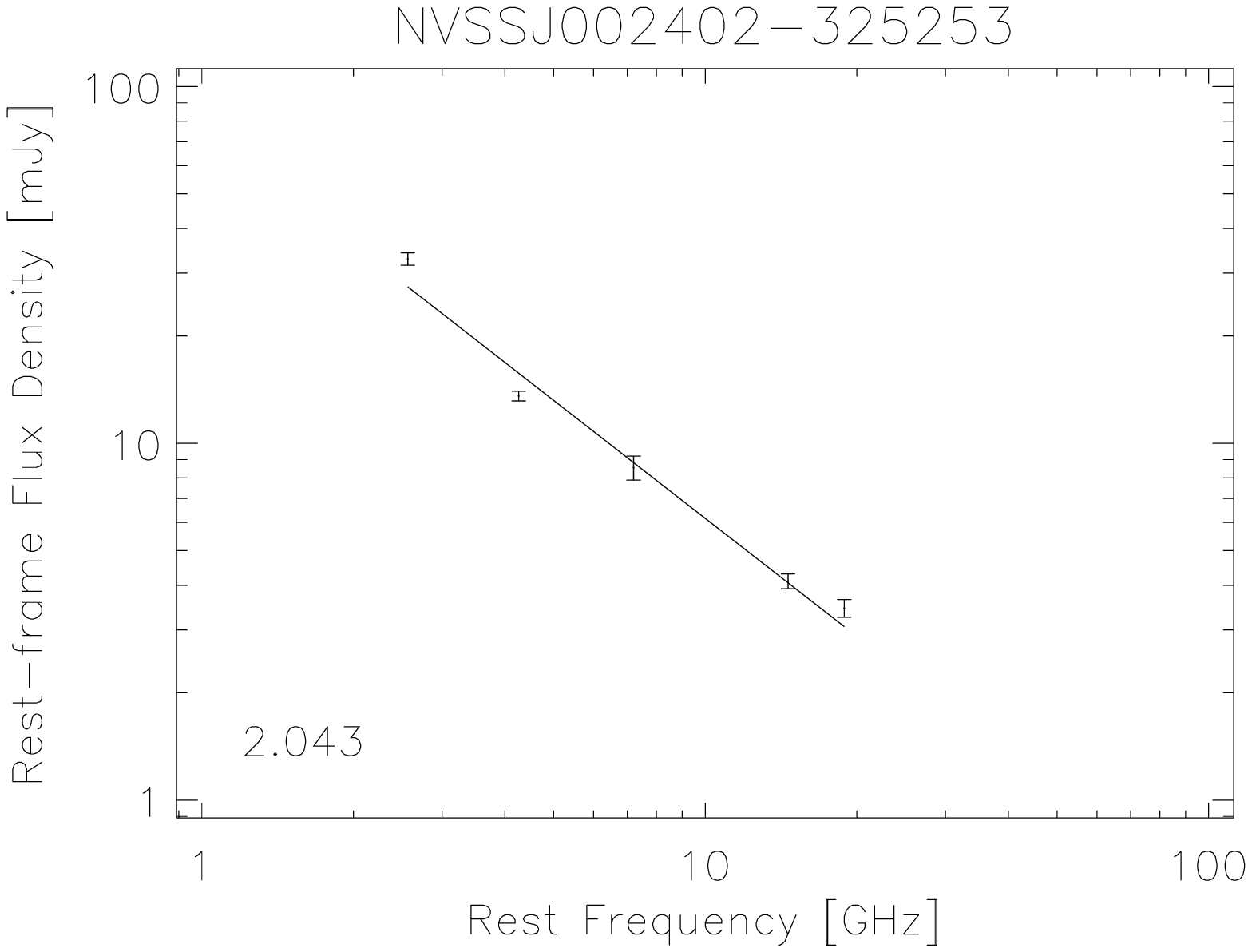} 

\includegraphics[width=8cm]{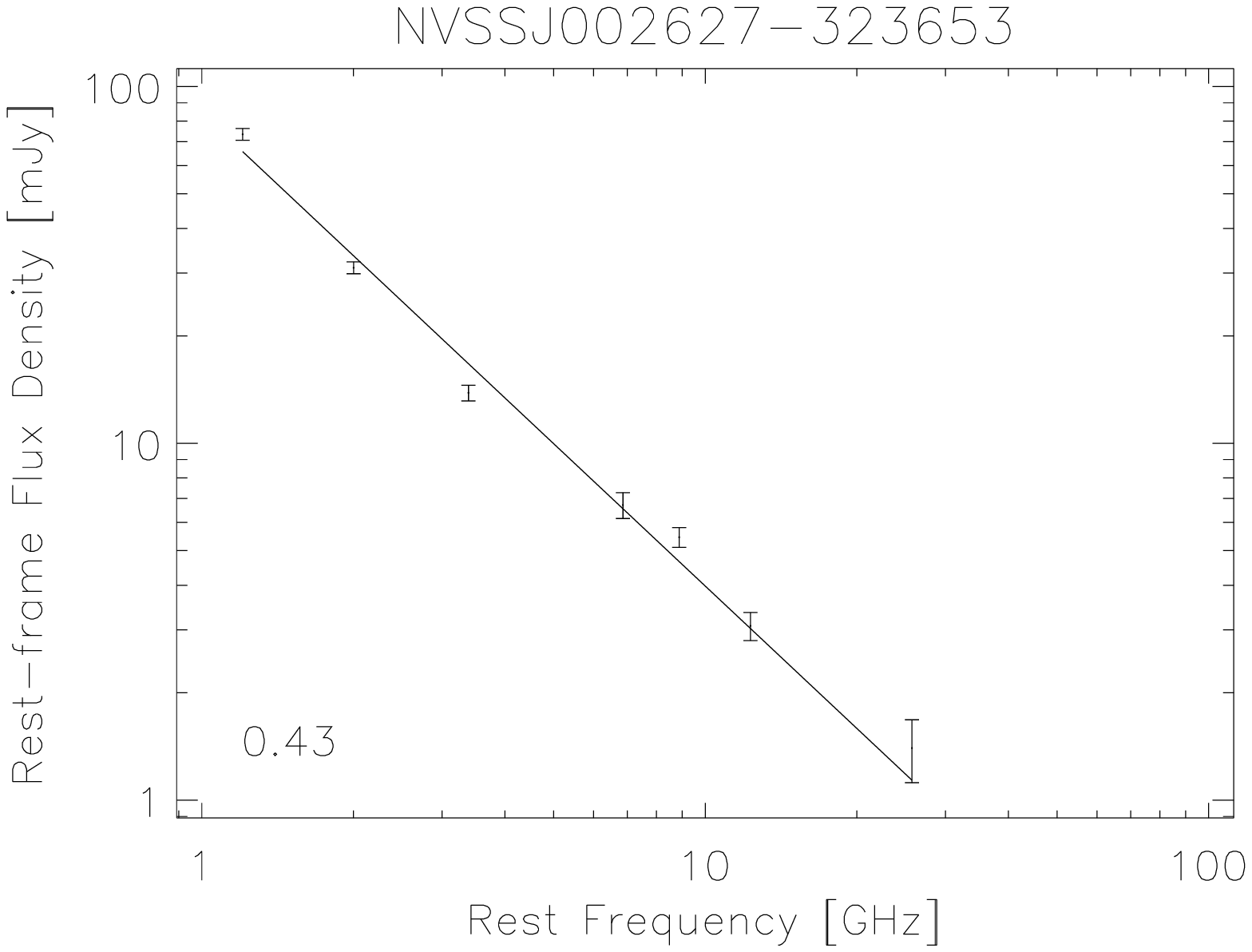} 
\includegraphics[width=8cm]{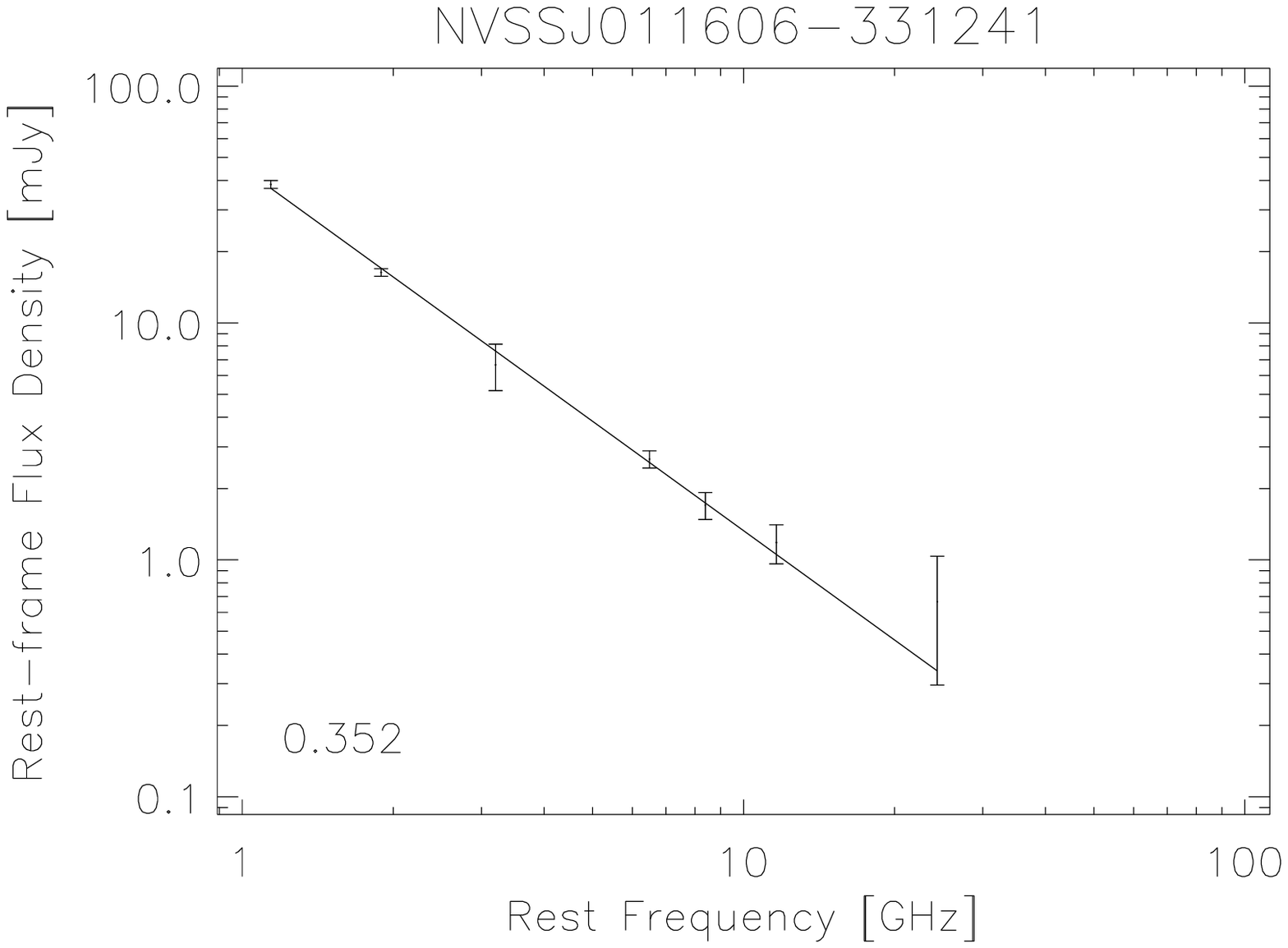} 

\includegraphics[width=8cm]{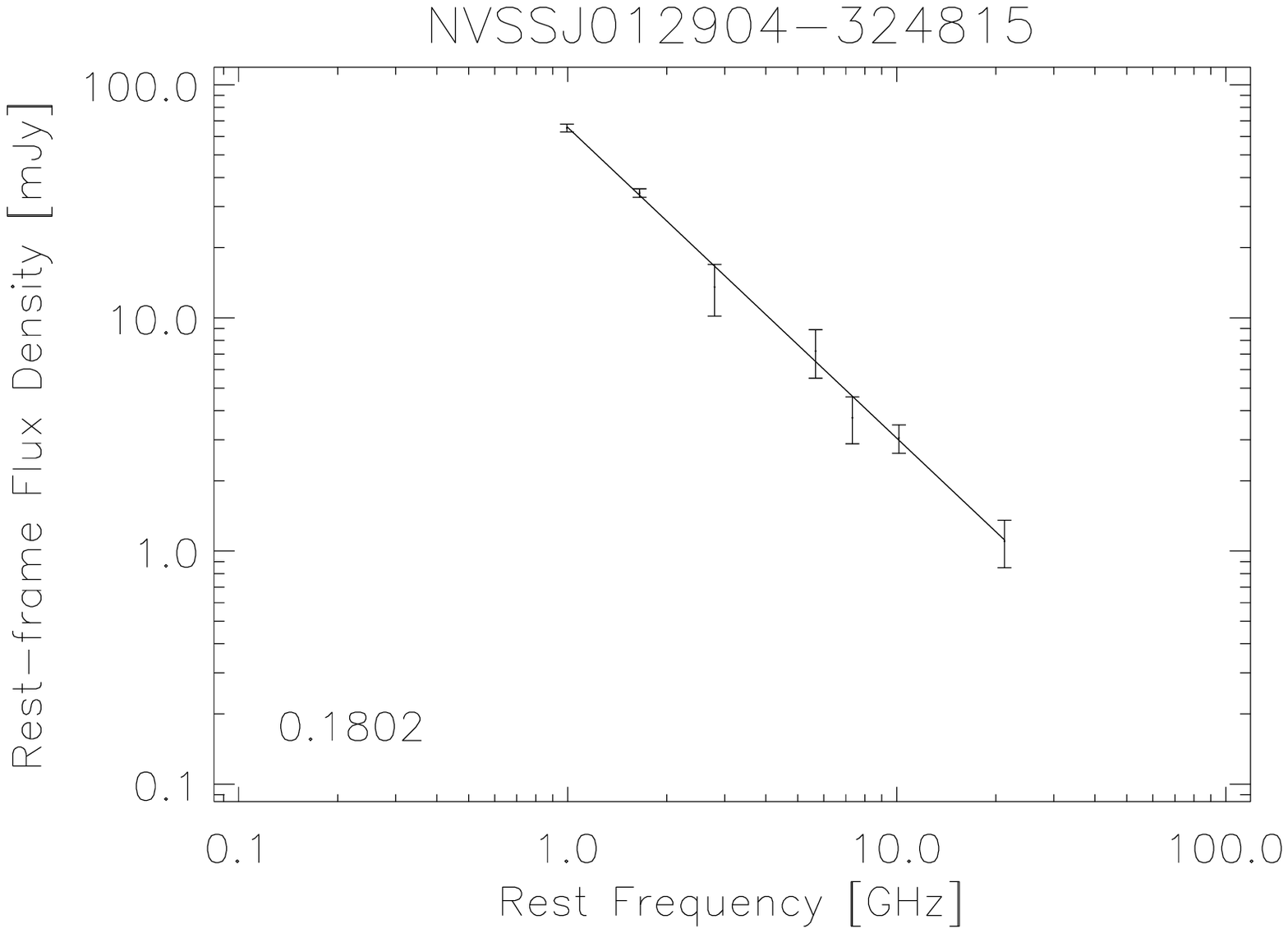} 
\includegraphics[width=8cm]{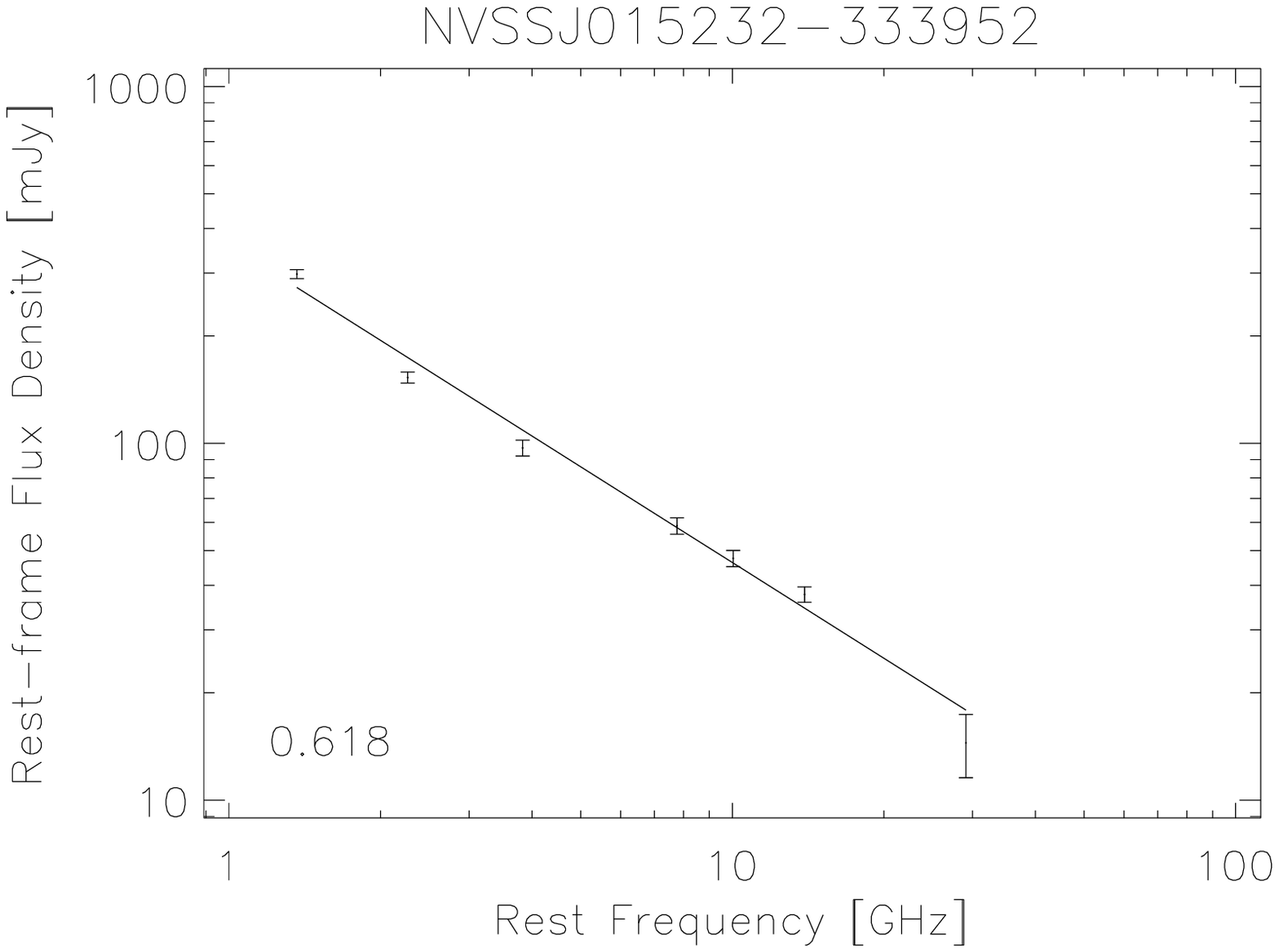} 

\caption{Rest-frame spectral energy distributions for the 37 SUMSS-NVSS USS radio galaxies in the sample, shown in order of increasing Right Ascension. The flux densities listed in Table~2 are plotted along with the functional form for the SED as given in Table~\ref{sedfits}. Source redshifts are labelled in the bottom left corner of each plot.}\label{sedplots}
\end{center}\end{figure*}\setcounter{figure}{0}\begin{figure*}\begin{center}  

\includegraphics[width=8cm]{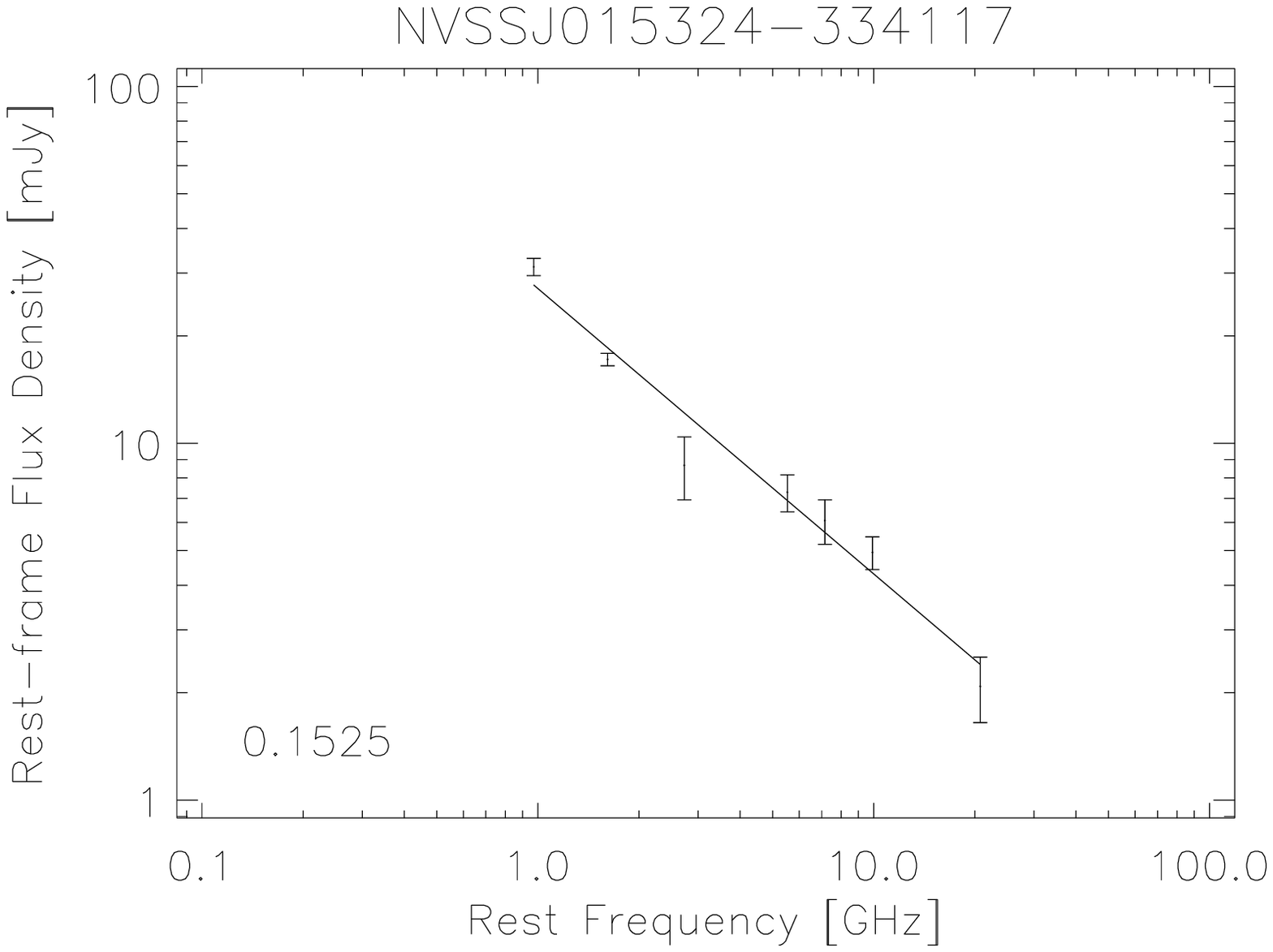} 
\includegraphics[width=8cm]{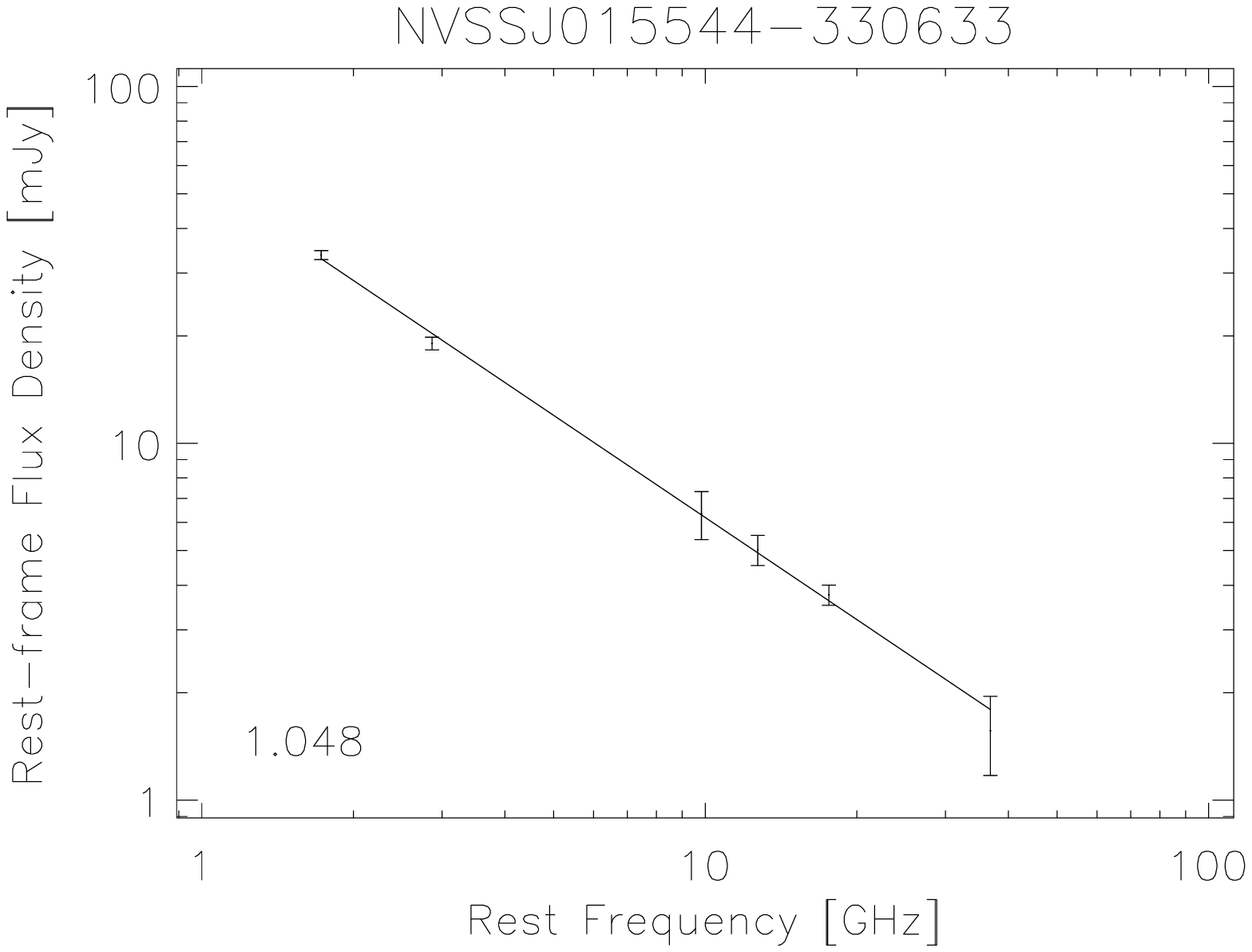} 

\includegraphics[width=8cm]{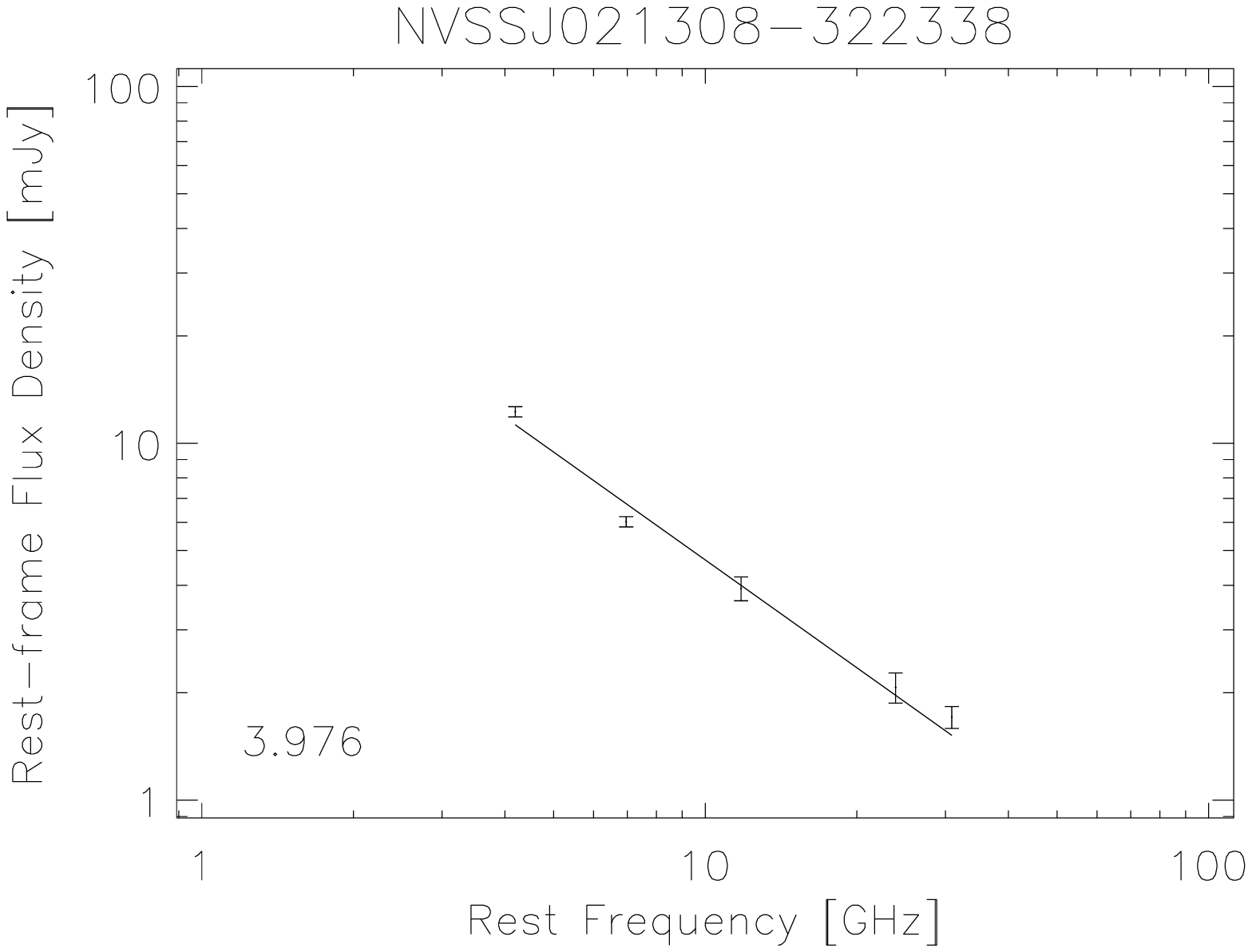} 
\includegraphics[width=8cm]{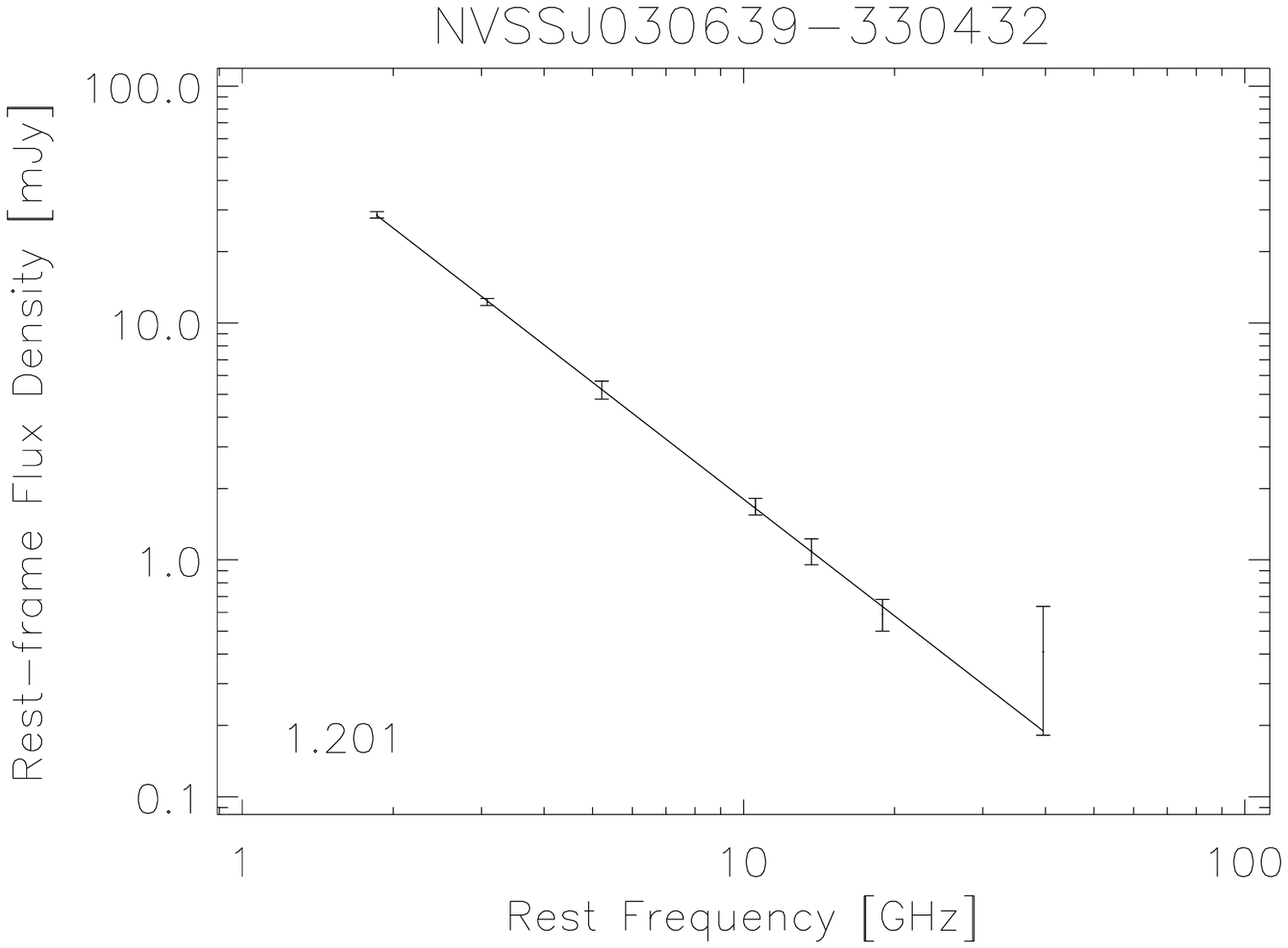} 

\includegraphics[width=8cm]{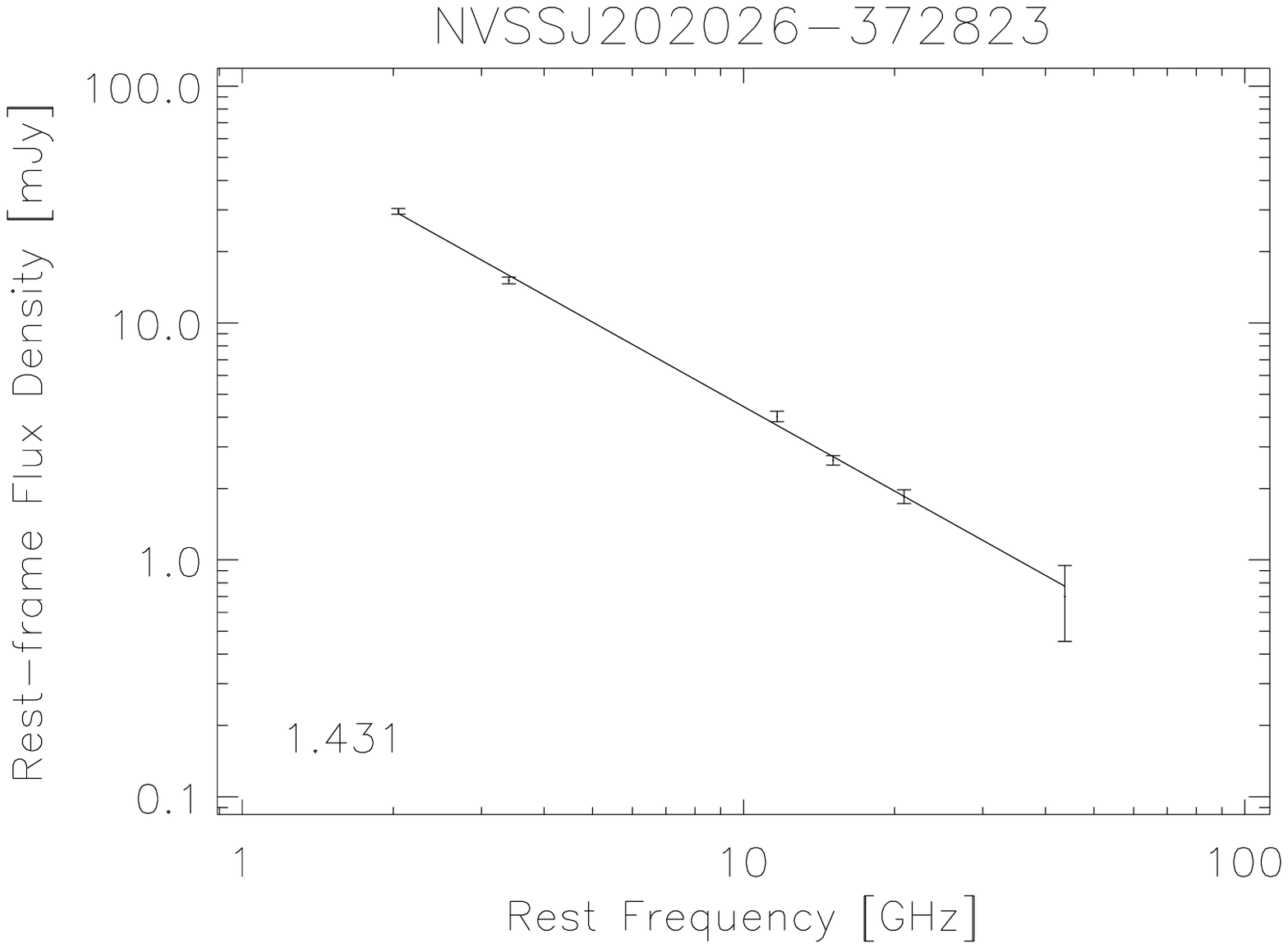} 
\includegraphics[width=8cm]{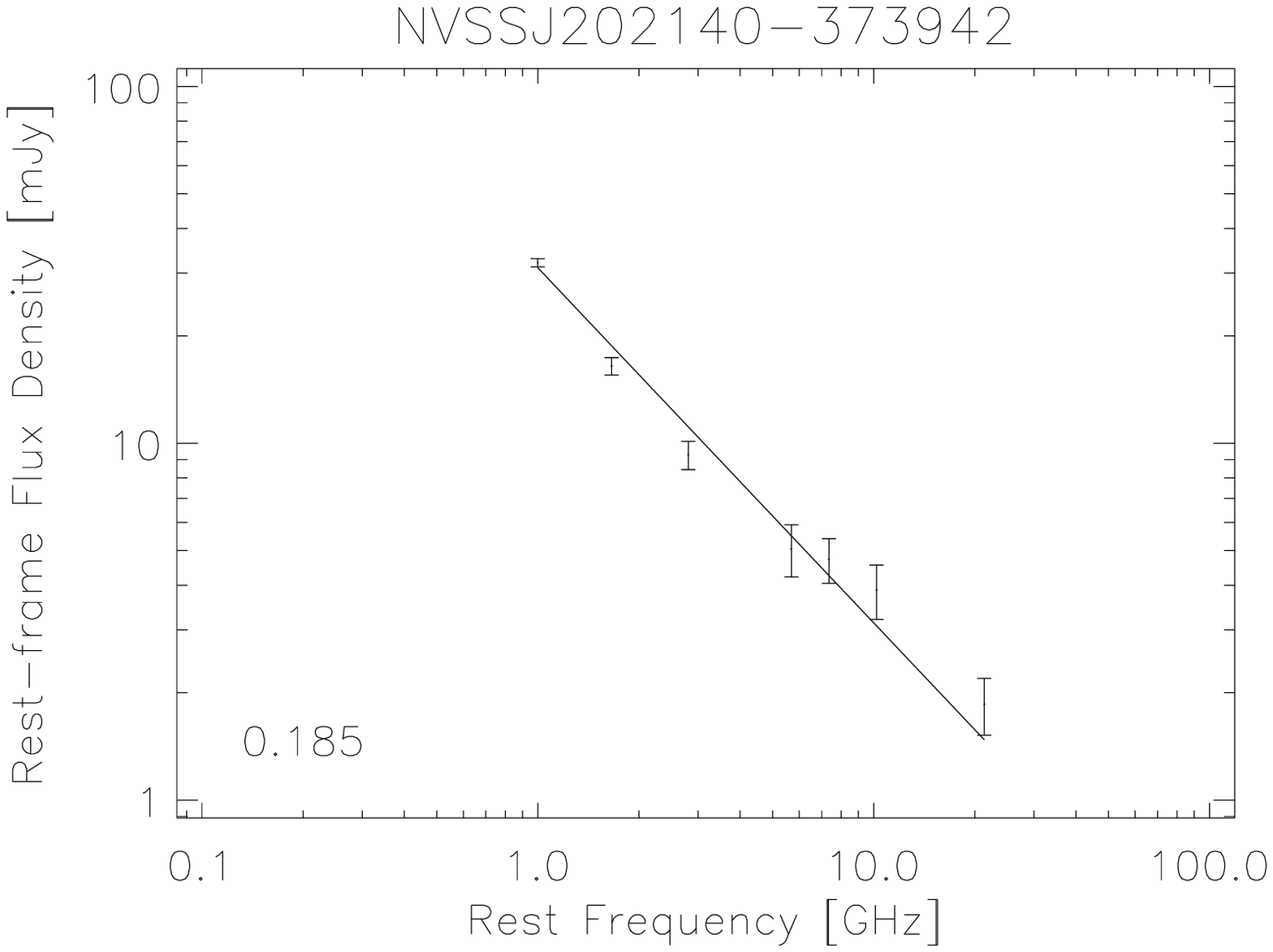} 

\includegraphics[width=8cm]{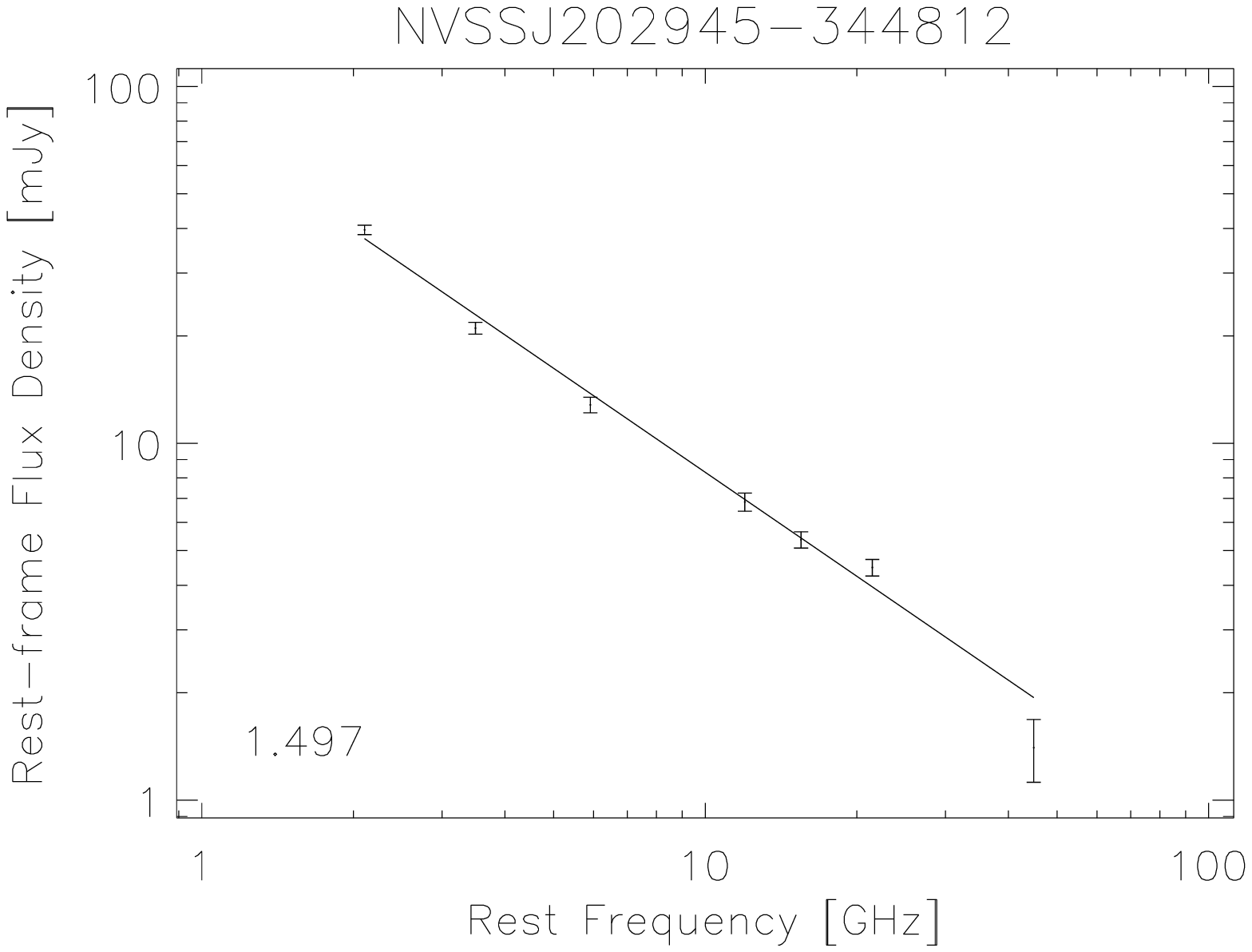} 
\includegraphics[width=8cm]{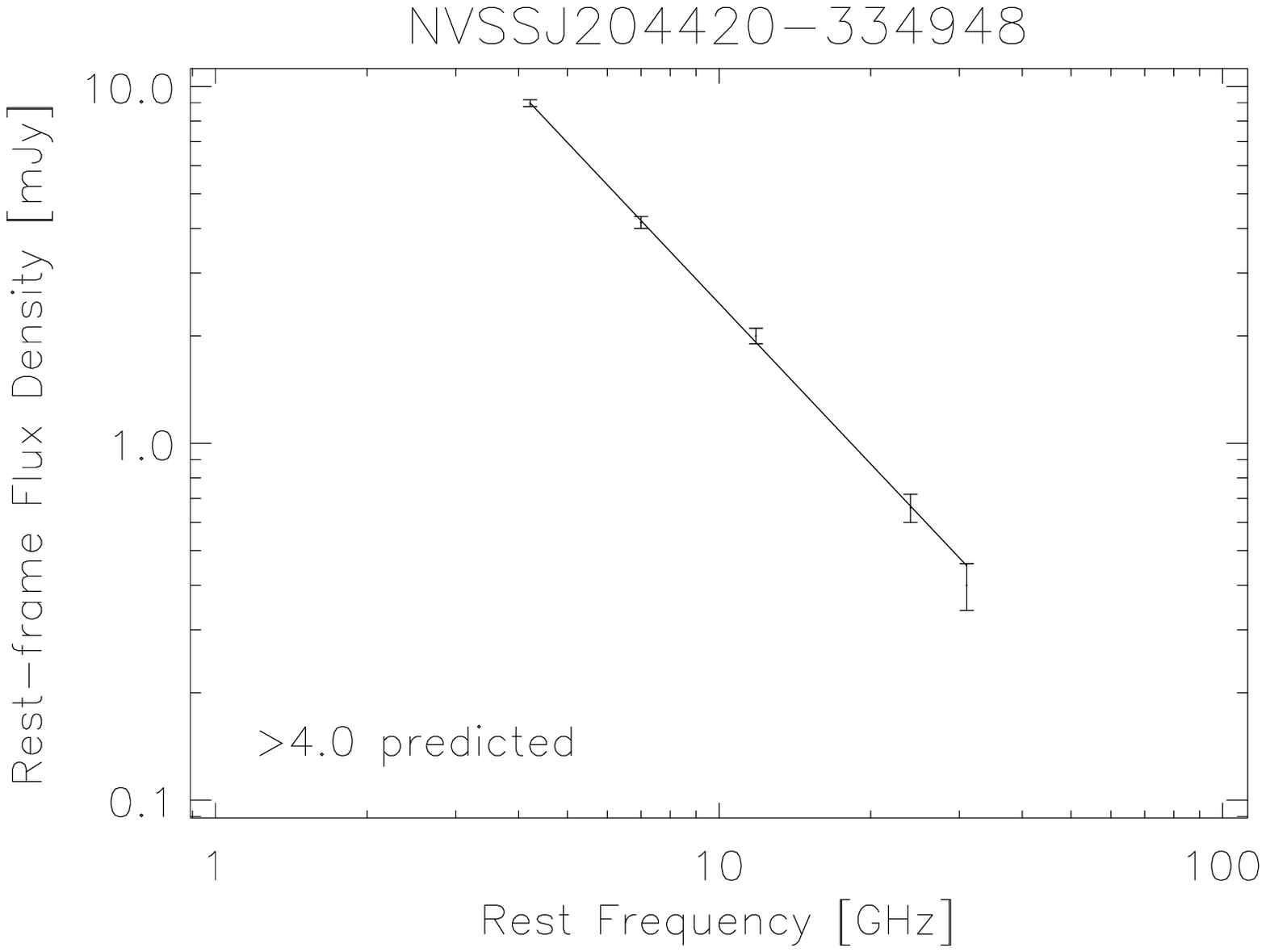} 
\caption{- continued}\end{center}\end{figure*}\setcounter{figure}{0}\begin{figure*}\begin{center}

\includegraphics[width=8cm]{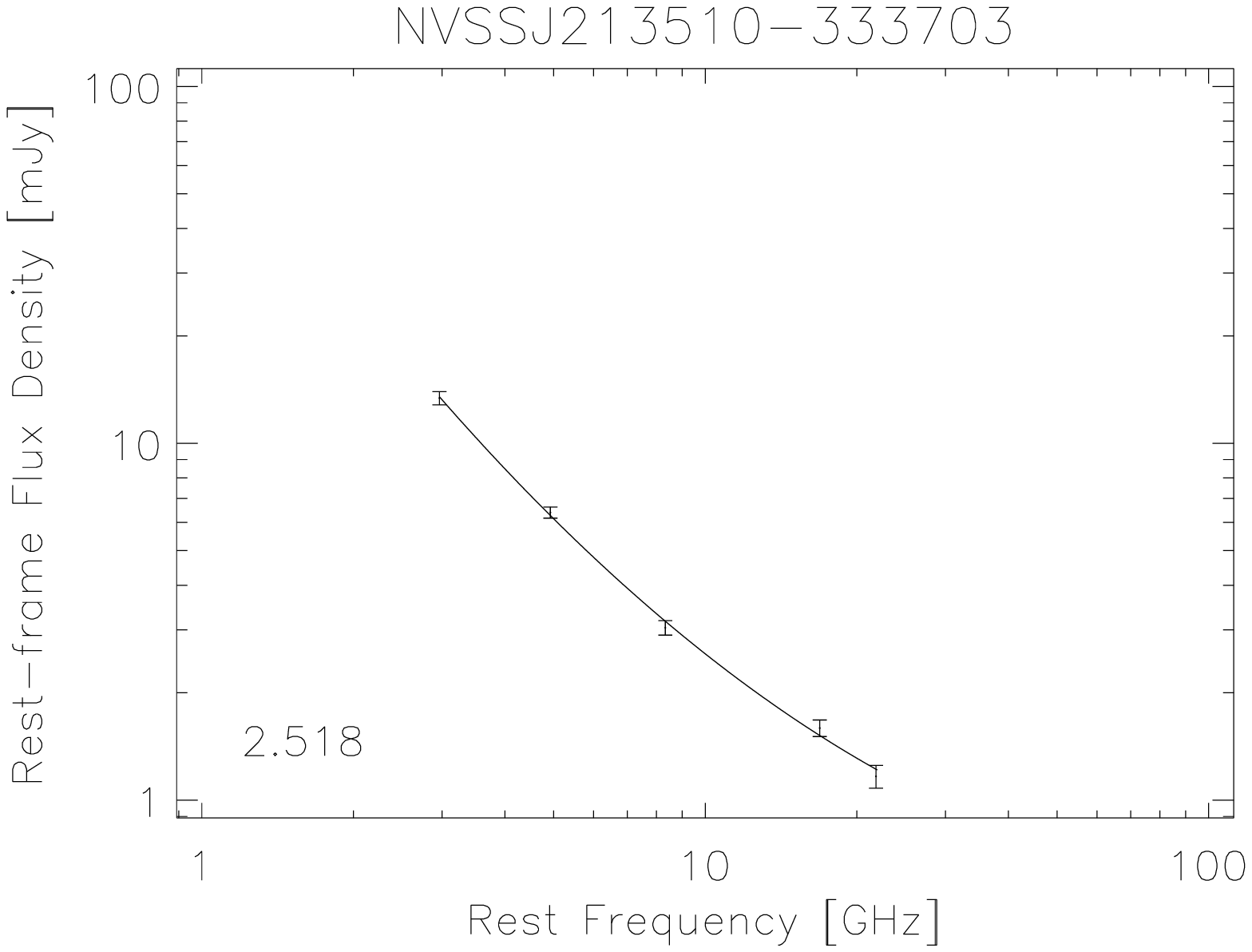} 
\includegraphics[width=8cm]{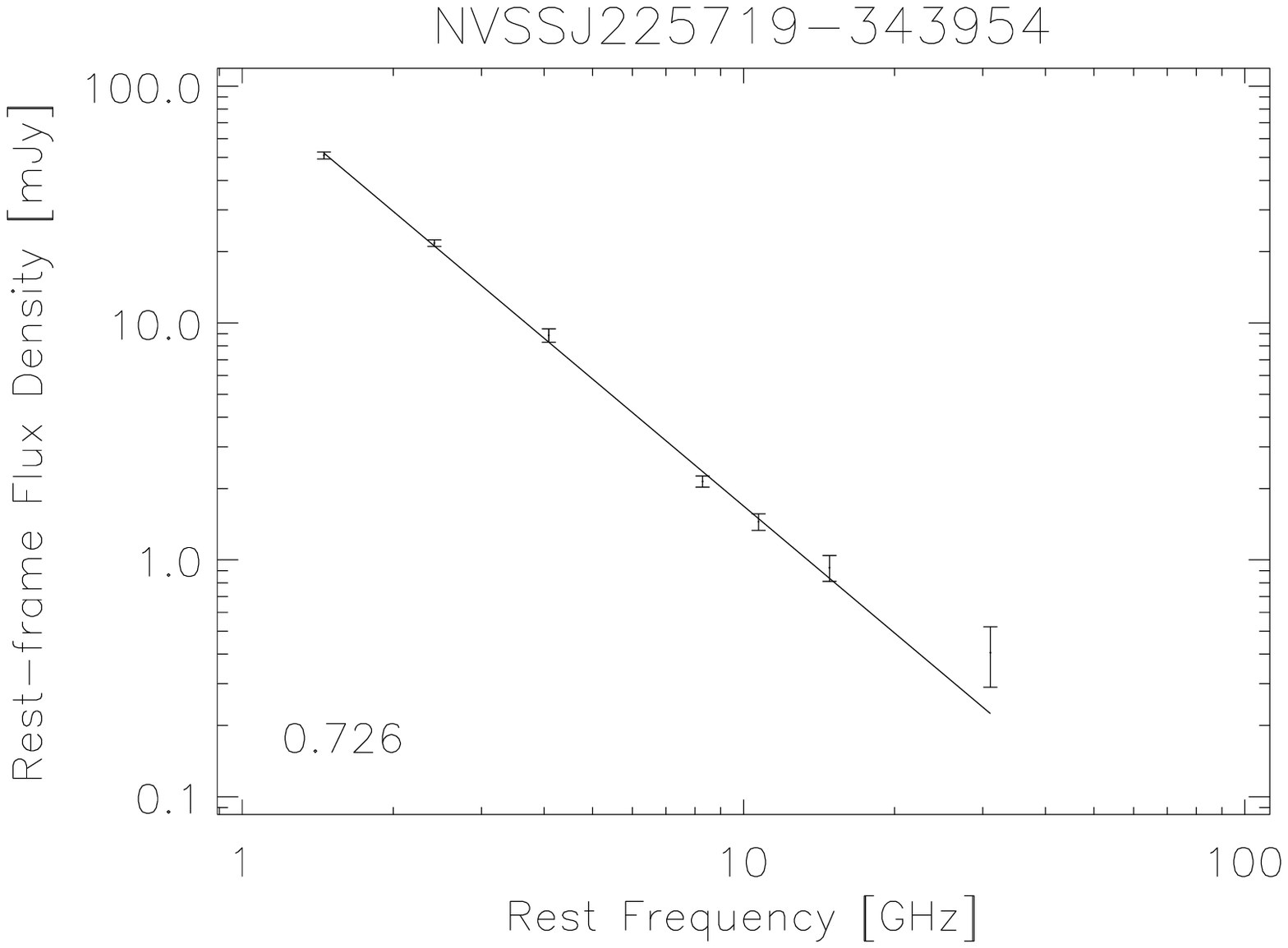} 

\includegraphics[width=8cm]{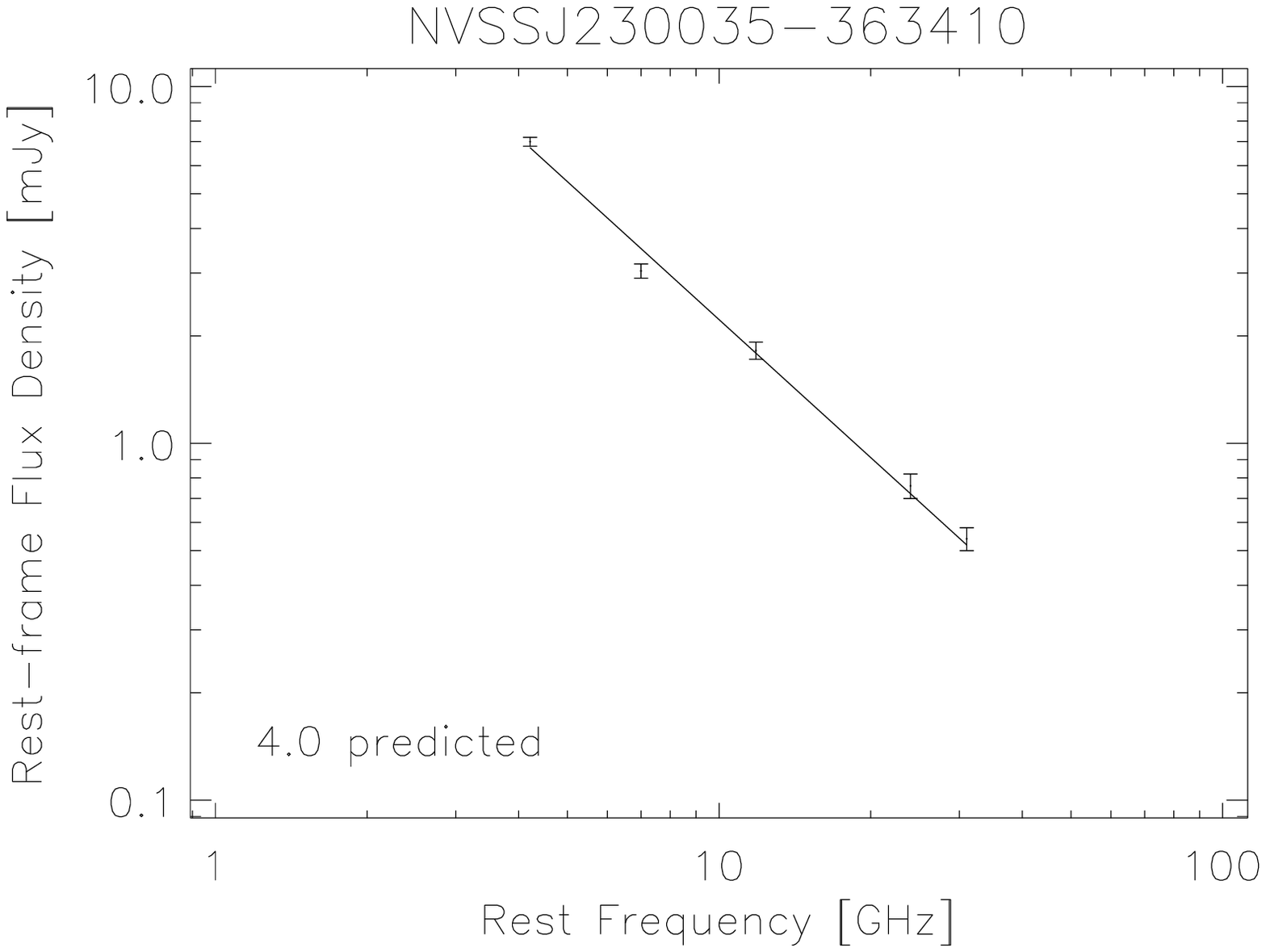} 
\includegraphics[width=8cm]{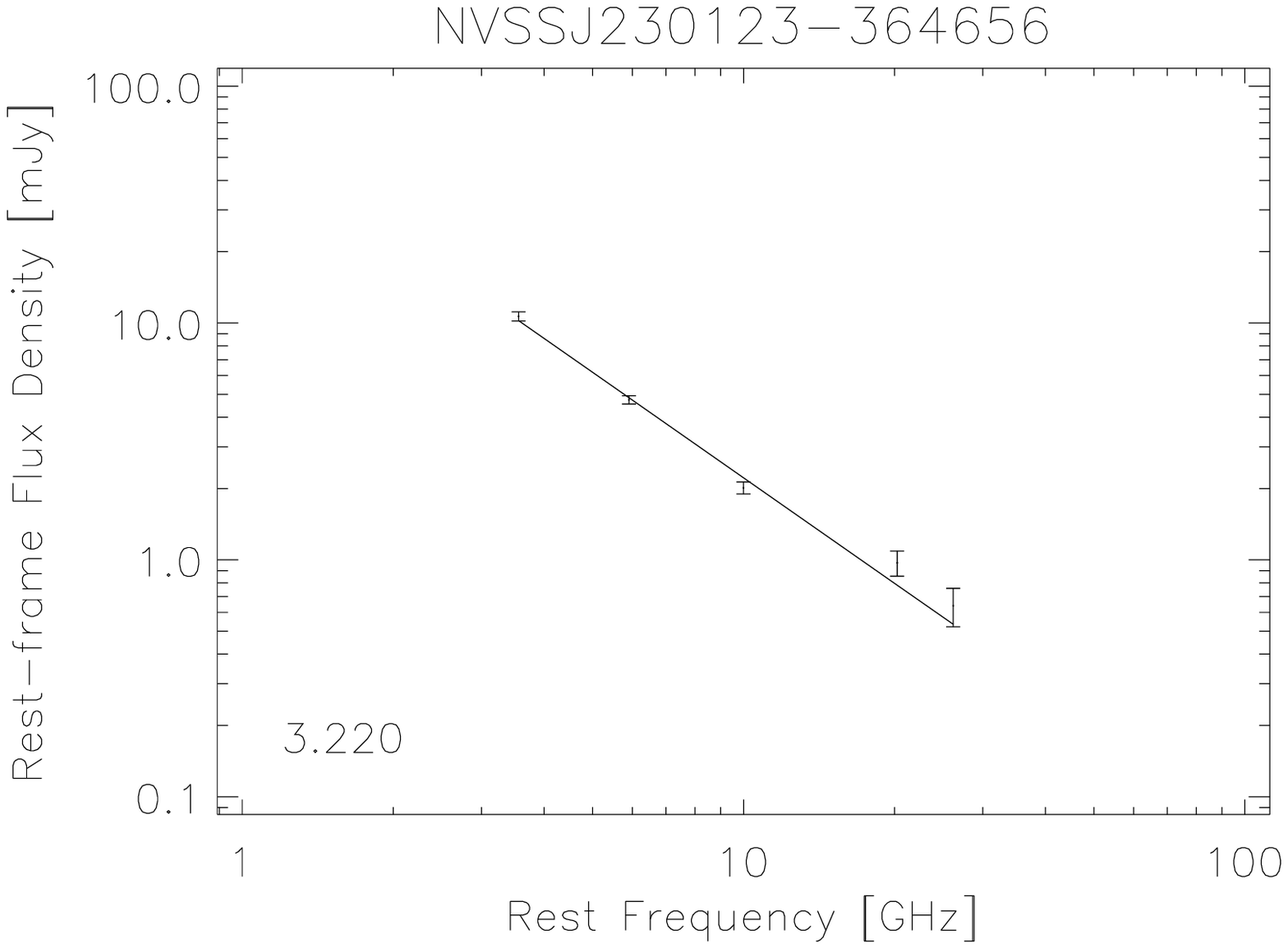} 

\includegraphics[width=8cm]{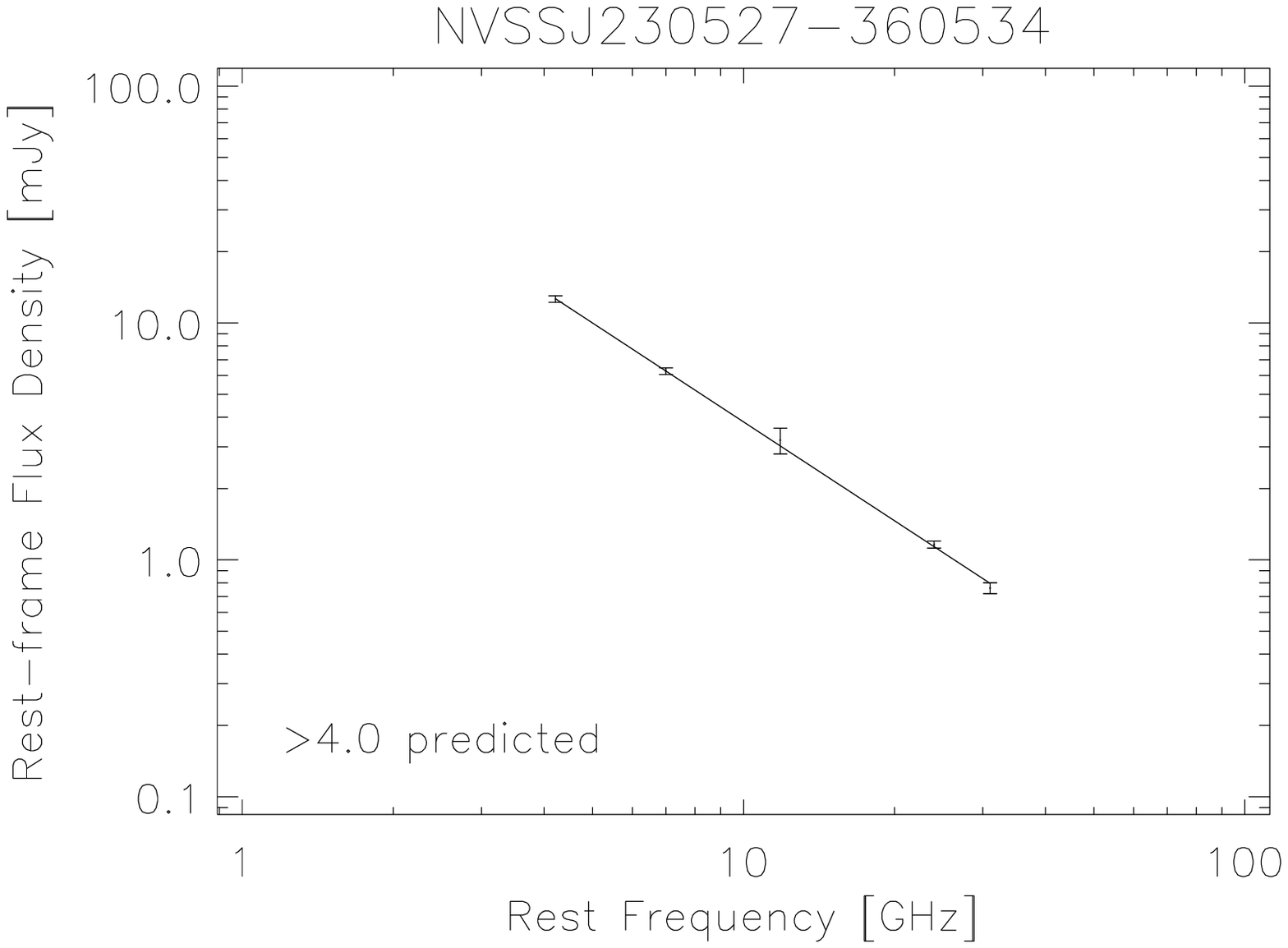} 
\includegraphics[width=8cm]{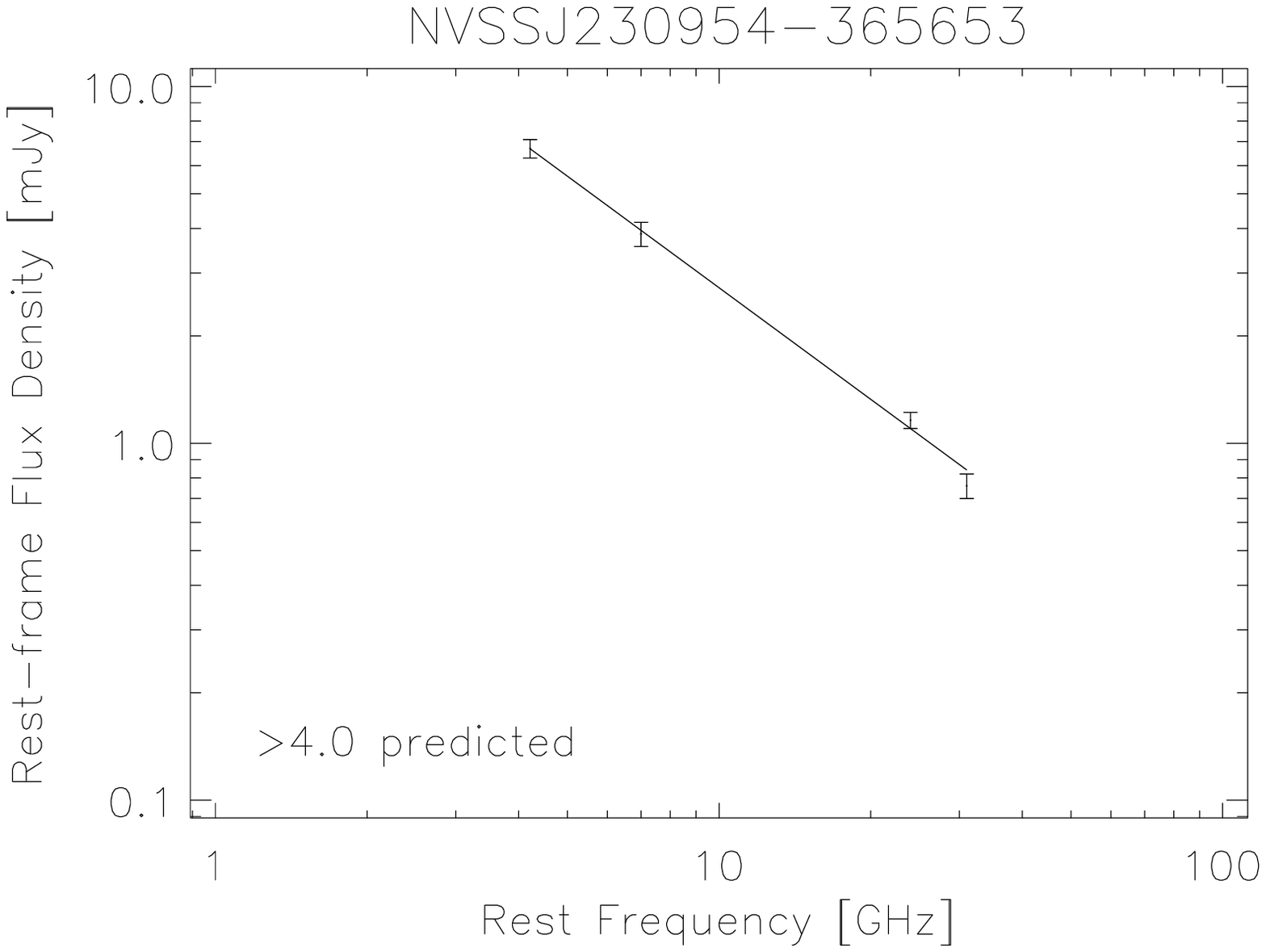} 

\includegraphics[width=8cm]{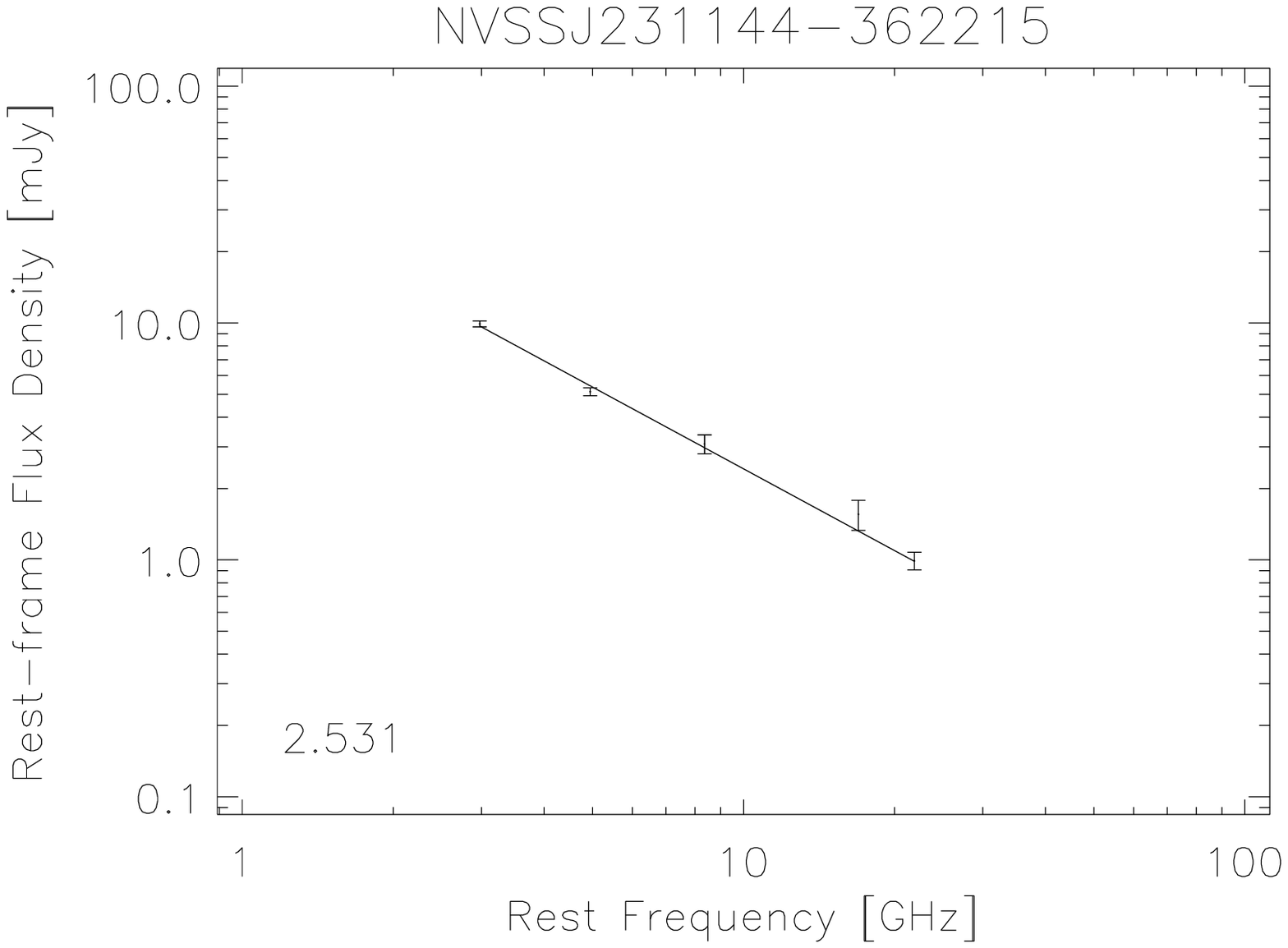} 
\includegraphics[width=8cm]{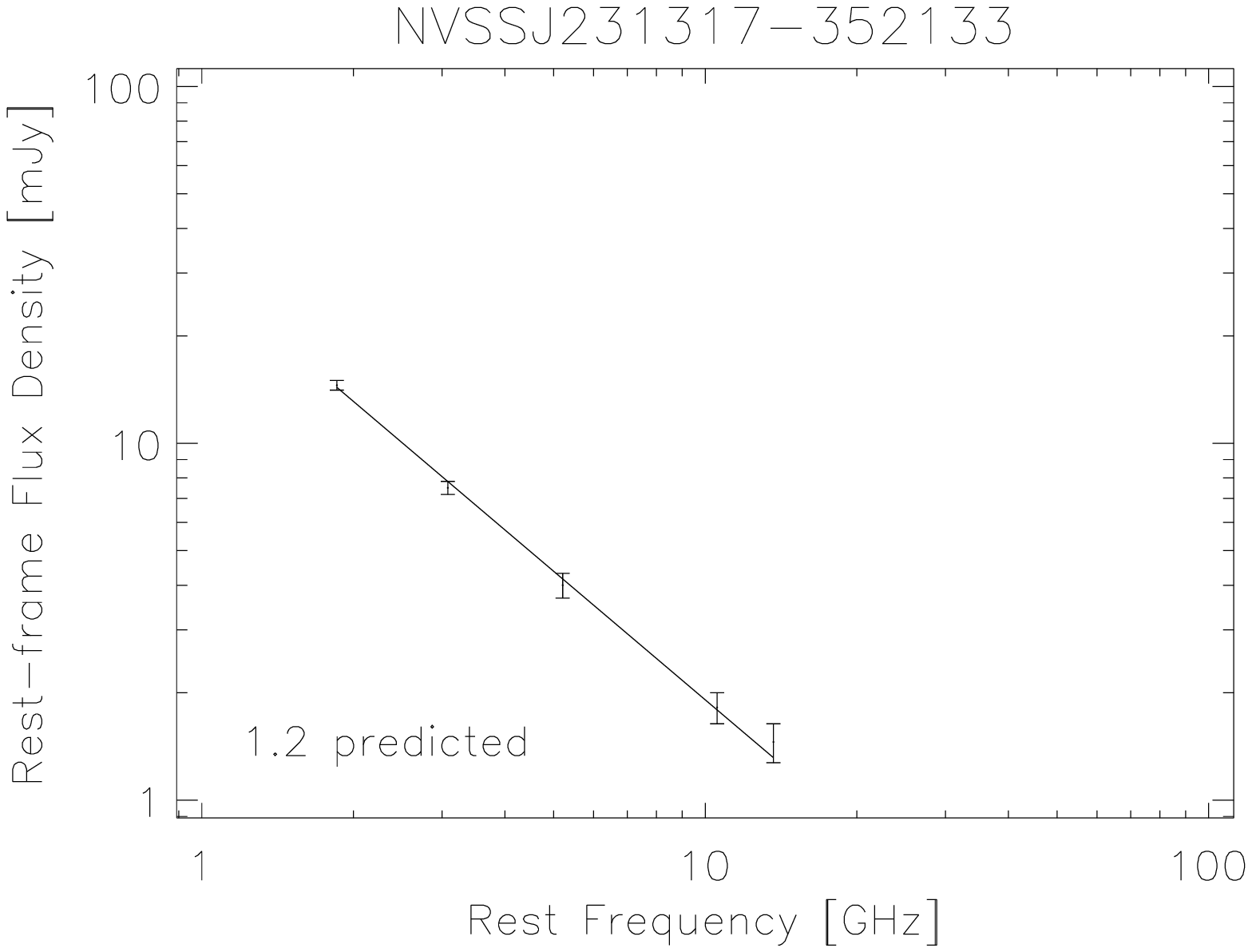} 
\caption{- continued}\end{center}\end{figure*}\setcounter{figure}{0}\begin{figure*}\begin{center}

\includegraphics[width=8cm]{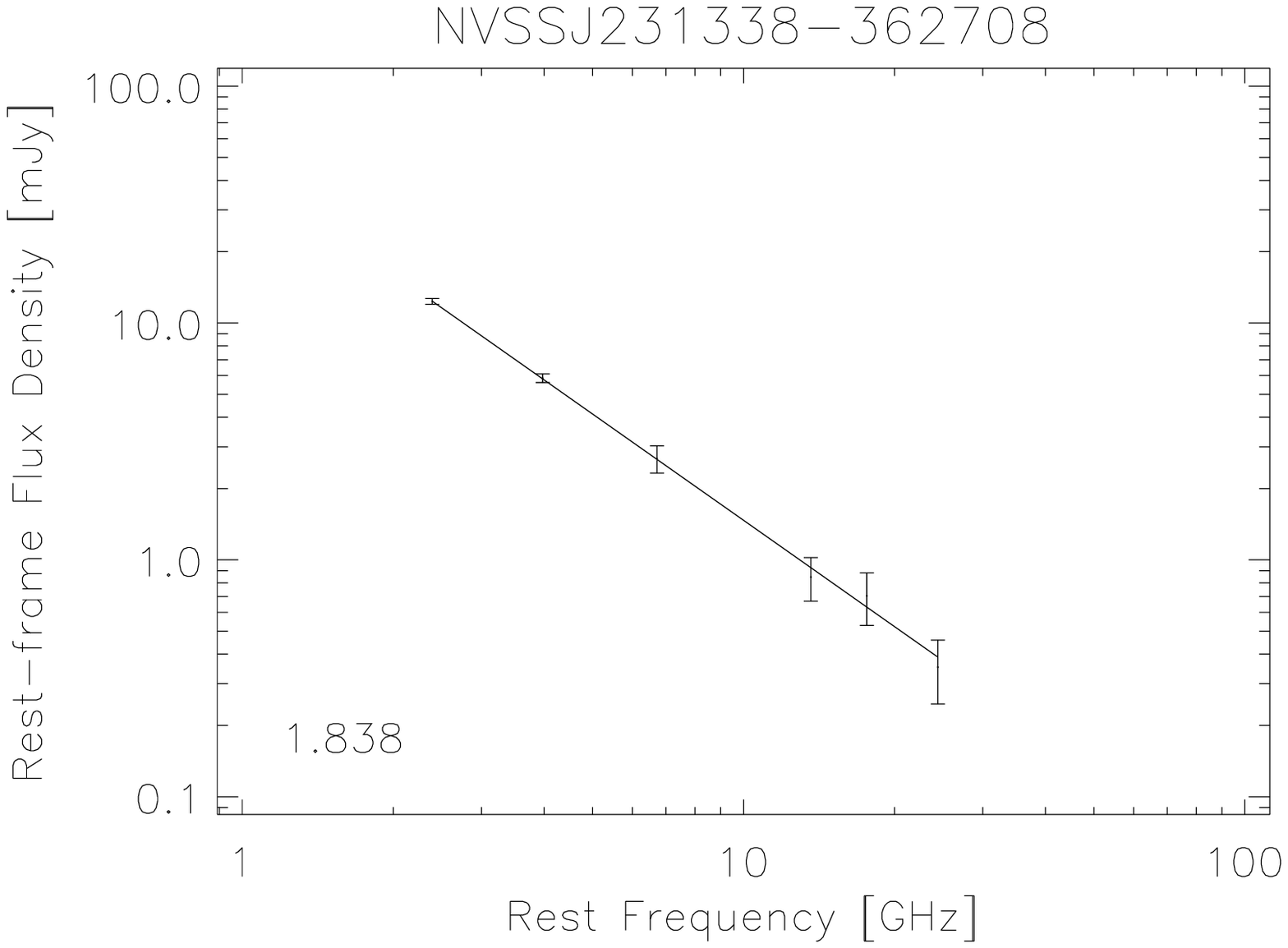} 
\includegraphics[width=8cm]{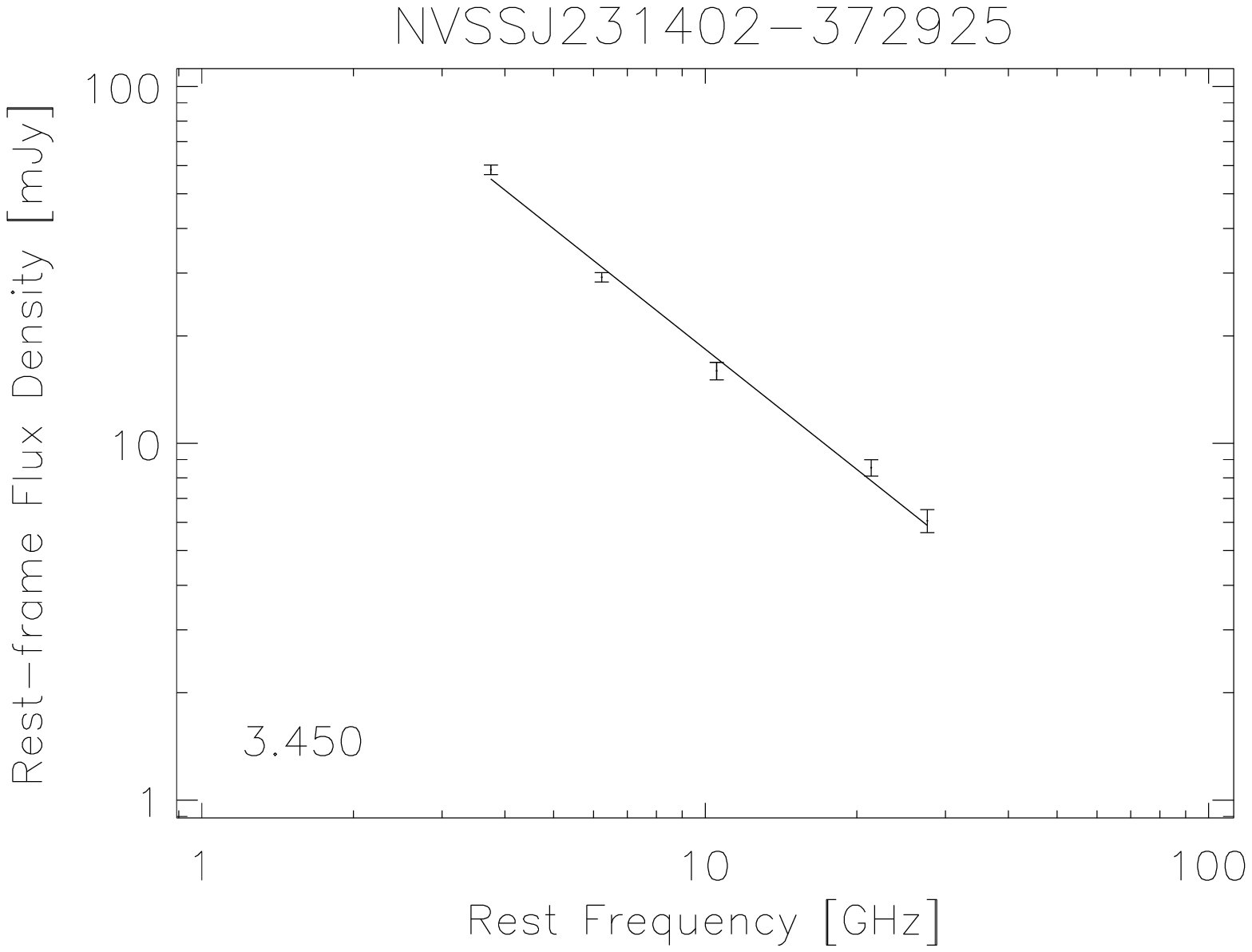} 

\includegraphics[width=8cm]{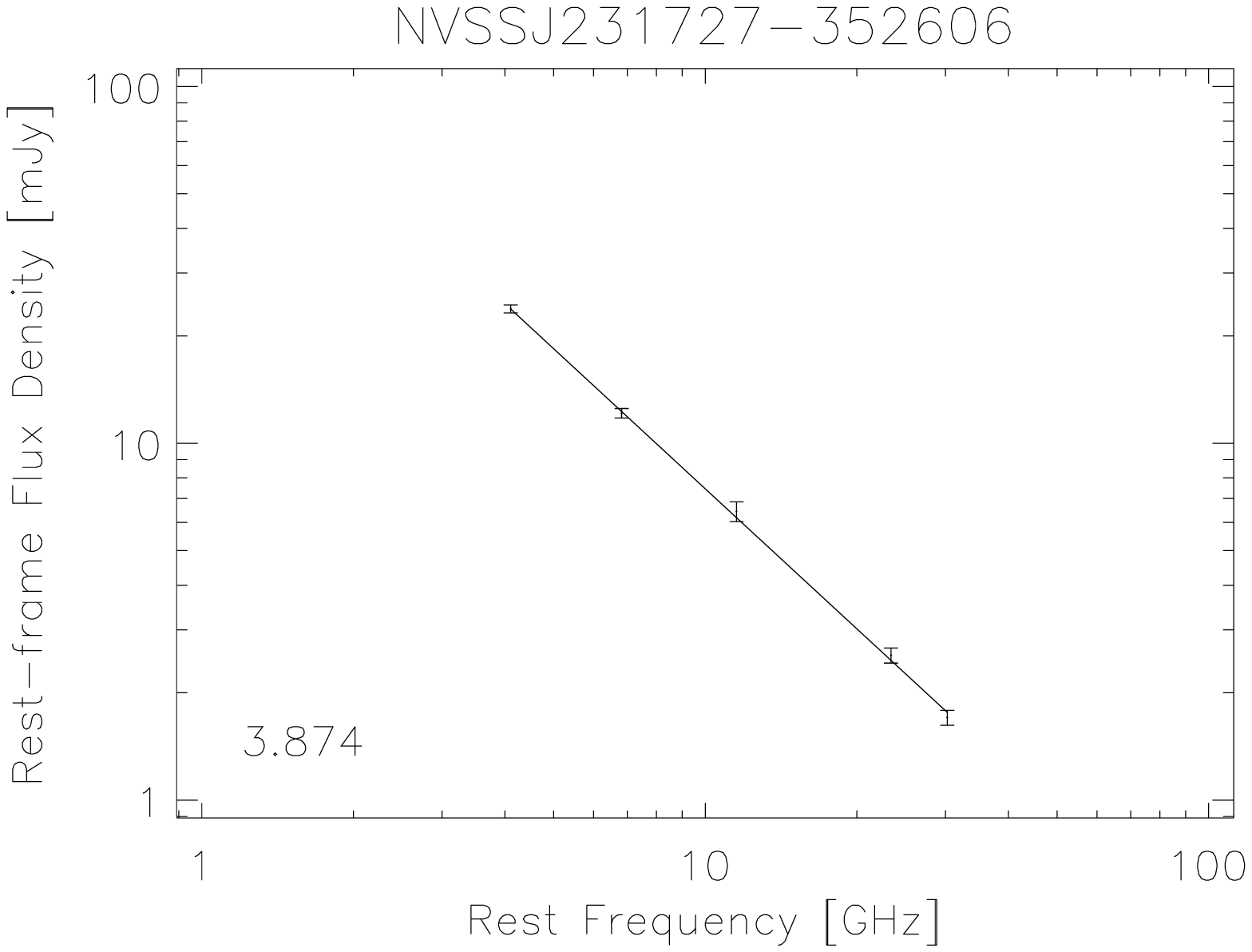} 
\includegraphics[width=8cm]{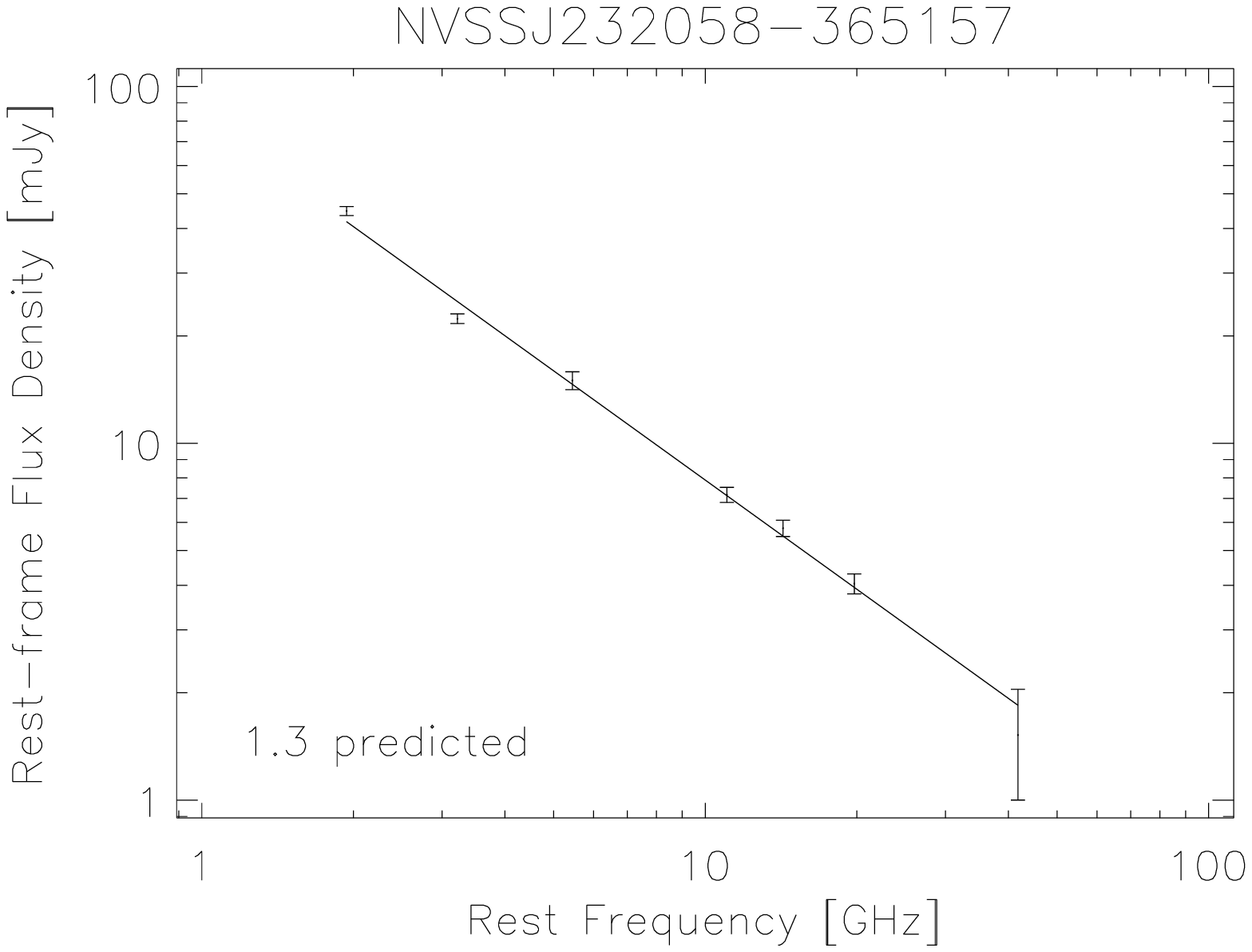}                    

\includegraphics[width=8cm]{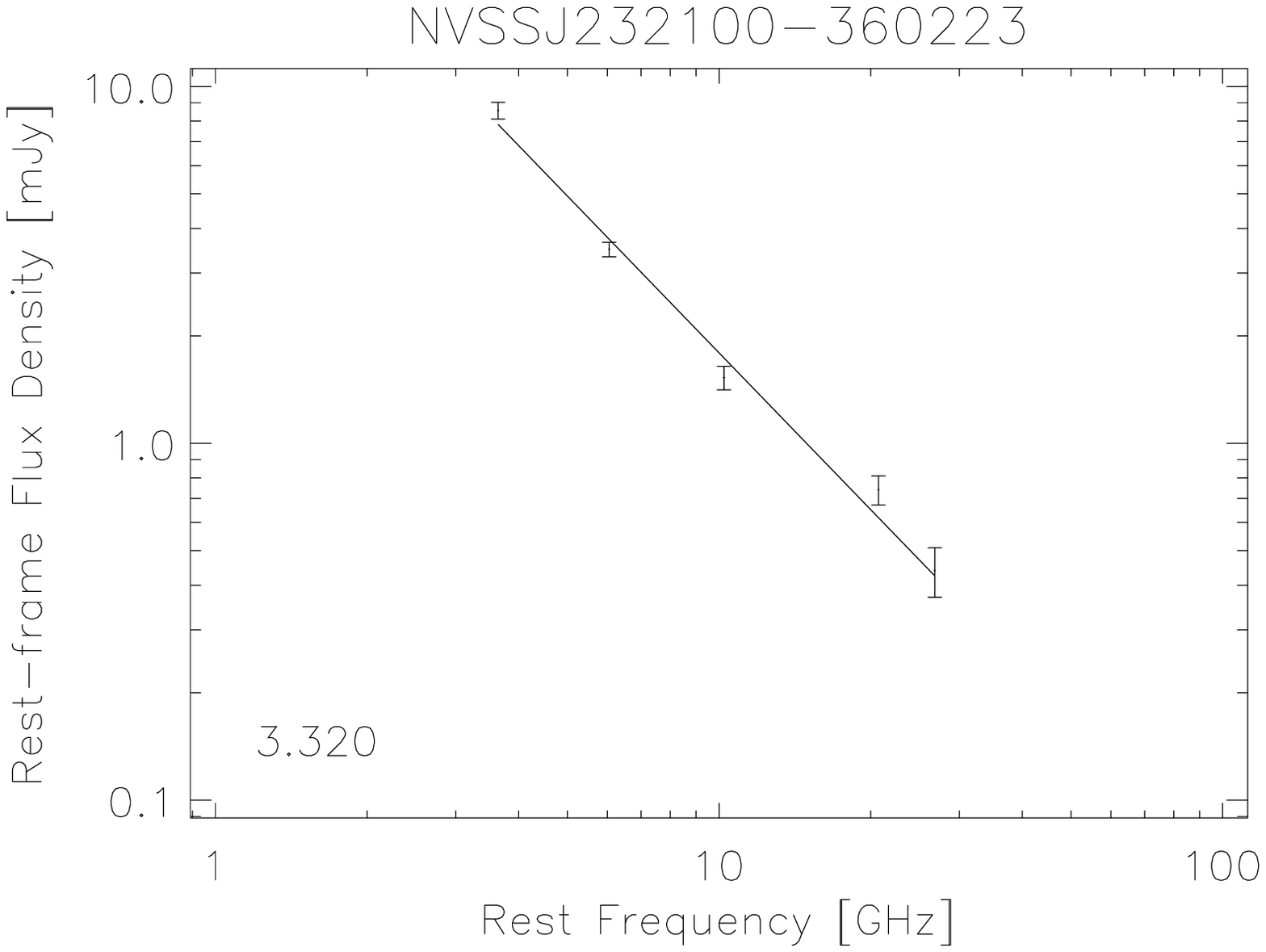} 
\includegraphics[width=8cm]{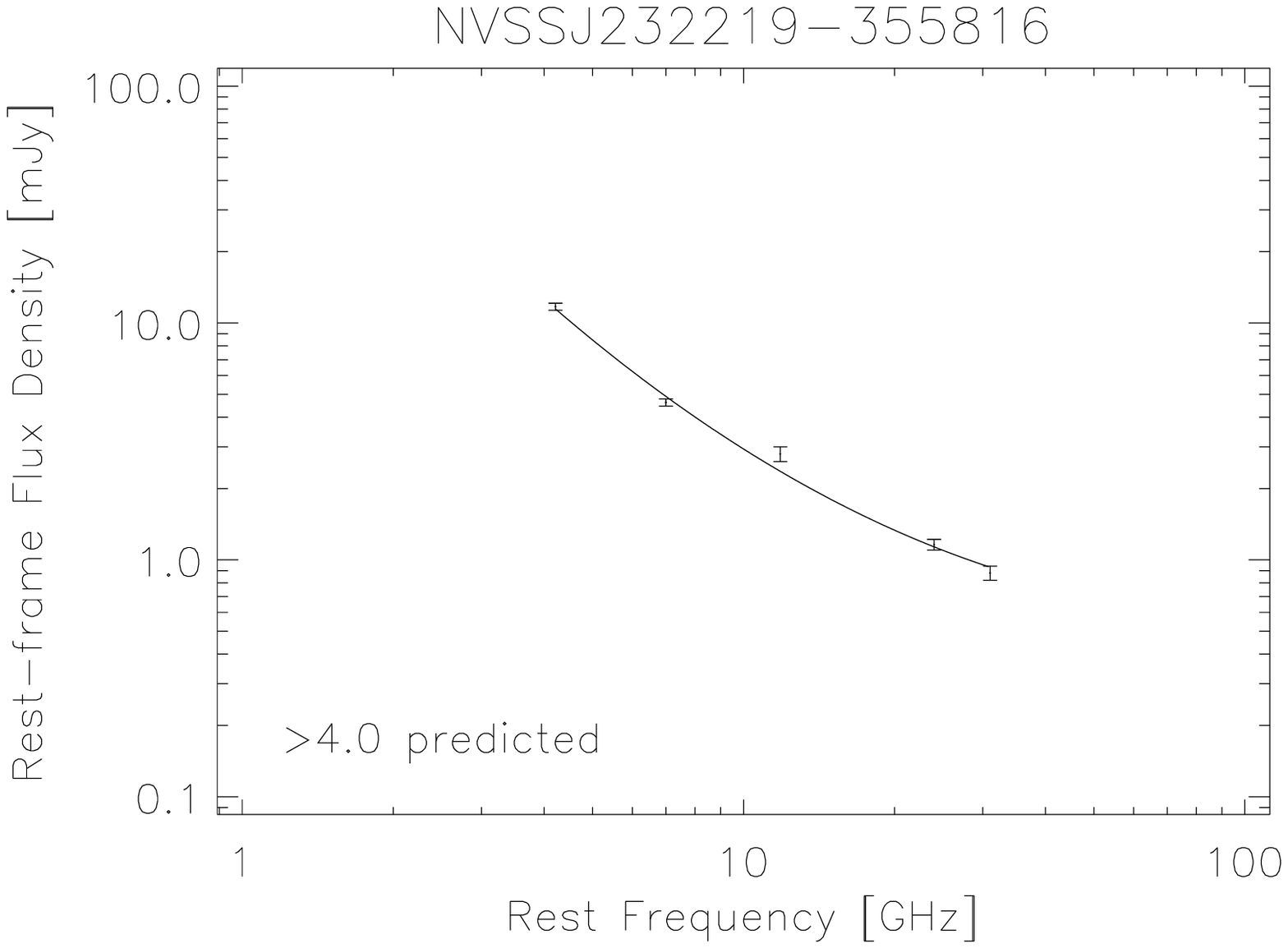} 

\includegraphics[width=8cm]{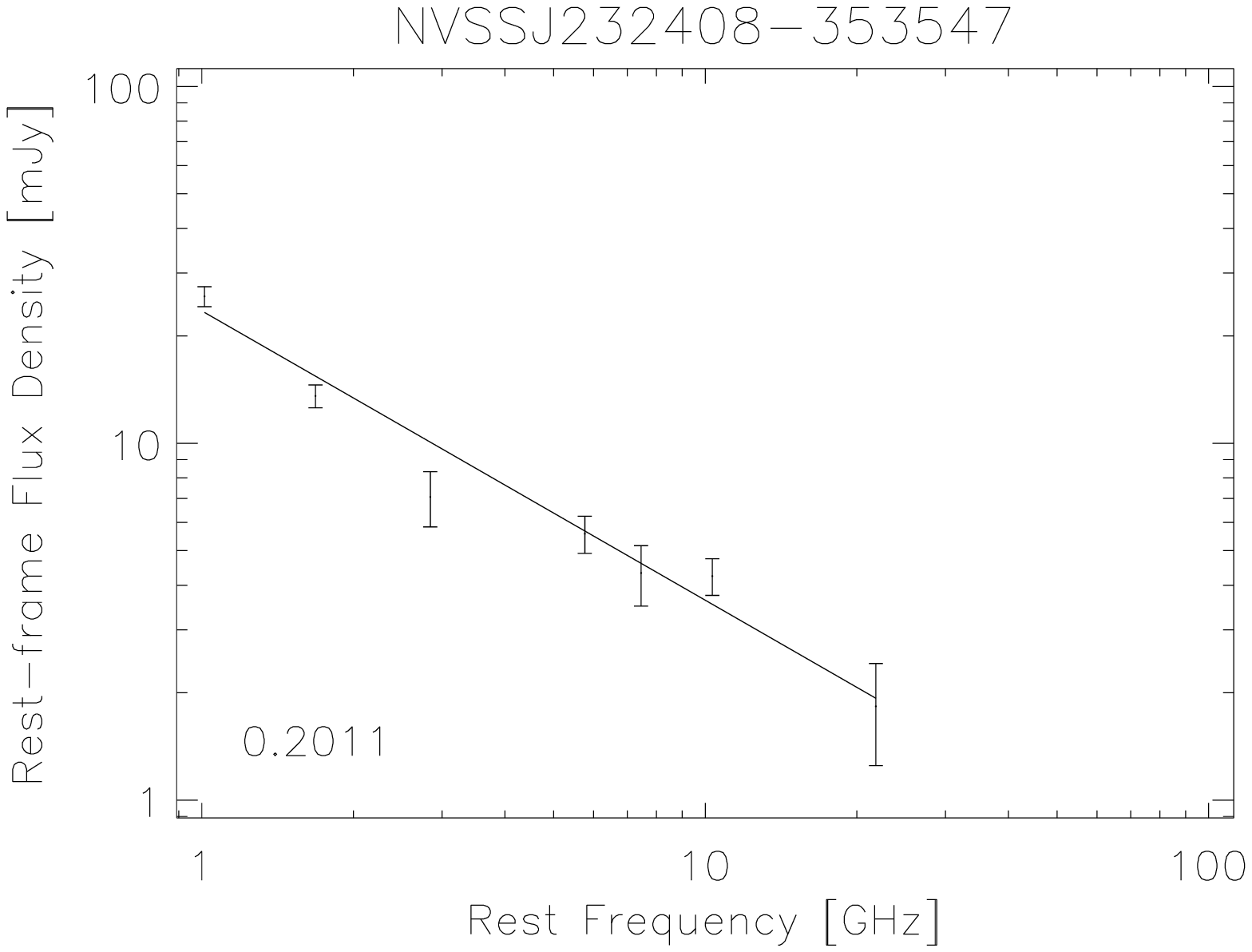} 
\includegraphics[width=8cm]{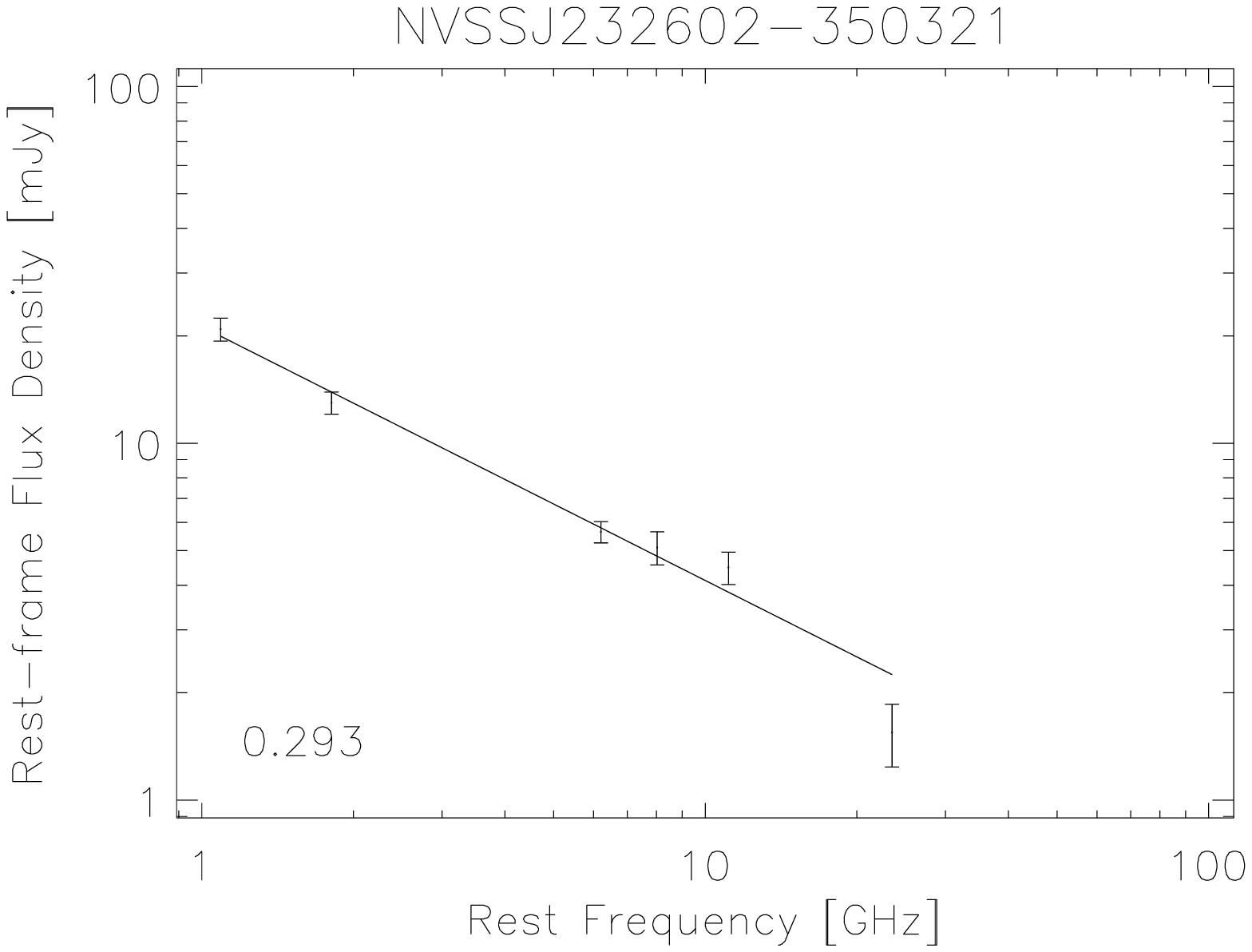} 

\caption{- continued}\end{center}\end{figure*}\setcounter{figure}{0}\begin{figure*}\begin{center}       

\includegraphics[width=8cm]{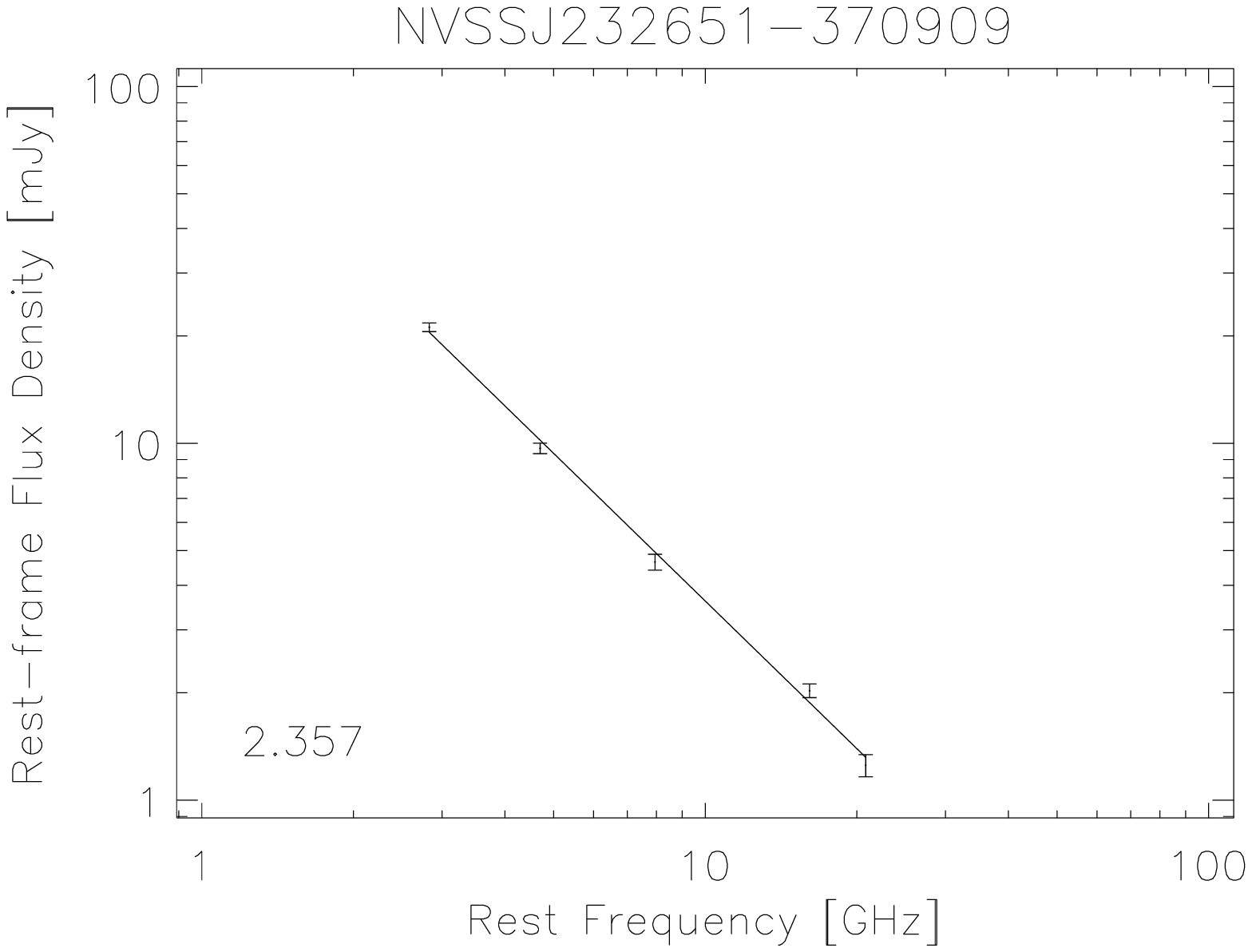} 
\includegraphics[width=8cm]{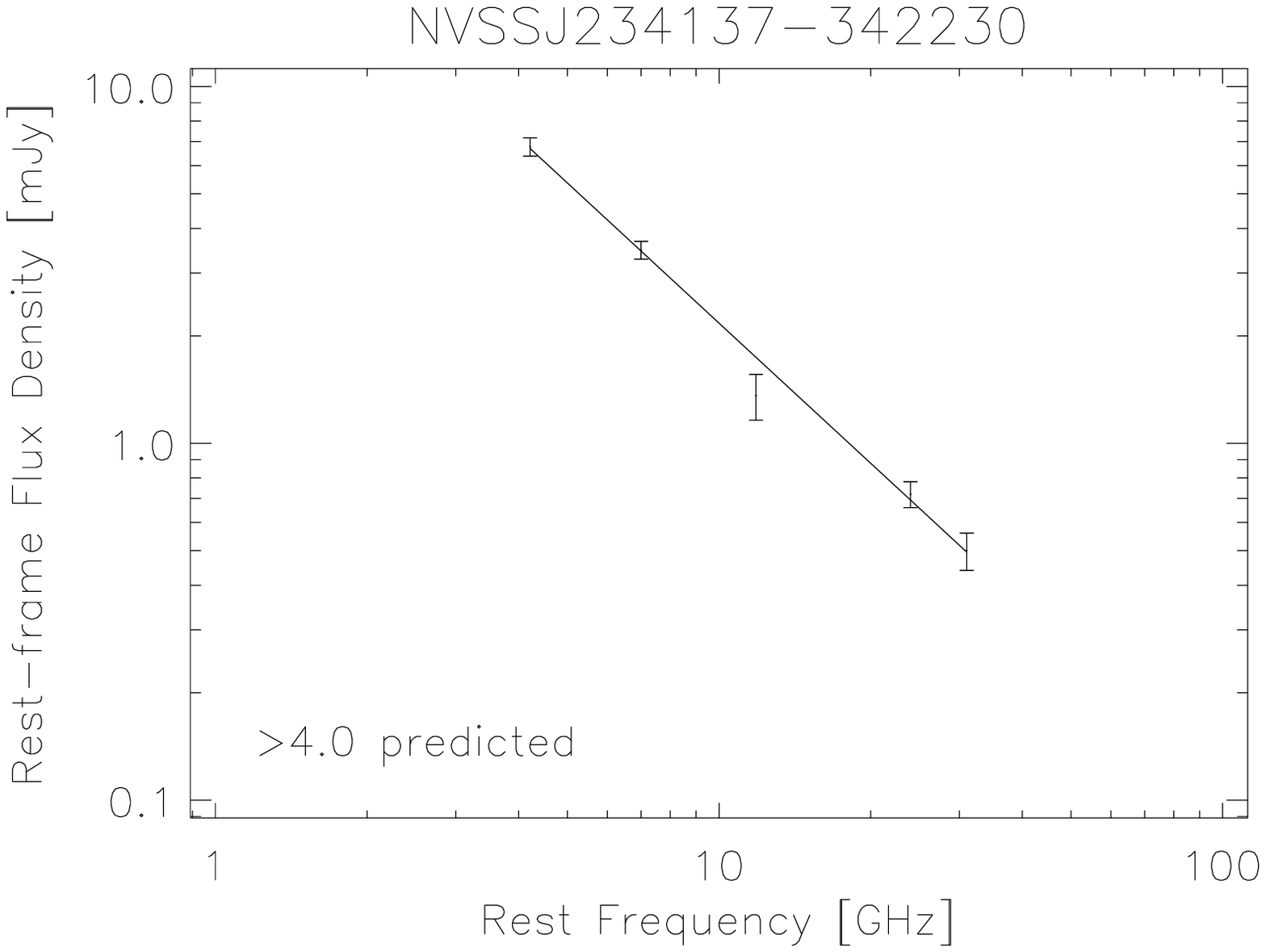}

\includegraphics[width=8cm]{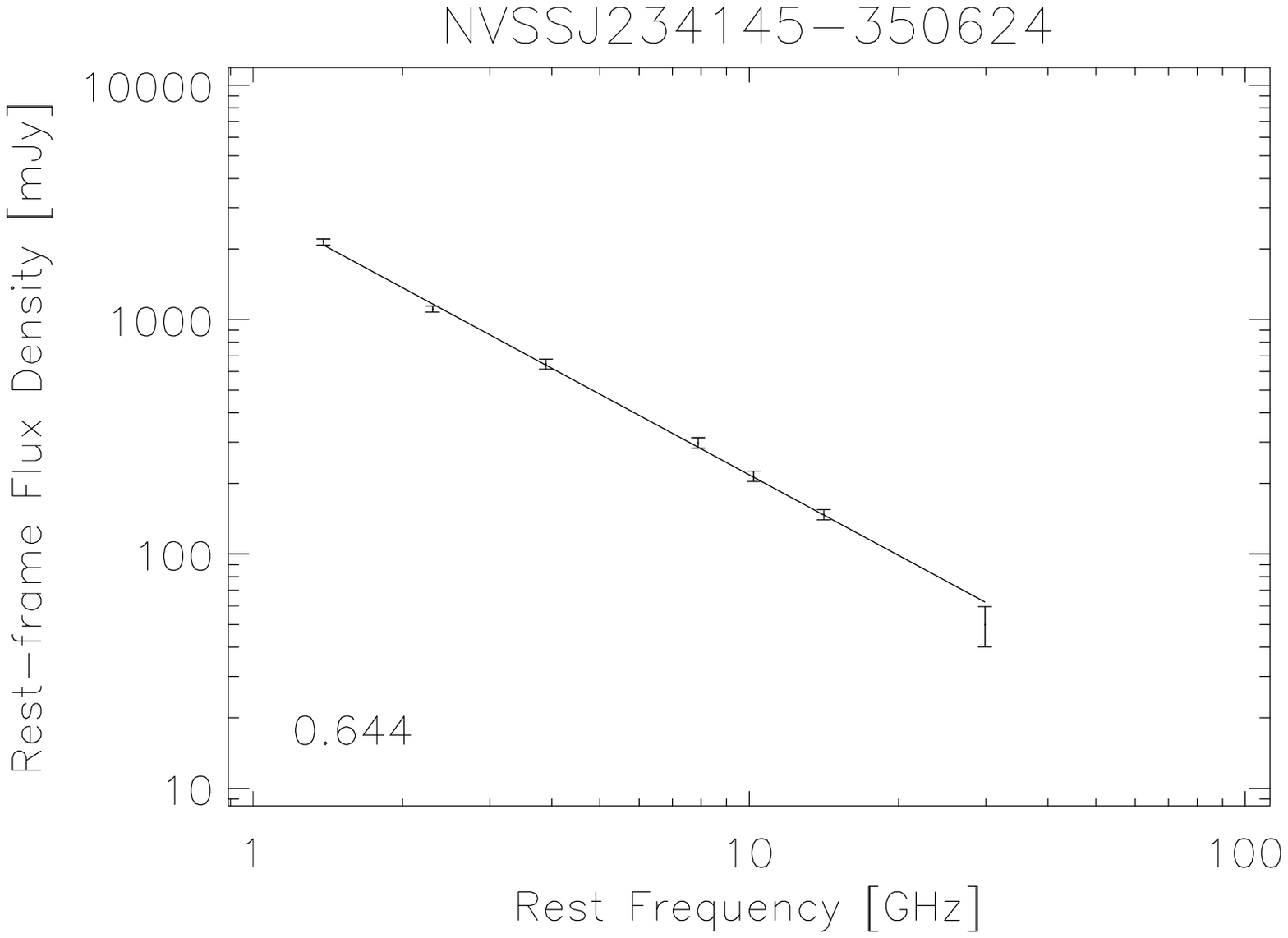} 
\includegraphics[width=8cm]{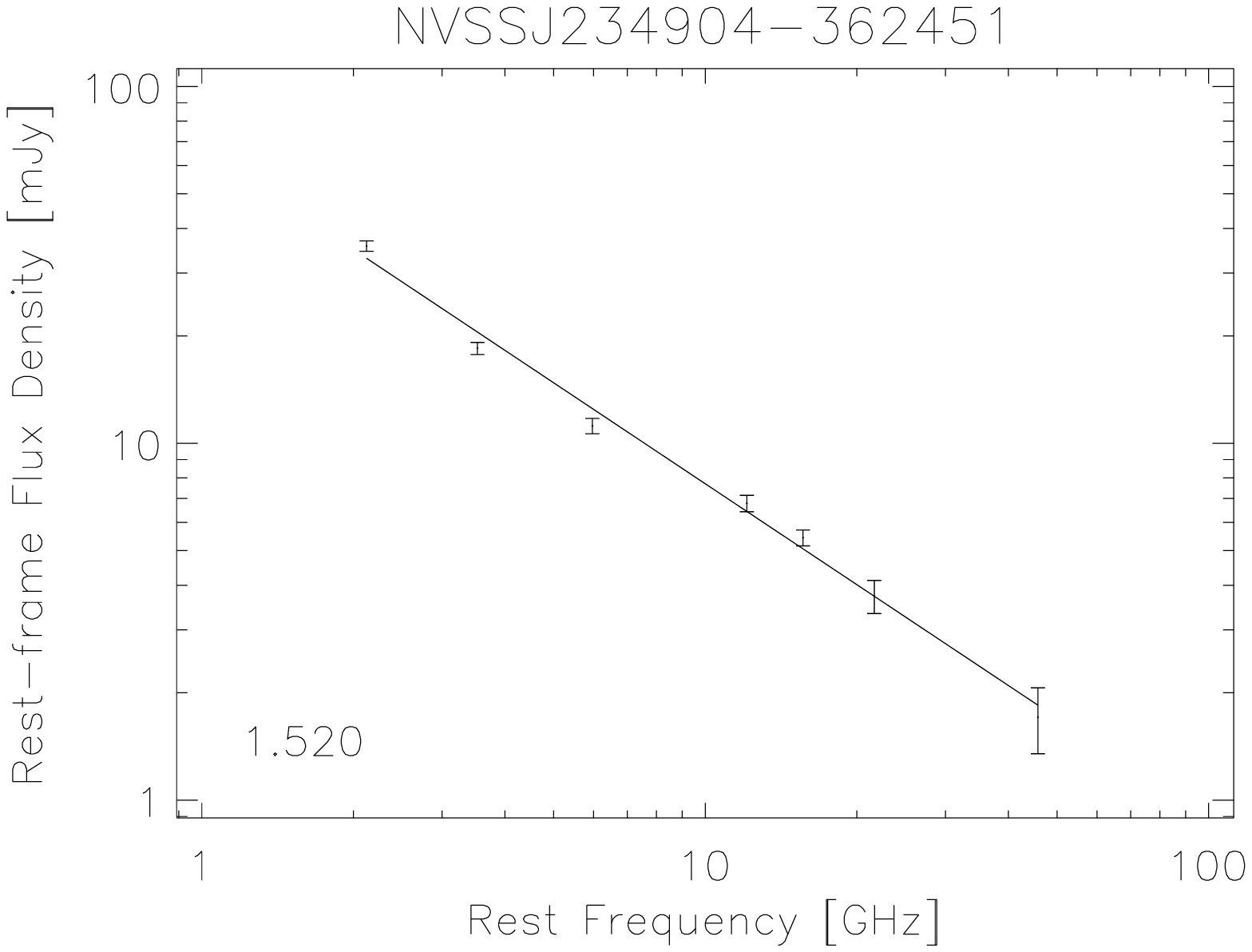} 

\includegraphics[width=8cm]{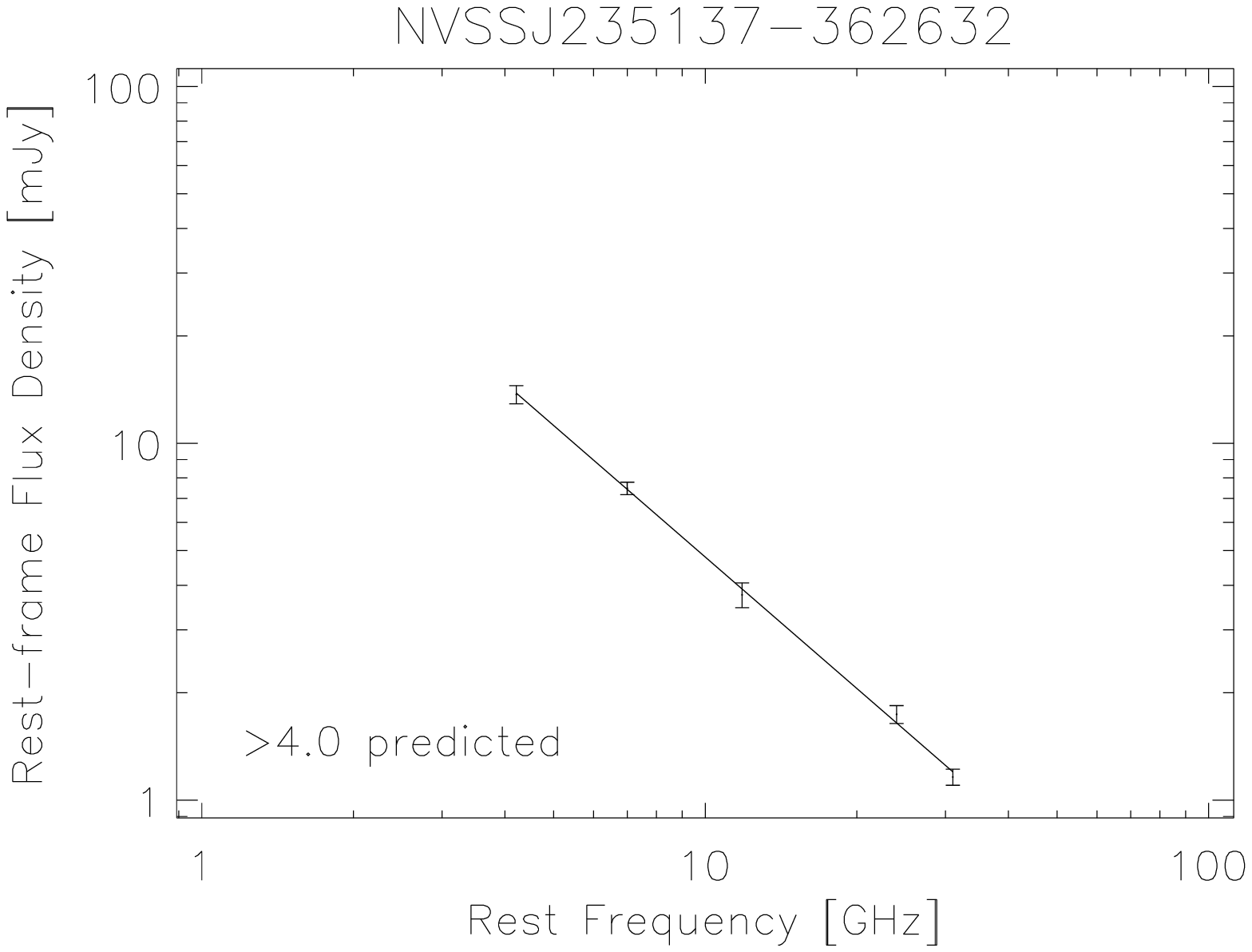} 
\caption{- continued}\end{center}\end{figure*}

\begin{table}\small
\caption{Functional form for the rest-frame SEDs fitted by $\rm log (\frac{S_\nu}{mJy})=\rm b + \alpha \rm log(\frac{\nu}{GHz}) + \beta (\rm log\frac{\nu}{GHz})^2$, and its associated $\chi^2$ goodness-of-fit. }
\label{sedfits}
\begin{center}
\begin{tabular}{lllll}\hline
Source  & b & $\alpha$ & $\beta$ & $\chi^2$\,$^{\dagger}$ \\
\hline
NVSS~J001339--322445$^*$&2.32& --0.97&0& 6.4\\
NVSS~J002131--342225&1.51&--1.59&0.57&4.1\\
NVSS~J002219--360728&1.42& --1.33 & 0.32& 12.4\\
NVSS~J002402--325253&1.79& --1.01&0&10.4\\
NVSS~J002627--323653$^*$&1.85& --1.23&0&12.0\\
NVSS~J011606--331241&1.60& --1.44&0&2.4\\
NVSS~J012904--324815&1.84& --1.37&0&1.9\\
NVSS~J015232--333952$^*$&2.49& --0.82&0&8.4\\
NVSS~J015324--334117$^*$&1.44& --0.81&0&14.6\\
NVSS~J015544--330633$^*$&1.70& --0.91&0&1.8\\
NVSS~J021308--322338&1.56& --0.90&0&3.8\\
NVSS~J030639--330432&1.89& --1.63&0&2.0\\
NVSS~J202026--372823$^*$&1.81& --1.16&0&4.6\\
NVSS~J202140--373942$^*$&1.44& --0.92&0&7.1\\
NVSS~J202945--344812$^*$&1.85& --0.94&0&13.5\\
NVSS~J204420--334948&1.89& --1.50&0&1.5\\
NVSS~J213510--333703&2.00& --2.07& 0.47&2.3\\
NVSS~J225719--343954&2.00& --1.77&0&11.1\\
NVSS~J230035--363410&1.64&--1.29&0& 13.3\\
NVSS~J230123--364656&1.81&--1.47&0& 7.3\\
NVSS~J230527--360534&1.97&--1.39&0& 1.3\\
NVSS~J230954--365653$^*$&1.47& --1.04&0& 2.8\\
NVSS~J231144--362215$^*$&1.49& --1.10& 0&2.1\\
NVSS~J231317--352133$^*$&1.45& --1.15& 0&2.1\\
NVSS~J231338--362708&1.69& --1.53& 0&1.8\\
NVSS~J231402--372925$^*$&2.32& --1.06&0& 4.6\\
NVSS~J231727--352606$^*$&2.15&--1.28&0& 2.5\\
NVSS~J232058--365157$^*$&1.87& --0.97& 0&9.2\\
NVSS~J232100--360223&1.67& --1.42&0& 7.8\\
NVSS~J232219--355816&2.45&--2.63&0.65&9.7\\
NVSS~J232408--353547$^*$&1.32& --0.75& 0&6.3\\
NVSS~J232602--350321$^*$&1.37&--0.76& 0&10.0\\
NVSS~J232651--370909&1.87& --1.32& 0& 3.9\\
NVSS~J234137--342230&1.64& --1.30& 0&3.1\\
NVSS~J234145--350624$^*$&3.45& --1.11&0& 3.3\\
NVSS~J234904--362451$^*$&1.79& --0.90&0& 13.2\\
NVSS~J235137--362632$^*$&1.91& --1.22 &0& 1.7\\

\hline
\end{tabular}\\
$*$These sources do not meet the original USS sample threshold of $\alpha\leq-1.3$ (see \S\ref{revised843}).\newline
$\dagger$The degrees of freedom for the $\chi^2$ values can be obtained directly from Table~2.
\end{center}
\end{table}

\section{Discussion}\label{discussion}
A remarkable feature of the 37 SEDs in our sample is that they can be described, in the majority (33 out of 37) of cases, by a single power law. For the four (11\%) spectra which do show curvature, these flatten rather than steepen toward higher frequencies. We note that these four flattening sources all have $\alpha^{1400}_{843}<-1.3$. This latter behaviour is indicative of either (i) the emergence of flatter spectrum core or hotspot components becoming important contributors to the total flux density at high frequencies (currently, we do not have
observations of these four sources at high enough angular resolution to test this), or (ii) preferential biasing toward sources with $\alpha_{843}^{1400}>-1.3$ which have scattered into the SUMSS-NVSS USS sample (see \S 5.2 in paper~I). Not one of the 37 objects in our sample shows evidence of steepening toward higher frequencies. We remind the reader that our sample was selected between 843~MHz and 1.4~GHz in the observed frame. The corresponding rest-frame frequencies range from $\sim4$~GHz for $z\geq3$ to $\sim1$~GHz at the lowest redshifts. We interpret this result as evidence that, regardless of its redshift, the SUMSS-NVSS USS sample cannot be described by SEDs which steepen at rest frequencies beyond about 1~GHz. Lower frequency observations of the highest redshift galaxies are, of course, required to determine the actual extent of any curvature between 1~GHz and 4~GHz in the rest frame.\newline 

This result is in striking contrast to complete samples such as 3C, 6C and 7C where approximately 70\% of the SEDs steepen toward higher frequencies \citep{blu99a}. The fraction seems to be dramatically reduced in steep-spectrum selected samples: down to 29\% for a sample selected at 151~MHz with $\alpha^{151}_{4850}<-0.981$ \citep{blu98}, and 0\% for a sample selected at 1400~MHz with $\alpha^{843}_{1400}\leq-1.3$ (this paper). The underlying reason for these different outcomes is unclear: possibilities include differences in flux density limit, finding frequency and the spectral index threshold which is applied to the samples. We have recently begun investigating these possibilities and will report our findings in a future paper. \newline

\section{The redshift -- spectral index correlation}\label{zedalphadiscussion}
As mentioned previously (\S\ref{intro}), the mechanisms invoked to explain the apparent correlation between redshift and observed radio spectral index are (i) k-corrections of steepening radio SEDs, (ii) enhanced spectral aging due to increased inverse Compton losses against the CMB at high redshift, and (iii) an intrinsic correlation between low frequency spectral index and radio luminosity coupled to a Malmquist bias. We now consider each of these possibilities as explanations of the observed $z-\alpha^{843}_{1400}$ correlation which has led to the discovery of a large fraction of $z>3$ SUMSS-NVSS USS radio galaxies (paper~II). 

\subsection{A k-correction}
\begin{figure*}\centering
\includegraphics[width=15cm]{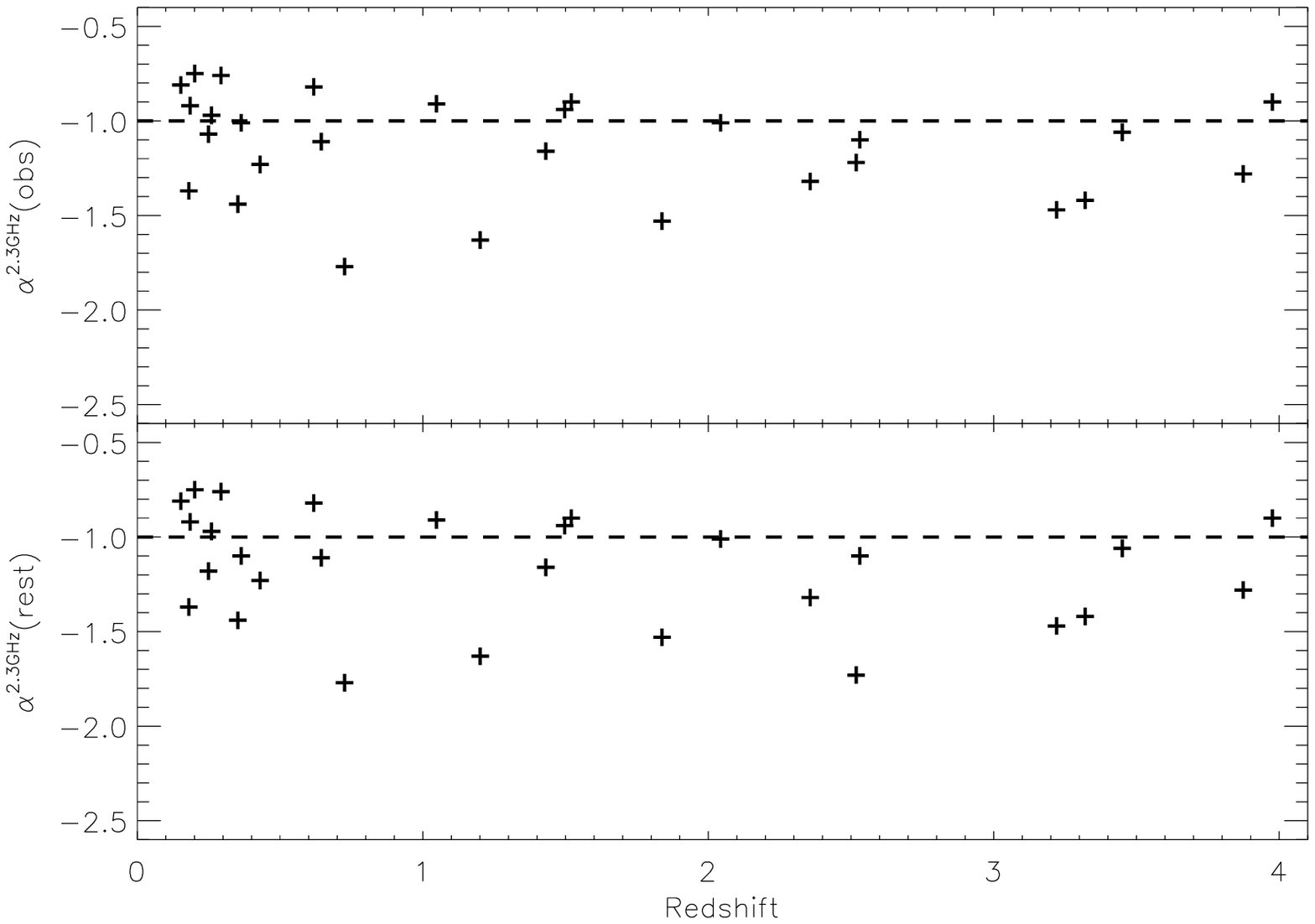}
\caption{Spectral index versus redshift, for the 28 galaxies in this sample with spectroscopically confirmed redshifts. {\textit{Upper panel:}} measured at 2.3~GHz in the observed frame. {\textit{Lower panel:}} measured at 2.3~GHz in the rest frame. Since all but three sources are characterised by a single power law, these plots are almost indistinguishable. The horizontal dashed line at $\alpha=-1.0$ helps to guide the eye.} \label{zalphaplots}
\end{figure*}

If a k-correction is mainly responsible for the large surface density of SUMSS-NVSS USS $z>3$ radio galaxies, then we expect those same sources to be characterised by SEDs which steepen toward higher frequencies. Clearly, our results are inconsistent with this requirement. Indeed, \citet{cha90} also dismissed a k-correction for the steep spectral index of the well-known $z=3.8$ radio galaxy 4C+41.17, after finding an essentially straight spectrum between 26\,MHz and 10\,GHz. In Figure~\ref{zalphaplots}, we show the spectral indices measured at 2.3~GHz in the observed frame ({\it upper panel)} and the rest frame ({\it lower panel)} for the 28 radio galaxies in this sample with spectroscopically confirmed redshifts. As expected, there is very little difference between the two distributions, with the only exception being a steepening in the rest frame for three sources in the sample with flattening SEDs. \newline

Whilst many authors have shown that a k-correction is a credible explanation for the $z-\alpha$ correlation \citep{gop88,lac93,car99preprint}, some of these same studies also show that an apparent --- yet weaker --- correlation still exists between redshift and {\it {rest-frame}} (k-corrected) radio spectral index \citep{lac93,car99preprint,blu99a}. These samples were all selected at frequencies in the range 38~MHz$\leq \nu\leq$408~MHz, at least a factor of three lower in finding frequency than the SUMSS-NVSS USS sample, and at least an order of magnitude brighter in radio luminosity. The differences in selection frequency and luminosity could explain our contrary result. However, at the same flux density limit, the surface density of SUMSS-NVSS USS radio galaxies at $z>3$ is similar to the WENSS-NVSS USS sample of \citet{cdb00sample} which was selected between 325~MHz and 1.4~GHz. Since a k-correction does not apply to our sample, there must be another mechanism contributing to a $z-\alpha$ correlation.  \newline

\subsection{Inverse Compton and synchrotron losses} 

Synchrotron physics predicts {\it equivalent} time-independent spectral changes due to IC and synchrotron losses. This means that cooling of the electron energy distribution due to both mechanisms manifests itself in the same way on the SED (see \S\ref{synchphysics}). Enhanced IC losses simply shorten the timescale over which this cooling occurs (Equation~\ref{10}). So, for example, a radio lobe with a given dynamical age will have suffered more radiative losses at high redshift due to the $(1+z)^4$ dependence. The critical issue here is that this will simply lead to a lower break frequency. It will not steepen the SED beyond the canonical $\Delta \alpha=0.5$, for a continuous injection model. Compared with a low-redshift radio galaxy which is sampled beyond its break frequency, enhanced IC losses will not lead to steeper radio spectral indices in high-redshift sources. \newline 

We also note that if equipartition between the magnetic field and relativistic electrons is a valid approximation at high redshift, then synchrotron losses will continue to dominate over IC losses out to $z\sim3$ in radio lobes (for typical lobe magnetic field strengths of $35\mu$G; \citealp{har02}) and out to $z\sim8$ in hotspots (for typical hotspot strengths of $150\mu$G; \citealp{har02}).

\subsection{Luminosity dependence} 

Because of Malmquist bias, any mechanism which produces a correlation between spectral index and radio luminosity will give rise to an indirect correlation between spectral index and redshift. \citet{blu99a} find that radio spectral indices (measured at rest-frame 151~MHz) are indeed steeper in more powerful radio galaxies. They posit that the correlation arises because radio galaxies with larger luminosities also have larger jet powers which, in turn, produce larger magnetic fields when the jet kinetic energies thermalise at the hotspots. The larger magnetic fields result in more rapid electron cooling times (Equations~\ref{4} and \ref{9}) which lead to steeper electron energy distributions injected into the lobes. The extent to which the spectral index -- radio luminosity correlation contributes to the large surface density of USS HzRGs in the SUMSS-NVSS sample remains unclear; in order to explore this further, the degeneracy between luminosity and redshift in this sample must be broken, as it was for the complete samples studied by \citet{blu99a}. \newline 

\section{What can we learn from nearby cluster galaxies?}\label{neighbours}

\begin{figure*}\centering
\includegraphics[width=15cm]{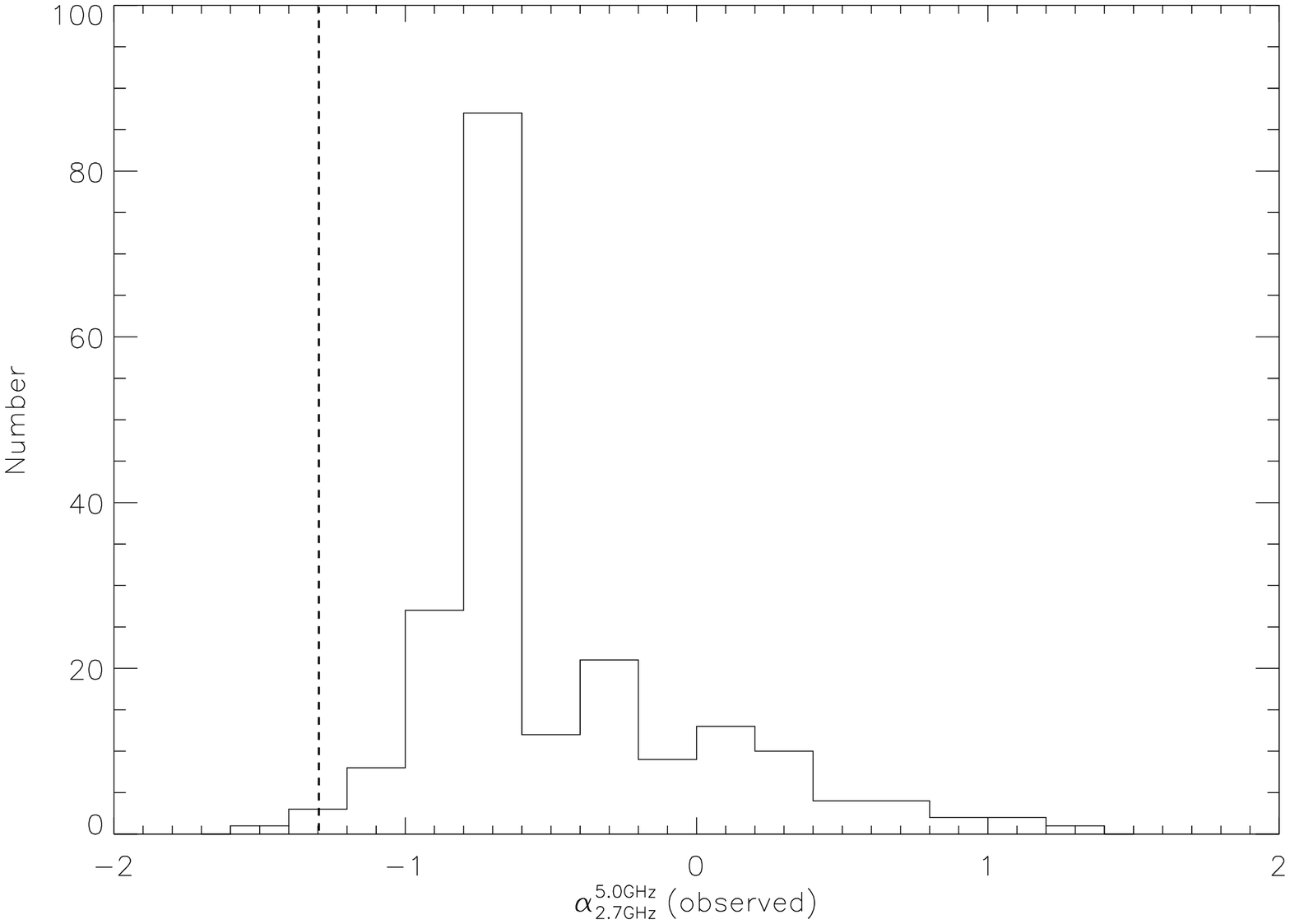}
\caption{Spectral index distribution between 2.7~GHz and 5~GHz for radio sources selected from high finding frequency samples and spectroscopically confirmed to have $z<0.5$ \citep{kue81,sti94}. Most sources have spectral indices around $\alpha=-0.85$. The vertical dashed line at $\alpha=-1.3$ indicates the USS cutoff we used to define our sample.} \label{kuehrdistr}
\end{figure*}

The spectral index distribution for a sample of nearby $z<0.5$ radio galaxies selected at 5~GHz \citep{kue81,sti94} is shown in Figure~\ref{kuehrdistr}. This nearby sample has been selected at a similar rest-frame finding frequency to the $z>3$ radio galaxies in the SUMSS-NVSS USS sample. The average spectral index of the nearby sample is $\alpha^{2.7}_{4.5}=-0.85$ with less than 1\% having indices as steep as $-1.3$ (the USS classification). If the rest-frame spectral index distributions of nearby (and less powerful) and distant radio galaxies are similar, this clearly implies that by using a spectral index cull as severe as $-1.3$, the high redshift radio galaxies being selected are not `typical' of the 5~GHz selected samples, but fall at the extreme edge of the spectral index distribution for radio sources.  \newline

The homogeneity of rest-frame SEDs for the 37 sources in the sample might imply that the physics leading to their steep spectral indices can also be explained by a uniform mechanism. However, as always the Malmquist bias inherent in single samples prevents us from breaking the redshift-luminosity degeneracy. Nonetheless, it is tempting to ask what self-consistent mechanism could at the same time explain the physics of USS radio galaxies, and demand that it correlates with increasing redshift. \newline

More than three decades ago it was realised that nearby radio sources in steep-spectrum-selected samples reside almost exclusively in rich clusters of galaxies \citep{bal73,sli74a,sli74b}. It is also known that cluster radio sources display steeper spectral indices than field radio sources, with the steepest spectrum sources residing closest to the cluster centres \citep{sle83}. This has been interpreted as the manifestation of pressure-confined radio lobes which slow adiabatic expansion losses \citep{bal73,kom94,jon01}. A radio lobe will expand adiabatically until gas pressure equilibrium is reached between the lobe and the ambient gas pressure. From this it follows that in the centres of rich clusters, and other similar environments where the ambient baryon densities are large, radio lobes will be pressure-confined and lose energy primarily via synchrotron and IC losses. In more rarefied environments, like those surrounding isolated galaxies, the luminosity of the radio lobes would fade out much faster due to adiabatic expansion energy losses and decreasing surface brightnesses. Note that these nearby sources are of the lower radio luminosity Fanaroff-Riley type~I (FR\,I) classification \citep{fan74}. The high-redshift sources being selected using USS techniques, on the other hand, invariably have to have radio luminosities comparable with with the more powerful FR\,II class to be above the flux selection limit. We note that previous studies of USS-selected radio galaxies find a large (30\%) fraction of compact steep-spectrum (CSS) sources \citep{cdb00sample}. In addition, USS radio galaxies with classical FR\,II morphologies at $z\sim2$ are more distorted compared with their low-redshift FR\,II counterparts \citep{car97,pen00}. So, although we acknowledge the danger in drawing analogies between local steep-spectrum FR\,Is and distant FR\,IIs, we believe that high ambient densities play an important role in the physics leading to the unusually steep radio spectral indices of both classes of source.

\subsection{A new explanation for the $z-\alpha$ correlation}

Nearby FR\,I radio galaxies with ultra-steep spectral indices are rare. Those which do exist reside overwhelmingly in regions of high baryonic densities. By analogy, we speculate that if there is evolution in the richness of the environments of powerful radio galaxies in the sense that these radio galaxies are increasingly more likely to occupy regions of higher ambient densities at higher redshifts, then we would expect to find, on average, more steep-spectrum radio galaxies as a function of redshift. This would be a natural, and physical, interpretation of the $z-\alpha$ correlation. We note here that almost a decade ago \citet{ath98b} also posited that the $z-\alpha$ correlation is driven by the propagation of radio jets into regions of higher gas density as a function of redshift. These authors asserted that the injection of steeper electron energy spectra occurs naturally as a function of redshift in first-order Fermi acceleration processes due to the decreased hotspot advance velocities in a progressively denser (and hotter) inter-galactic medium. \newline

There are some good reasons we believe that radio galaxies are increasingly more likely to occupy regions of higher ambient densities at higher redshifts. Most importantly, cosmological expansion of the universe means that the average gas density of the intergalactic medium was greater in the past by a factor of $(1+z)^3$. In addition, there is abundant observational evidence to support the notion that distant radio galaxies reside in high density environments:
\begin{enumerate}
\item{Firstly, the environments of many powerful radio galaxies within the range $1\lesssim z \lesssim 2$ show an excess of companion galaxies within a 1~Mpc radius (\citealp{bor04,bes03,bes00}, and references therein). The direct detection of star-forming companion galaxies surrounding $z>2$ radio galaxies, with narrow-band Ly\,$\alpha$ or H\,$\alpha$ emission line searches, also indicates that steep-spectrum radio galaxies live in regions of galaxy overdensity comparable with the richest galaxy clusters \citep{kur00,pen00b,ven02,ven04,mil04,ven04,ove06}.}\newline

\item{Secondly, radio polarimetric imaging of $z>2$ radio galaxies and radio-loud quasars has shown that many have rotation measures which, if intrinsic to the source environment, are in excess of 1000~rad~m$^{-2}$ \citep{car97,pen00a}. The rotation measure ($RM$) is given by 
\begin{equation}
RM = 811.9 \int_0^L n_e\, \vec{\pmb{B}} \cdot \vec{\pmb{dl}},
\end{equation}
where $n_e$ is the electron density in cm$^{-3}$, $\vec{\pmb{B}}$ is the magnetic field in $\mu G$ and $L$ is the path length in kpc. Unless the magnetic fields are somehow unusually large at high redshift (a situation that seems unlikely), this implies that the extreme RMs found in HzRGs are the result of dense environments similar to nearby X-ray cooling-flow clusters \citep[and references therein]{car02araa}. Extended X-ray emission has been attributed to thermal emission from dense (0.05~cm$^{-3}$) gas in at least one HzRG with a large rotation measure \citep{car02xray}.}\newline

\item {Thirdly, observations of the fields surrounding high-redshift radio galaxies have revealed that many of these sources are enveloped by extended (up to 100\,kpc) Ly\,$\alpha$ haloes (e.g. \citealp{dey05,wei05,reu03b,vil03}). Others have been shown to be immersed in massive reservoirs of molecular gas and dust (e.g. \citealp{ste03,reu04,pap00,pap01}). These observations are all consistent with a scenario in which radio jets in high-redshift radio galaxies are propagating through denser environments than at low redshift. }\newline

\item {In addition, radio imaging of some high-redshift radio galaxies reveal unusual knotty radio emission along the radio axis \citep{car97,pen00a}. This has been interpreted as evidence of ``frustrated jets'' propagating through a medium which is dense and clumpy on scales of between 50 and 100~kpc.}\newline

\item{And finally, the total (but unvirialised) mass contained within high-redshift protoclusters is estimated to be in the range $2-9\times10^{14}$~\mdot\ \citep{venthesis05}, similar to the virialised masses of the richest present-day galaxy clusters.}
\end{enumerate}

We have recently begun an observational campaign to test our hypothesis that high-redshift USS radio galaxies are located in environments with higher gas densities compared with their low-redshift FR\,II counterparts. This remains a work in progress. The results of our investigation will be given elsewhere. 

\section{Concluding Remarks}\label{summary}
In this paper we have presented multi-frequency radio observations for a sample of 37 steep-spectrum radio galaxies. The sample was originally designed to find distant radio galaxies by exploiting a well-known correlation between redshift and radio spectral index. We determined rest-frame SEDs for the sources in our sample, which cover a broad range of redshifts from $0.1<z<4.0$. We have found that 34 (89\%) of the SEDs are well-characterised by a single power law, with the remaining 4 (11\%) sources showing evidence for spectral flattening toward higher frequencies. This result appears in stark contrast to complete samples of radio sources in which approximately 70\% of SEDs progressively steepen toward high frequencies. We have begun investigating whether differences in the sample selection frequencies, flux density limits or spectral index thresholds are responsible for these differences. We also note that the Malmquist bias inherent in our sample currently prevents us from separating out the contribution from a luminosity-spectral index correlation. \newline 

Coupled to the rarity of USS sources in nearby GHz-selected samples, we are not persuaded that a k-correction plays a dominant role in our SUMSS-NVSS USS sample. Instead, we believe that USS HzRGs are intrinsically steeper than their nearby counterparts. Since the surface density of $z>3$ radio galaxies in our sample is up to three times higher than in lower-frequency-selected USS samples, and since the derived space density of $z>3$ radio galaxies in our sample is consistent with representing the majority of all powerful radio galaxies at high redshift (paper~II), we speculate that a mechanism other than a k-correction or luminosity-induced correlation is driving the $z-\alpha$ correlation. \newline

There is a well-known trend between the spectral index and the ambient surroundings of nearby, low-luminosity radio galaxies, in the sense that the steepest spectrum sources reside at the centre of the richest galaxy clusters. By analogy, the $z-\alpha$ correlation could be driven by evolution in the richness of the environment of radio galaxies {\textit{if}} a higher fraction of radio galaxies reside within cluster-like ambient densities as a function of redshift. Further investigations are now underway to test this hypothesis. 

\section{Acknowledgements}
We thank Ann Burgess for her spectral fitting routine, Katherine Blundell, Brendon Brewer and Thomas Mauch for advice and suggestions, and the anonymous referee for meticulously proof-reading the manuscript. We gratefully acknowledge the use of Edward (Ned) Wright's online Cosmology Calculator. The Australia Telescope Compact Array is part of the Australia Telescope which is funded by the Commonwealth of Australia for operation as a National Facility managed by CSIRO. 

\small
 
\bibliography{mnemonic,mnemonic-simple,bibliography}

\begin{thebibliography}{}

\bibitem[\protect\citeauthoryear{{Athreya} \& {Kapahi}}{{Athreya} \&
  {Kapahi}}{1998}]{ath98b}
{Athreya} R.~M.,  {Kapahi} V.~K.,  1998, Journal of Astrophysics and Astronomy,
  19, 63

\bibitem[\protect\citeauthoryear{{Baldwin} \& {Scott}}{{Baldwin} \&
  {Scott}}{1973}]{bal73}
{Baldwin} J.~E.,  {Scott} P.~F.,  1973, MNRAS, 165, 259

\bibitem[\protect\citeauthoryear{{Best}}{{Best}}{2000}]{bes00}
{Best} P.~N.,  2000, MNRAS, 317, 720

\bibitem[\protect\citeauthoryear{{Best}, {Lehnert}, {Miley} \& {R{\"
  o}ttgering}}{{Best} et~al.}{2003}]{bes03}
{Best} P.~N.,  {Lehnert} M.~D.,  {Miley} G.~K.,    {R{\" o}ttgering} H.~J.~A.,
  2003, MNRAS, 343, 1

\bibitem[\protect\citeauthoryear{{Blumenthal} \& {Miley}}{{Blumenthal} \&
  {Miley}}{1979}]{blu79}
{Blumenthal} G.,  {Miley} G.,  1979, A\&A, 80, 13

\bibitem[\protect\citeauthoryear{{Blundell}, {Rawlings}, {Eales}, {Taylor} \&
  {Bradley}}{{Blundell} et~al.}{1998}]{blu98}
{Blundell} K.~M.,  {Rawlings} S.,  {Eales} S.~A.,  {Taylor} G.~B.,    {Bradley}
  A.~D.,  1998, MNRAS, 295, 265

\bibitem[\protect\citeauthoryear{{Blundell}, {Rawlings} \&
  {Willott}}{{Blundell} et~al.}{1999}]{blu99a}
{Blundell} K.~M.,  {Rawlings} S.,    {Willott} C.~J.,  1999, AJ, 117, 677

\bibitem[\protect\citeauthoryear{{Bornancini}, {Mart{\'{\i}}nez}, {Lambas}, {de
  Vries}, {van Breugel}, {De Breuck} \& {Minniti}}{{Bornancini}
  et~al.}{2004}]{bor04}
{Bornancini} C.~G.,  {Mart{\'{\i}}nez} H.~J.,  {Lambas} D.~G.,  {de Vries} W.,
  {van Breugel} W.,  {De Breuck} C.,    {Minniti} D.,  2004, AJ, 127, 679

\bibitem[\protect\citeauthoryear{{Carilli}, {Harris}, {Pentericci}, {R{\"
  o}ttgering}, {Miley}, {Kurk} \& {van Breugel}}{{Carilli}
  et~al.}{2002}]{car02xray}
{Carilli} C.~L.,  {Harris} D.~E.,  {Pentericci} L.,  {R{\" o}ttgering}
  H.~J.~A.,  {Miley} G.~K.,  {Kurk} J.~D.,    {van Breugel} W.,  2002, ApJ,
  567, 781

\bibitem[\protect\citeauthoryear{{Carilli}, {Perley}, {Dreher} \&
  {Leahy}}{{Carilli} et~al.}{1991}]{car91}
{Carilli} C.~L.,  {Perley} R.~A.,  {Dreher} J.~W.,    {Leahy} J.~P.,  1991,
  ApJ, 383, 554

\bibitem[\protect\citeauthoryear{{Carilli}, {R{\"o}ttgering}, {Miley},
  {Pentericci} \& {Harris}}{{Carilli} et~al.}{1999}]{car99preprint}
{Carilli} C.~L.,  {R{\"o}ttgering} H.~J.~A.,  {Miley} G.~K.,  {Pentericci}
  L.~H.,    {Harris} D.~E.,  1999, in {R{\"o}ttgering} H.~J.~A.,  {Best} P.~N.,
    {Lehnert} M.~D.,  eds, The Most Distant Radio Galaxies {Radio Observations
  of High Redshift Radio Galaxies}.
p.~123

\bibitem[\protect\citeauthoryear{{Carilli}, {R{\"o}ttgering}, {van Ojik},
  {Miley} \& {van Breugel}}{{Carilli} et~al.}{1997}]{car97}
{Carilli} C.~L.,  {R{\"o}ttgering} H.~J.~A.,  {van Ojik} R.,  {Miley} G.~K.,
  {van Breugel} W.~J.~M.,  1997, ApJS, 109, 1

\bibitem[\protect\citeauthoryear{{Carilli} \& {Taylor}}{{Carilli} \&
  {Taylor}}{2002}]{car02araa}
{Carilli} C.~L.,  {Taylor} G.~B.,  2002, ARA\&A, 40, 319

\bibitem[\protect\citeauthoryear{{Chambers}, {Miley} \& {van
  Breugel}}{{Chambers} et~al.}{1990}]{cha90}
{Chambers} K.~C.,  {Miley} G.~K.,    {van Breugel} W.~J.~M.,  1990, ApJ, 363,
  21

\bibitem[\protect\citeauthoryear{{Cohen}, {R{\" o}ttgering}, {Jarvis}, {Kassim}
  \& {Lazio}}{{Cohen} et~al.}{2004}]{coh04}
{Cohen} A.~S.,  {R{\" o}ttgering} H.~J.~A.,  {Jarvis} M.~J.,  {Kassim} N.~E.,
   {Lazio} T.~J.~W.,  2004, ApJS, 150, 417

\bibitem[\protect\citeauthoryear{{Condon}, {Cotton}, {Greisen}, {Yin},
  {Perley}, {Taylor} \& {Broderick}}{{Condon} et~al.}{1998}]{con98}
{Condon} J.~J.,  {Cotton} W.~D.,  {Greisen} E.~W.,  {Yin} Q.~F.,  {Perley}
  R.~A.,  {Taylor} G.~B.,    {Broderick} J.~J.,  1998, AJ, 115, 1693

\bibitem[\protect\citeauthoryear{{De Breuck}, {Hunstead}, {Sadler},
  {Rocca-Volmerange} \& {Klamer}}{{De Breuck} et~al.}{2004}]{cdb04}
{De Breuck} C.,  {Hunstead} R.~W.,  {Sadler} E.~M.,  {Rocca-Volmerange} B.,
  {Klamer} I.,  2004, MNRAS, 347, 837

\bibitem[\protect\citeauthoryear{{De Breuck}, {Klamer}, {Johnston}, {Hunstead},
  {Bryant}, {Rocca-Volmerange} \& {Sadler}}{{De Breuck} et~al.}{2006}]{cdb06}
{De Breuck} C.,  {Klamer} I.,  {Johnston} H.,  {Hunstead} R.~W.,  {Bryant} J.,
  {Rocca-Volmerange} B.,    {Sadler} E.~M.,  2006, MNRAS, 366, 58

\bibitem[\protect\citeauthoryear{{De Breuck}, {Tang}, {de Bruyn}, {R{\"
  o}ttgering} \& {van Breugel}}{{De Breuck} et~al.}{2002}]{cdb02wish}
{De Breuck} C.,  {Tang} Y.,  {de Bruyn} A.~G.,  {R{\" o}ttgering} H.,    {van
  Breugel} W.,  2002, A\&A, 394, 59

\bibitem[\protect\citeauthoryear{{De Breuck}, {van Breugel}, {R{\" o}ttgering}
  \& {Miley}}{{De Breuck} et~al.}{2000}]{cdb00sample}
{De Breuck} C.,  {van Breugel} W.,  {R{\" o}ttgering} H.~J.~A.,    {Miley} G.,
  2000, A\&AS, 143, 303

\bibitem[\protect\citeauthoryear{{De Breuck}, {van Breugel}, {Stanford}, {R{\"
  o}ttgering}, {Miley} \& {Stern}}{{De Breuck} et~al.}{2002}]{cdb02a}
{De Breuck} C.,  {van Breugel} W.,  {Stanford} S.~A.,  {R{\" o}ttgering} H.,
  {Miley} G.,    {Stern} D.,  2002, AJ, 123, 637

\bibitem[\protect\citeauthoryear{{Dey}, {Bian}, {Soifer}, {Brand}, {Brown},
  {Chaffee}, {Le Floc'h}, {Hill}, {Houck}, {Jannuzi}, {Rieke}, {Weedman},
  {Brodwin} \& {Eisenhardt}}{{Dey} et~al.}{2005}]{dey05}
{Dey} A.,  {Bian} C.,  {Soifer} B.~T.,  {Brand} K.,  {Brown} M.~J.~I.,
  {Chaffee} F.~H.,  {Le Floc'h} E.,  {Hill} G.,  {Houck} J.~R.,  {Jannuzi}
  B.~T.,  {Rieke} M.,  {Weedman} D.,  {Brodwin} M.,    {Eisenhardt} P.,  2005,
  ApJ, 629, 654

\bibitem[\protect\citeauthoryear{{Fanaroff} \& {Riley}}{{Fanaroff} \&
  {Riley}}{1974}]{fan74}
{Fanaroff} B.~L.,  {Riley} J.~M.,  1974, MNRAS, 167, 31P

\bibitem[\protect\citeauthoryear{{Gopal-Krishna}}{{Gopal-Krishna}}{1988}]{gop8%
8}
{Gopal-Krishna} 1988, A\&A, 192, 37

\bibitem[\protect\citeauthoryear{{Gunn}, {Hoessel}, {Westphal}, {Perryman} \&
  {Longair}}{{Gunn} et~al.}{1981}]{gun81}
{Gunn} J.~E.,  {Hoessel} J.~G.,  {Westphal} J.~A.,  {Perryman} M.~A.~C.,
  {Longair} M.~S.,  1981, MNRAS, 194, 111

\bibitem[\protect\citeauthoryear{{Hardcastle}, {Birkinshaw}, {Cameron},
  {Harris}, {Looney} \& {Worrall}}{{Hardcastle} et~al.}{2002}]{har02}
{Hardcastle} M.~J.,  {Birkinshaw} M.,  {Cameron} R.~A.,  {Harris} D.~E.,
  {Looney} L.~W.,    {Worrall} D.~M.,  2002, ApJ, 581, 948

\bibitem[\protect\citeauthoryear{{Jarvis}, {Cruz}, {Cohen}, {R{\" o}ttgering}
  \& {Kassim}}{{Jarvis} et~al.}{2004}]{jar04}
{Jarvis} M.~J.,  {Cruz} M.~J.,  {Cohen} A.~S.,  {R{\" o}ttgering} H.~J.~A.,
  {Kassim} N.~E.,  2004, MNRAS, 355, 20

\bibitem[\protect\citeauthoryear{{Jones} \& {Preston}}{{Jones} \&
  {Preston}}{2001}]{jon01}
{Jones} D.~L.,  {Preston} R.~A.,  2001, AJ, 122, 2940

\bibitem[\protect\citeauthoryear{{Kardashev}}{{Kardashev}}{1962}]{kar62}
{Kardashev} N.~S.,  1962, {Soviet Astronomy}, 6, 317

\bibitem[\protect\citeauthoryear{{Komissarov} \& {Gubanov}}{{Komissarov} \&
  {Gubanov}}{1994}]{kom94}
{Komissarov} S.~S.,  {Gubanov} A.~G.,  1994, A\&A, 285, 27

\bibitem[\protect\citeauthoryear{{Kristian}, {Sandage} \& {Katem}}{{Kristian}
  et~al.}{1974}]{kri74}
{Kristian} J.,  {Sandage} A.,    {Katem} B.,  1974, ApJ, 191, 43

\bibitem[\protect\citeauthoryear{{Kristian}, {Sandage} \& {Katem}}{{Kristian}
  et~al.}{1978}]{kri78}
{Kristian} J.,  {Sandage} A.,    {Katem} B.,  1978, ApJ, 219, 803

\bibitem[\protect\citeauthoryear{{Krolik} \& {Chen}}{{Krolik} \&
  {Chen}}{1991}]{kro91}
{Krolik} J.~H.,  {Chen} W.,  1991, AJ, 102, 1659

\bibitem[\protect\citeauthoryear{{Kuehr}, {Witzel}, {Pauliny-Toth} \&
  {Nauber}}{{Kuehr} et~al.}{1981}]{kue81}
{Kuehr} H.,  {Witzel} A.,  {Pauliny-Toth} I.~I.~K.,    {Nauber} U.,  1981,
  A\&AS, 45, 367

\bibitem[\protect\citeauthoryear{{Kurk}, {R{\" o}ttgering}, {Pentericci},
  {Miley}, {van Breugel}, {Carilli}, {Ford}, {Heckman}, {McCarthy} \&
  {Moorwood}}{{Kurk} et~al.}{2000}]{kur00}
{Kurk} J.~D.,  {R{\" o}ttgering} H.~J.~A.,  {Pentericci} L.,  {Miley} G.~K.,
  {van Breugel} W.,  {Carilli} C.~L.,  {Ford} H.,  {Heckman} T.,  {McCarthy}
  P.,    {Moorwood} A.,  2000, A\&A, 358, L1

\bibitem[\protect\citeauthoryear{{Lacy}, {Hill}, {Kaiser} \& {Rawlings}}{{Lacy}
  et~al.}{1993}]{lac93}
{Lacy} M.,  {Hill} G.~J.,  {Kaiser} M.~E.,    {Rawlings} S.,  1993, MNRAS, 263,
  707

\bibitem[\protect\citeauthoryear{{Laing}, {Riley} \& {Longair}}{{Laing}
  et~al.}{1983}]{lai83}
{Laing} R.~A.,  {Riley} J.~M.,    {Longair} M.~S.,  1983, MNRAS, 204, 151

\bibitem[\protect\citeauthoryear{{Mauch}, {Murphy}, {Buttery}, {Curran},
  {Hunstead}, {Piestrzynski}, {Robertson} \& {Sadler}}{{Mauch}
  et~al.}{2003}]{mau03}
{Mauch} T.,  {Murphy} T.,  {Buttery} H.~J.,  {Curran} J.,  {Hunstead} R.~W.,
  {Piestrzynski} B.,  {Robertson} J.~G.,    {Sadler} E.~M.,  2003, MNRAS, 342,
  1117

\bibitem[\protect\citeauthoryear{{Miley}, {Overzier}, {Tsvetanov}, {Bouwens},
  {Ben{\'{\i}}tez}, {Blakeslee}, {Ford}, {Illingworth}, {Postman}, {Rosati},
  {Clampin}, {Hartig}, {Zirm}, {R{\" o}ttgering} \& {Venemans}}{{Miley}
  et~al.}{2004}]{mil04}
{Miley} G.~K.,  {Overzier} R.~A.,  {Tsvetanov} Z.~I.,  {Bouwens} R.~J.,
  {Ben{\'{\i}}tez} N.,  {Blakeslee} J.~P.,  {Ford} H.~C.,  {Illingworth} G.~D.,
   {Postman} M.,  {Rosati} P.,  {Clampin} M.,  {Hartig} G.~F.,  {Zirm} A.~W.,
  {R{\" o}ttgering} H.~J.~A.,    {Venemans} B.~P.,  2004, Nat, 427, 47

\bibitem[\protect\citeauthoryear{{Overzier}}{{Overzier}}{2006}]{ove06}
{Overzier} R.~A. e.~a.,  2006, ApJ, 637, 58

\bibitem[\protect\citeauthoryear{{Papadopoulos}, {Ivison}, {Carilli} \&
  {Lewis}}{{Papadopoulos} et~al.}{2001}]{pap01}
{Papadopoulos} P.,  {Ivison} R.,  {Carilli} C.,    {Lewis} G.,  2001, Nat, 409,
  58

\bibitem[\protect\citeauthoryear{{Papadopoulos}, {R{\" o}ttgering}, {van der
  Werf}, {Guilloteau}, {Omont}, {van Breugel} \& {Tilanus}}{{Papadopoulos}
  et~al.}{2000}]{pap00}
{Papadopoulos} P.~P.,  {R{\" o}ttgering} H.~J.~A.,  {van der Werf} P.~P.,
  {Guilloteau} S.,  {Omont} A.,  {van Breugel} W.~J.~M.,    {Tilanus} R.~P.~J.,
   2000, ApJ, 528, 626

\bibitem[\protect\citeauthoryear{{Pedani}}{{Pedani}}{2003}]{ped03}
{Pedani} M.,  2003, New Astron. Rev., 8, 805

\bibitem[\protect\citeauthoryear{{Pentericci}, {Kurk}, {R{\" o}ttgering},
  {Miley}, {van Breugel}, {Carilli}, {Ford}, {Heckman}, {McCarthy} \&
  {Moorwood}}{{Pentericci} et~al.}{2000}]{pen00b}
{Pentericci} L.,  {Kurk} J.~D.,  {R{\" o}ttgering} H.~J.~A.,  {Miley} G.~K.,
  {van Breugel} W.,  {Carilli} C.~L.,  {Ford} H.,  {Heckman} T.,  {McCarthy}
  P.,    {Moorwood} A.,  2000, A\&A, 361, L25

\bibitem[\protect\citeauthoryear{{Pentericci}, {Van Reeven}, {Carilli}, {R{\"
  o}ttgering} \& {Miley}}{{Pentericci} et~al.}{2000a}]{pen00}
{Pentericci} L.,  {Van Reeven} W.,  {Carilli} C.~L.,  {R{\" o}ttgering}
  H.~J.~A.,    {Miley} G.~K.,  2000a, A\&AS, 145, 121

\bibitem[\protect\citeauthoryear{{Pentericci}, {Van Reeven}, {Carilli}, {R{\"
  o}ttgering} \& {Miley}}{{Pentericci} et~al.}{2000b}]{pen00a}
{Pentericci} L.,  {Van Reeven} W.,  {Carilli} C.~L.,  {R{\" o}ttgering}
  H.~J.~A.,    {Miley} G.~K.,  2000b, A\&AS, 145, 121

\bibitem[\protect\citeauthoryear{{Perryman}, {Lilly}, {Longair} \&
  {Downes}}{{Perryman} et~al.}{1984}]{per84}
{Perryman} M.~A.~C.,  {Lilly} S.~J.,  {Longair} M.~S.,    {Downes} A.~J.~B.,
  1984, MNRAS, 209, 159

\bibitem[\protect\citeauthoryear{{Reuland}, {R{\" o}ttgering}, {van Breugel} \&
  {De Breuck}}{{Reuland} et~al.}{2004}]{reu04}
{Reuland} M.,  {R{\" o}ttgering} H.,  {van Breugel} W.,    {De Breuck} C.,
  2004, MNRAS, 353, 377

\bibitem[\protect\citeauthoryear{{Reuland}, {van Breugel}, {R{\" o}ttgering},
  {de Vries}, {Stanford}, {Dey}, {Lacy}, {Bland-Hawthorn}, {Dopita} \&
  {Miley}}{{Reuland} et~al.}{2003}]{reu03b}
{Reuland} M.,  {van Breugel} W.,  {R{\" o}ttgering} H.,  {de Vries} W.,
  {Stanford} S.~A.,  {Dey} A.,  {Lacy} M.,  {Bland-Hawthorn} J.,  {Dopita} M.,
    {Miley} G.,  2003, ApJ, 592, 755

\bibitem[\protect\citeauthoryear{{Sault}, {Teuben} \& {Wright}}{{Sault}
  et~al.}{1995}]{miriad}
{Sault} R.~J.,  {Teuben} P.~J.,    {Wright} M.~C.~H.,  1995, in {Shaw} R.~A.,
  {Payne} H.~E.,   {Hayes} J.~J.~E.,  eds, ASP Conf. Ser. 77: Astronomical Data
  Analysis Software and Systems IV {A Retrospective View of MIRIAD}.
p.~433

\bibitem[\protect\citeauthoryear{{Sault} \& {Wieringa}}{{Sault} \&
  {Wieringa}}{1994}]{sau94}
{Sault} R.~J.,  {Wieringa} M.~H.,  1994, A\&AS, 108, 585

\bibitem[\protect\citeauthoryear{{Scheuer} \& {Williams}}{{Scheuer} \&
  {Williams}}{1968}]{sch68}
{Scheuer} P.~A.~G.,  {Williams} P.~J.~S.,  1968, ARA\&A, 6, 321

\bibitem[\protect\citeauthoryear{{Slee}, {Siegman} \& {Wilson}}{{Slee}
  et~al.}{1983}]{sle83}
{Slee} O.~B.,  {Siegman} C.~B.,    {Wilson} I.~R.~G.,  1983, Australian Journal
  of Physics, 36, 101

\bibitem[\protect\citeauthoryear{{Slingo}}{{Slingo}}{1974a}]{sli74a}
{Slingo} A.,  1974a, MNRAS, 166, 101

\bibitem[\protect\citeauthoryear{{Slingo}}{{Slingo}}{1974b}]{sli74b}
{Slingo} A.,  1974b, MNRAS, 168, 307

\bibitem[\protect\citeauthoryear{{Smith} \& {Spinrad}}{{Smith} \&
  {Spinrad}}{1980}]{smi80}
{Smith} H.~E.,  {Spinrad} H.,  1980, PASP, 92, 553

\bibitem[\protect\citeauthoryear{{Spergel}, {Verde}, {Peiris}, {Komatsu},
  {Nolta}, {Bennett}, {Halpern}, {Hinshaw}, {Jarosik}, {Kogut}, {Limon},
  {Meyer}, {Page}, {Tucker}, {Weiland}, {Wollack} \& {Wright}}{{Spergel}
  et~al.}{2003}]{spe03}
{Spergel} D.~N.,  {Verde} L.,  {Peiris} H.~V.,  {Komatsu} E.,  {Nolta} M.~R.,
  {Bennett} C.~L.,  {Halpern} M.,  {Hinshaw} G.,  {Jarosik} N.,  {Kogut} A.,
  {Limon} M.,  {Meyer} S.~S.,  {Page} L.,  {Tucker} G.~S.,  {Weiland} J.~L.,
  {Wollack} E.,    {Wright} E.~L.,  2003, ApJS, 148, 175

\bibitem[\protect\citeauthoryear{{Spinrad}}{{Spinrad}}{1982}]{spin82}
{Spinrad} H.,  1982, PASP, 94, 397

\bibitem[\protect\citeauthoryear{{Spinrad} \& {Djorgovski}}{{Spinrad} \&
  {Djorgovski}}{1984}]{spin84}
{Spinrad} H.,  {Djorgovski} S.,  1984, ApJ, 285, L49

\bibitem[\protect\citeauthoryear{{Spinrad}, {Marr}, {Aguilar} \&
  {Djorgovski}}{{Spinrad} et~al.}{1985}]{spin85}
{Spinrad} H.,  {Marr} J.,  {Aguilar} L.,    {Djorgovski} S.,  1985, PASP, 97,
  932

\bibitem[\protect\citeauthoryear{{Stevens}, {Ivison}, {Dunlop}, {Smail},
  {Percival}, {Hughes}, {R{\" o}ttgering}, {van Breugel} \&
  {Reuland}}{{Stevens} et~al.}{2003}]{ste03}
{Stevens} J.~A.,  {Ivison} R.~J.,  {Dunlop} J.~S.,  {Smail} I.~R.,  {Percival}
  W.~J.,  {Hughes} D.~H.,  {R{\" o}ttgering} H.~J.~A.,  {van Breugel} W.~J.~M.,
     {Reuland} M.,  2003, Nat, 425, 264

\bibitem[\protect\citeauthoryear{{Stickel}, {Meisenheimer} \&
  {Kuehr}}{{Stickel} et~al.}{1994}]{sti94}
{Stickel} M.,  {Meisenheimer} K.,    {Kuehr} H.,  1994, A\&AS, 105, 211

\bibitem[\protect\citeauthoryear{{Tielens}, {Miley} \& {Willis}}{{Tielens}
  et~al.}{1979}]{tie79}
{Tielens} A.~G.~G.~M.,  {Miley} G.~K.,    {Willis} A.~G.,  1979, A\&AS, 35, 153

\bibitem[\protect\citeauthoryear{{Venemans}}{{Venemans}}{2005}]{venthesis05}
{Venemans} B.,  2005, PhD thesis, {Universiteit Leiden}

\bibitem[\protect\citeauthoryear{{Venemans}, {Kurk}, {Miley}, {R{\"
  o}ttgering}, {van Breugel}, {Carilli}, {De Breuck}, {Ford}, {Heckman},
  {McCarthy} \& {Pentericci}}{{Venemans} et~al.}{2002}]{ven02}
{Venemans} B.~P.,  {Kurk} J.~D.,  {Miley} G.~K.,  {R{\" o}ttgering} H.~J.~A.,
  {van Breugel} W.,  {Carilli} C.~L.,  {De Breuck} C.,  {Ford} H.,  {Heckman}
  T.,  {McCarthy} P.,    {Pentericci} L.,  2002, ApJ, 569, L11

\bibitem[\protect\citeauthoryear{{Venemans}, {R{\" o}ttgering}, {Overzier},
  {Miley}, {De Breuck}, {Kurk}, {van Breugel}, {Carilli}, {Ford}, {Heckman},
  {McCarthy} \& {Pentericci}}{{Venemans} et~al.}{2004}]{ven04}
{Venemans} B.~P.,  {R{\" o}ttgering} H.~J.~A.,  {Overzier} R.~A.,  {Miley}
  G.~K.,  {De Breuck} C.,  {Kurk} J.~D.,  {van Breugel} W.,  {Carilli} C.~L.,
  {Ford} H.,  {Heckman} T.,  {McCarthy} P.,    {Pentericci} L.,  2004, A\&A,
  424, L17

\bibitem[\protect\citeauthoryear{{Villar-Mart{\'{\i}}n}, {Vernet}, {di Serego
  Alighieri}, {Fosbury}, {Humphrey} \& {Pentericci}}{{Villar-Mart{\'{\i}}n}
  et~al.}{2003}]{vil03}
{Villar-Mart{\'{\i}}n} M.,  {Vernet} J.,  {di Serego Alighieri} S.,  {Fosbury}
  R.,  {Humphrey} A.,    {Pentericci} L.,  2003, MNRAS, 346, 273

\bibitem[\protect\citeauthoryear{{Weidinger}, {M{\o}ller}, {Fynbo} \&
  {Thomsen}}{{Weidinger} et~al.}{2005}]{wei05}
{Weidinger} M.,  {M{\o}ller} P.,  {Fynbo} J.~P.~U.,    {Thomsen} B.,  2005,
  A\&A, 436, 825

\bibitem[\protect\citeauthoryear{{Willott}, {Rawlings}, {Jarvis} \&
  {Blundell}}{{Willott} et~al.}{2003}]{wil03a}
{Willott} C.~J.,  {Rawlings} S.,  {Jarvis} M.~J.,    {Blundell} K.~M.,  2003,
  MNRAS, 339, 173

\end{thebibliography}
 
\bibliographystyle{mn2e}

\bsp
 
\label{lastpage}

\end{document}